\documentclass[prb,showpacs,twocolumn,superscriptaddress]{revtex4}

\usepackage{amsmath,graphicx,latexsym,bm}

\begin{document}

\title{Flexoelectricity from density-functional perturbation theory}

\author{Massimiliano Stengel}
\affiliation{ICREA - Instituci\'o Catalana de Recerca i Estudis Avan\c{c}ats, 08010 Barcelona, Spain}
\affiliation{Institut de Ci\`encia de Materials de Barcelona 
(ICMAB-CSIC), Campus UAB, 08193 Bellaterra, Spain}

\date{\today}

\begin{abstract} 
We derive the complete flexoelectric tensor, including electronic and 
lattice-mediated effects, of an arbitrary insulator in terms of the 
microscopic linear response of the crystal to atomic displacements.
The basic ingredient, which can be readily calculated from first principles
in the framework of density-functional perturbation theory, is
the quantum-mechanical probability current response to a long-wavelength 
acoustic phonon.
Its second-order Taylor expansion in the  wavevector ${\bf q}$ around the
$\Gamma$ (${\bf q}=0$) point in the Brillouin zone naturally yields 
the flexoelectric tensor.
At order one in ${\bf q}$ we recover Martin's theory of piezoelectricity 
[R. M. Martin, Phys. Rev. B \textbf{5}, 1607 (1972)],
thus providing an alternative derivation thereof.
To put our derivations on firm theoretical grounds, we perform a thorough
analysis of the nonanalytic behavior of the dynamical matrix and other
response functions in a vicinity of $\Gamma$.
Based on this analysis, we find that there is an ambiguity in the specification
of the ``zero macroscopic field'' condition in the flexoelectric case;
such arbitrariness can be related to an analytic band-structure term,
in close analogy to the theory of deformation potentials.
As a byproduct, we derive a rigorous generalization of the 
%lattice-dynamical theory of Pick, Cohen and Martin [R. M. Pick, 
%M. H. Cohen and R. M. Martin, Phys. Rev. B {\bf 1}, 910 (1970)]
Cochran-Cowley
formula [W. Cochran and R. A. Cowley, J. Phys. Chem. Solids ${\bf 23}$,
447 (1962)] 
to higher orders in ${\bf q}$. This can be 
of great utility in building reliable atomistic models of electromechanical 
phenomena, as well as for improving the accuracy of the calculation of phonon 
dispersion curves.
Finally, we discuss the physical interpretation of the various contributions 
to the flexoelectric response, either in the static or dynamic regime, and
we relate our findings to earlier theoretical works on the subject.

\end{abstract}

 %Lattice dynamics: first-principles theory
\maketitle

%%%% Introduction %%%

\section{Introduction}

Flexoelectricity, the electric polarization linearly induced by
an inhomogeneous deformation,~\cite{Kogan} has become a popular topic in 
material science during the past few years~\cite{Cross,Pavlo,Noh_flexo,Gustau1,Gustau2,advmat,pavlo_review}.
The interest is motivated by the universality of the flexoelectric 
effect, which, unlike piezoelectricity, is present in insulators 
of any symmetry and composition (including simple solids such as 
crystalline Si and NaCl).~\footnote{We exclude from this discussion 
  phenomena that occur in biological systems and liquid crystals, of slightly different 
  nature.} 
While flexoelectricity is not a new discovery
(the effect was predicted by Kogan in 1964~\cite{Kogan}; the 
bending of a parallel-plate capacitor induced by an applied voltage
was experimentally demonstrated in 1968 by Bursian {\em et al.}~\cite{bursian}),
it has traditionally been regarded as a very weak effect, hardly 
detectable in macroscopic samples.
Only in the past few years research has taken off on this front, thanks
to the breathtaking progress in the design and control of nanoscale
structures. 
From an application point of view, reducing the size of the active elements
is crucial to obtaining a sufficiently large response:
The uniform strain gradient that can be sustained by a sample before 
material failure is \emph{inversely proportional} to its lateral dimensions.
Ironically, in the context of perovskite thin films,
%
%As the flexoelectric response is known to be highest in materials
%with high static permittivity~\cite{Tagantsev,ma/cross-02}, thin epitaxial films
%of ferroelectric perovskites have naturally received a lot of 
%attention.
%
%There are historical reasons as well: 
strain gradients (e.g. occurring during epitaxial growth) have long been regarded 
as harmful to the operation of ferroelectric memories,~\cite{Gustau-05,Dawber:2005}  
and only later explored as a potentially useful functional property.
%
%While the flexoelectric response is tiny in macroscopic samples, it 
%can be enhanced dramatically at the nanoscale~\cite{Noh_flexo}, where 
%huge strain gradients can occur thanks to the reduced feature size~\cite{Cross}.
%
Several recent experimental breakthroughs~\cite{Pavlo,Gustau1,Gustau2} have 
convincingly demonstrated that the effect can indeed be giant~\cite{Noh_flexo} in thin 
films, large enough to rotate~\cite{Gustau1} and/or switch~\cite{Gustau2} ferroelectric 
domains, or to replace conventional piezoelectric materials~\cite{Fousek,Zhu} 
in sensors and transducers.

At the level of the theory, advances have been comparatively slow. 
For a long time, the main reference in the field was the seminal
work by Tagantsev~\cite{Tagantsev}, which focused on lattice-mediated
responses only, and from a phenomenological perspective. Maranganti and
Sharma~\cite{mara/sha} have later applied the method of 
Ref.~\onlinecite{Tagantsev} to the calculation of the flexoelectric 
coefficients in selected materials.
Unfortunately, a considerable spread emerged between the predictions 
of different microscopic models, hence the need for a more fundamental treatment.
It has taken many years before a full first-principles calculation of 
the flexoelectric coefficients was attempted~\cite{Hong-10}. 
More recently Resta~\cite{Resta-10}, and Hong and Vanderbilt~\cite{Hong-11} 
have established the basis for a general formulation of the problem in the 
context of electronic-structure density functional theory, but a 
unified approach, encompassing both electronic and lattice-mediated effects, 
has not emerged yet.
Note that most theoretical treatments to date have defined
the flexoelectric tensors starting from the \emph{real-space
moments} of localized response functions (either atomic forces
induced on neighboring atoms~\cite{Tagantsev} or multipolar
expansions of the charge response to atomic displacements~\cite{Resta-10}).
This is a drawback in the context of electronic-structure calculations,
where working with periodic functions would be preferable,
as it would eliminate the need for expensive supercell geometries.

Given the incomplete state of the theory, there are pressing questions 
coming from the experiments that are still unresolved to date.
First, whether the flexoelectric tensor is a well-defined bulk property 
has been a matter of debate for several years;~\cite{Resta-10,Tagantsev}
consequently, it is currently unclear if it is at all possible to separate the 
surface and bulk contributions in a typical experiment.
Next, it has been pointed out~\cite{Pavlo} that static measurements
alone leave the flexoelectric tensor undetermined -- in order to 
solve for all the independent components one needs to combine static 
with dynamic data. Is it, however, physically justified to ``mix'' the two?
What do we get as a result, a static or a dynamic quantity?
%
%
%of the flexoelectric 
%of the full tensor required a combination of static and dynamic data to 
%deduce all the independent components, it is currently unclear whether flexoelectricity is a static or 
%dynamic effect:
%
Finally, of particular importance in the area of perovskite
oxides (which are by far the best studied and most promising materials
for flexoelectric applications) is the interplay of inhomogeneous 
deformations with the main order parameters (either ferroelectric 
polarization or antiferrodistortive tilts of the oxygen octahedral 
network). 
This has been the subject of several studies in the
context of phenomenological~\cite{morozovska-12a,morozovska-12b}
and effective Hamiltonian~\cite{ponomareva} approaches, but a
systematic way to calculate the coupling coefficients (to be used 
as an input to the higher-level simulations) is still missing.
%how to calculate the coupling constants? there have been attempts 
%(Bellaiche) to include flexoelectricity in Heff, but more 

%While these results are certainly impressive, it is important to keep in mind that the 
In a broader context, it worth noting that the
interest of flexoelectricity is by no means limited to perovskites:
For example, curvature-induced effects are of outstanding relevance in the physics of 
two-dimensional nanostructures~\cite{Ortix} such as $sp^2$-bonded crystals 
(e.g. graphene~\cite{Kalinin_graphene} or boron nitride~\cite{Naumov_BN}).
Also, the electrostatic potential induced by deformation fields is a known 
concern in the performance of optoelectronic quantum-well devices,~\cite{review_semi} 
especially in the promising area of foldable inorganic light-emitting diodes.~\cite{iled}
The theory of absolute deformation potentials,~\cite{vandeWalle} intimately
related to flexoelectricity,~\cite{Resta-DP,Resta-10} is an invaluable
tool in the band-gap engineering of these (and other) semiconductor-based
systems.
Rationalizing these diverse and technologically important phenomena into a 
unified theory would be, of course, highly desirable from a modeling perspective.

Here we show how to consistently address the above issues by 
using density functional perturbation theory~\cite{Baroni/deGironcoli/DalCorso:2001} 
(DFPT) as a methodological framework.
By taking the long-wavelength limit of acoustic phonons, %at zero
%macroscopic field, 
we derive the electromechanical tensors
(both piezoelectric and flexoelectric) in terms of standard
lattice-periodic response functions, which can be readily 
calculated by means of publicly available first-principles codes.
We demonstrate the consistency of our formalism by
rederiving already established results, such as 
Martin's theory~\cite{Martin} of piezoelectricity and existing
theories~\cite{Tagantsev,Resta-10,Hong-11} of flexoelectricity.
To substantiate our arguments, we carefully study the
nonanalyticities, due to the long-range character of
the electrostatic interactions, that plague the 
electronic response functions in the long-wave regime.
%, which stems from the long-range 
%character of the electrostatic interactions.
%
In particular, we devise a rigorous strategy to dealing with this
issue by suppressing the macroscopic (${\bf G}=0$) component of 
the self-consistent electrostatic potential in the linear
response calculations.
We find, however, that such a procedure is not unique --
there is an inherent ambiguity in the specification
of the ``zero macroscopic field'' condition in the flexoelectric
case, which can be traced back to the choice of an arbitrary 
reference energy in the periodic crystal.
We rationalize such ambiguity by establishing a formal link 
between the present theory of flexoelectricity and the preexisting theory
of absolute deformation potentials.~\cite{Resta-DP}
%Distinct ``versions'' of the piezoelectric tensor are anyway related 
%via a trivial band-structure term (i.e. a relative deformation 
%potential). Here we choose the cell-average of the electrostatic
%potential as a reference, both for simplicity and continuity with 
%earlier theoretical works.
%perform a thorough
%analysis of the nonanalytic behavior of the relevant
%response functions in a vicinity of $\Gamma$.
%
%Based on this analysis, we find that there is an ambiguity in the specification
%of the ``zero macroscopic field'' condition in the flexoelectric case;
%such arbitrariness can be related to an analytic band-structure term,
%in close analogy to the theory of deformation potentials.
%
In addition to providing a solid formal basis to our derivations,
our treatment of macroscopic electrostatics also 
yields a rigorous generalization of the Cochran-Cowley 
formula to higher orders in ${\bf q}$, which can be 
of great utility in future lattice-dynamical studies.
Finally, based on our findings, (i) we derive an exact sum rule, relating 
the flexoelectric coefficients to the macroscopic elastic tensor;
(ii) we use such a sum rule to demonstrate that the %re is a unique 
same definition of the flexoelectric tensor is equally well suited to describing
static or dynamic phenomena; (iii) we discuss the physical interpretation of the various
physical contributions to the flexoelectric tensor, relating them to 
earlier first-principles~\cite{Resta-DP} and 
phenomenological~\cite{morozovska-12a,morozovska-12b} studies.
%

%In this context, we show that the flexoelectric
%tensor is not well defined in piezoelectric materials, where it
%suffers from an ambiguity that can be interpreted as an origin dependence.
%dynamical nature of the 
%flexoelectric effect, and the impossibility of defining ``static'' 
%counterparts of the individual tensor elements; (iii) 
%

This work is structured as follows. In Section~\ref{prelim}
we introduce some useful basic concepts of continuum mechanics, and the general 
strategy that we use to attack the flexoelectric problem. In Section~\ref{dfpt}
we introduce the formalism of density-functional perturbation theory,
and the basic ingredients that will be used in the remainder of this work.
In Section~\ref{longw} we proceed to performing the long-wave analysis of an 
acoustic phonon, deriving the piezoelectric and flexoelectric response tensors
in terms of the basic ingredients defined above. 
%respectively to the first and second order terms of the Taylor expansion in the wavevector 
%${\bf q}$.
%
In Section~\ref{elec} we discuss several important properties of the
electronic response functions (polarization and charge density), and use them
to draw a formal connection to Martin's theory of piezoelectricity,~\cite{Martin} 
and to earlier theories~\cite{Hong-11,Resta-10,Tagantsev} of flexoelectricity.
In Section~\ref{cochran} we study the nonanalytic properties of the 
aforementioned response functions, obtaining (among other results) a 
higher-order generalization of the lattice-dynamical theory of Pick,
Cohen and Martin.~\cite{rmm_thesis}
%Cochran-Cowley~\cite{Cochran/Cowley} 
%formula.
%
%In Section~\ref{earlier} we use these properties to 
%
Finally, Section~\ref{discuss} %and Section~\ref{earlier} are 
is devoted to discussing the physical implications of the derived 
formulas, while 
%and to establishing the 
%formal links between the present theory and earlier theoretical works.
%
%Finally, 
in Section~\ref{concl} we briefly summarize our main results and
conclusions.
%
%Finally, in the Appendix, we discuss the nonanalytic behavior
%of the force-constant matrix at the zone center, and relate
%it to the long-range multipolar interactions up to the dipole-octupole 
%term.

\begin{figure}
\begin{tabular}{c @{\hspace{30pt}} c}
\includegraphics[width=1.2in]{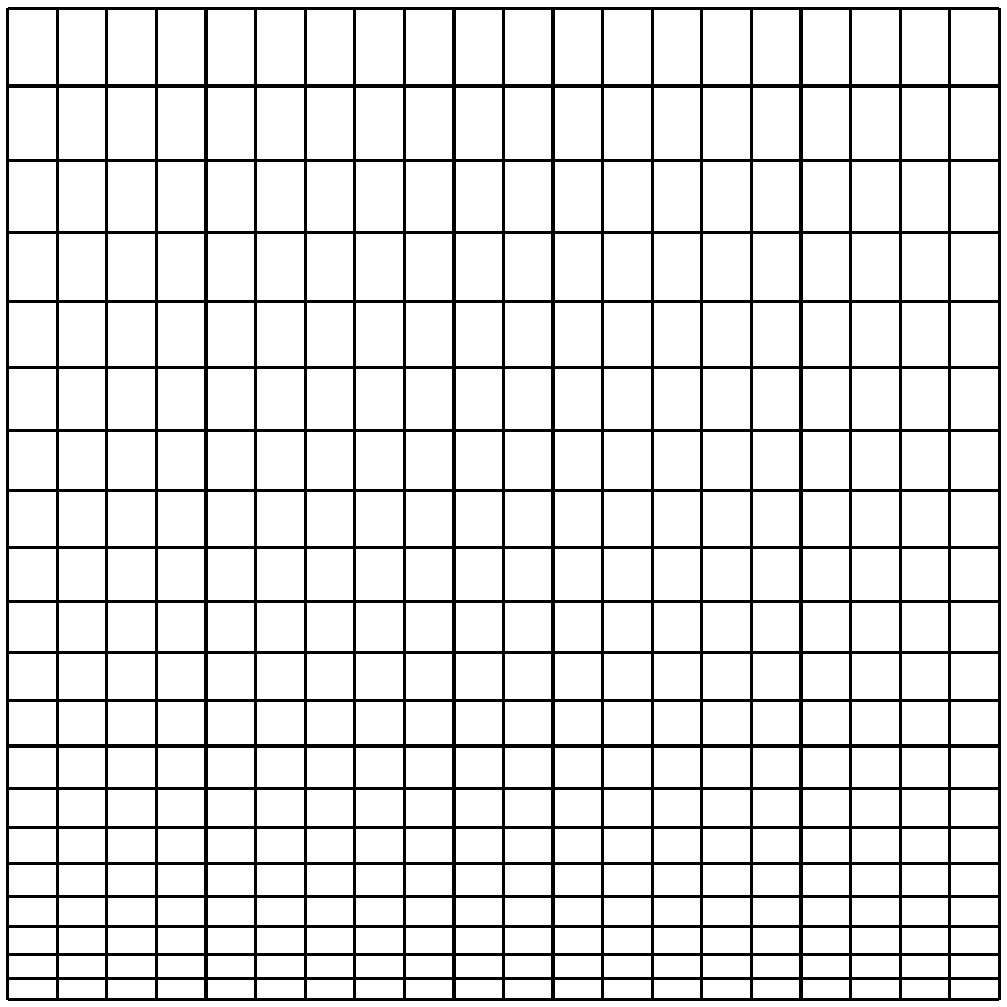} &
\includegraphics[width=1.2in]{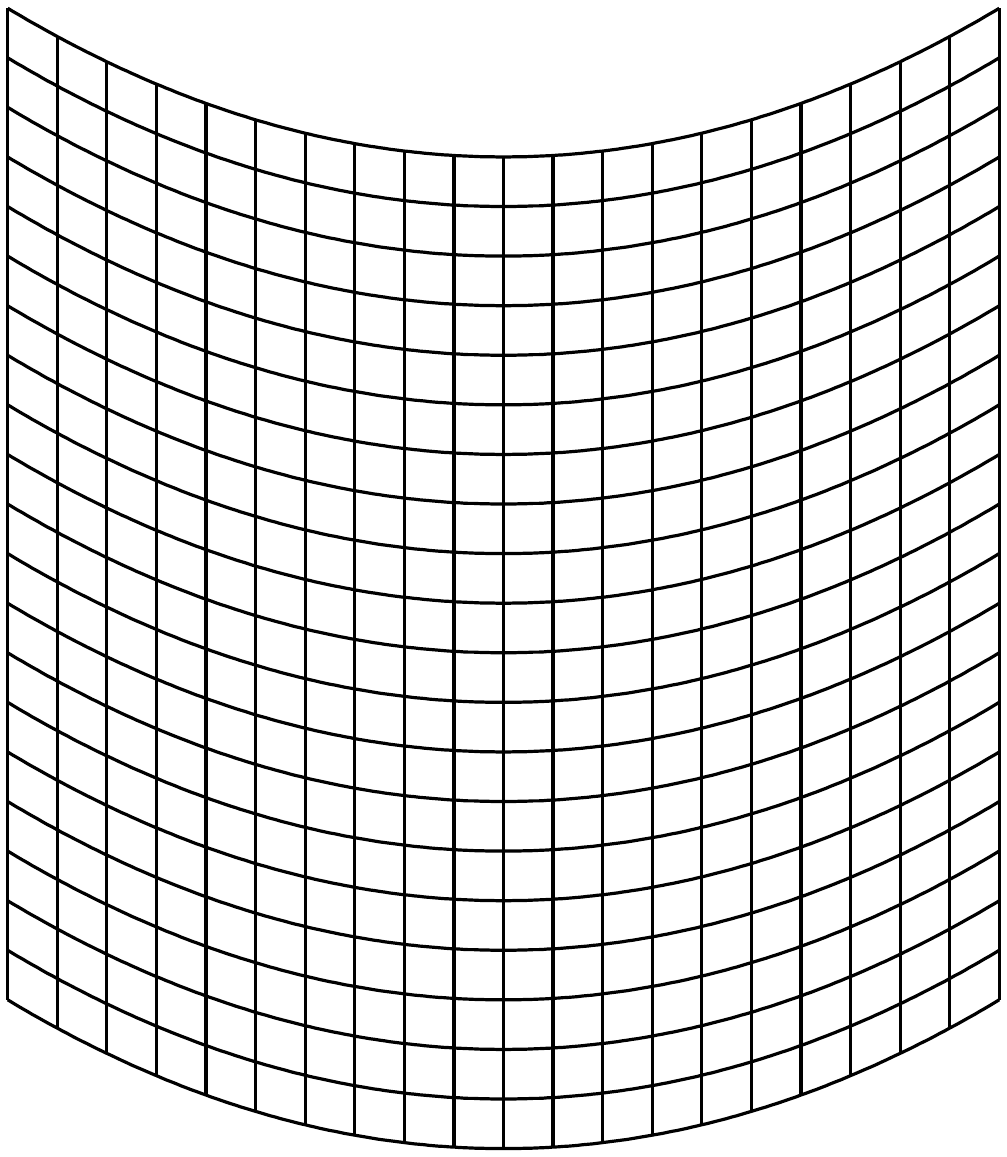} \\
$\eta_{1,11}$ & $\eta_{2,11}$ \\
$\varepsilon_{11,1}$ & $\varepsilon_{12,1} = \varepsilon_{21,1}$ \\[20pt]
%\end{tabular}
%\vspace{20pt}
%\begin{tabular}{c @{\hspace{30pt}} c}
%\vspace{20pt}
\includegraphics[width=1.4in]{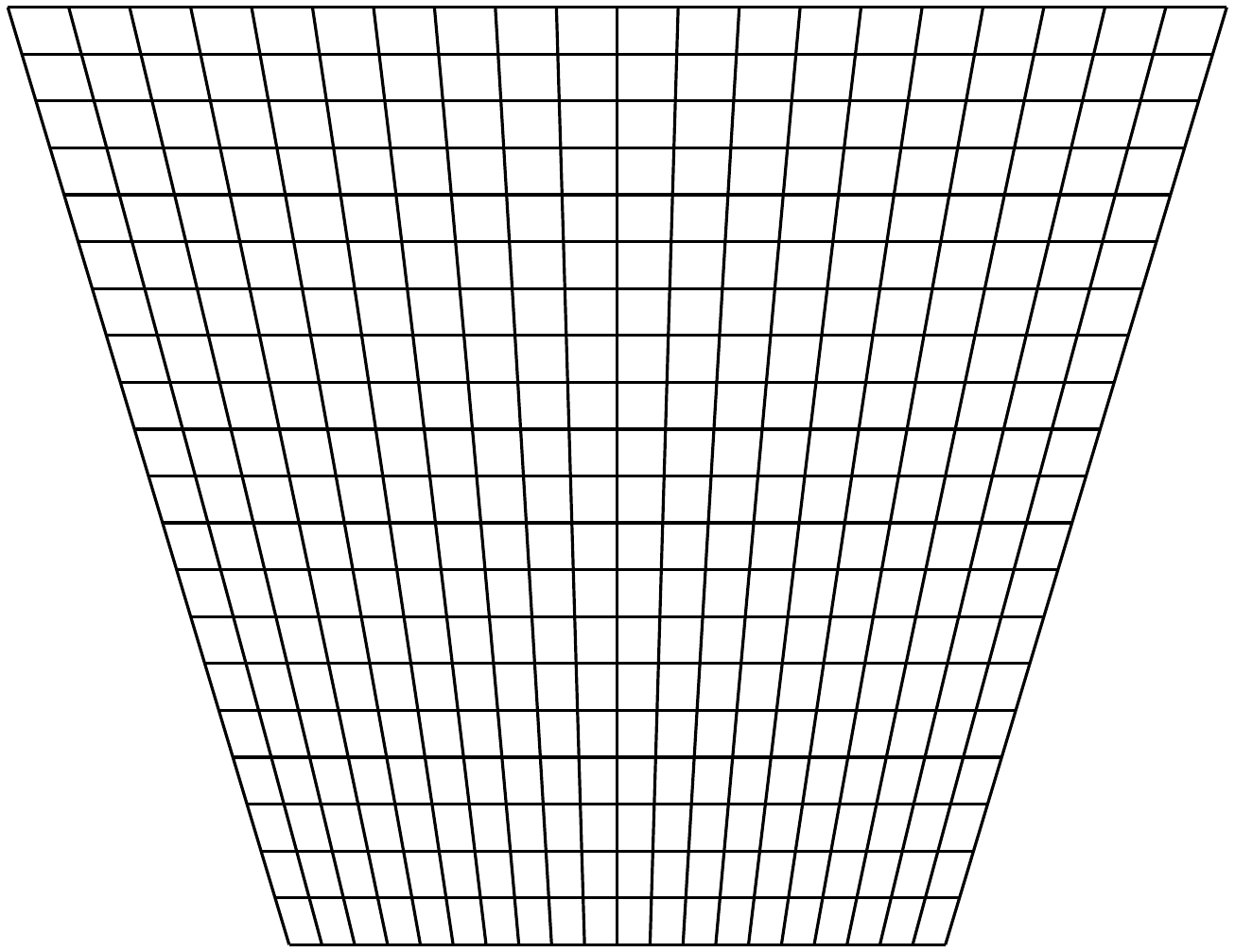} &
\includegraphics[width=1.4in]{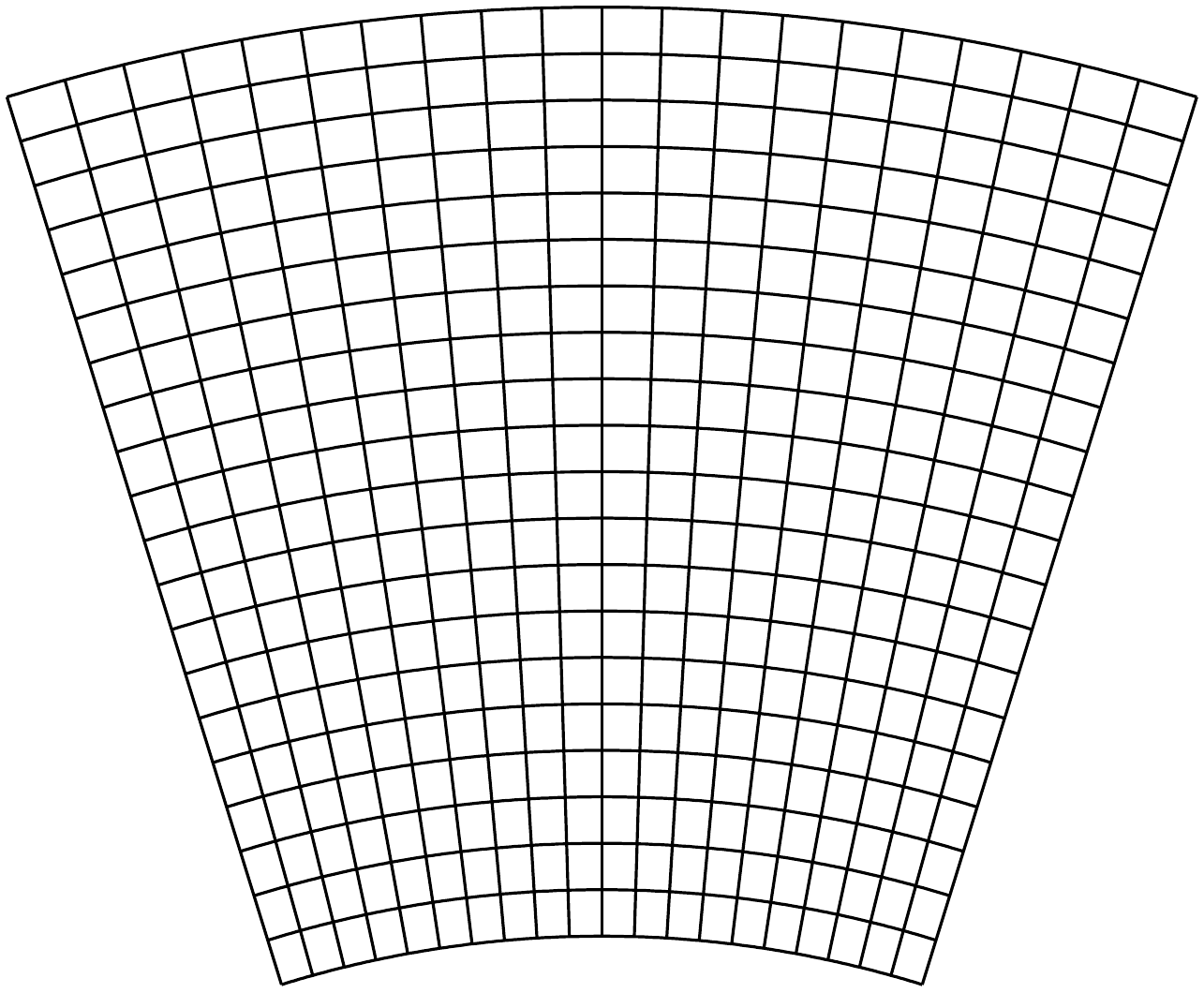} \\
$\eta_{1,12} = \eta_{1,21}$ &
$\varepsilon_{11,2}$ \\
\end{tabular}
\caption{ Individual components of the strain gradient tensor in
a square two-dimensional lattice. 
The ``longitudinal'' ($\varepsilon_{1,11}$), ``shear'' ($\varepsilon_{12,1}$) and
``transverse'' (or ``bending'', $\varepsilon_{11,2}$) components are linearly independent; a fourth pattern 
of less obvious physical interpretation ($\eta_{1,12}$) is a combination of $\varepsilon_{12,1}$ and 
$\varepsilon_{11,2}$. Direction 1 corresponds to the horizontal axis, 2 to the vertical; the polarization vector 
is oriented along 2. \label{fig1} }
\end{figure}

%%%%% Theory %%%%%

\section{Preliminaries}

\label{prelim}

\subsection{Strain and strain gradients}

In continuum mechanics, a deformation can be expressed as a three-dimensional (3D)
vector function, $u_\alpha ({\bf r})$, describing the displacement of a material
point from its reference position at ${\bf r}$ to its current location ${\bf r}'$,
$$%\begin{equation}
r_\alpha'({\bf r}) = r_\alpha + u_\alpha ({\bf r}),
$$%\end{equation}
The \emph{deformation gradient} is defined as
the gradient of $u_\alpha$ taken in the reference configuration,
\begin{equation}
\tilde{\varepsilon}_{\alpha \beta}({\bf r}) = u_{\alpha,\beta}({\bf r}) = \frac{\partial u_\alpha({\bf r})}{\partial r_\beta}.
\label{defgrad}
\end{equation}
$\tilde{\varepsilon}_{\alpha \beta}({\bf r})$ is often indicated in the literature as  
``\emph{unsymmetrized} strain tensor'', as it generally contains a proper strain plus a rotation.
By symmetrizing its indices one can remove the rotational component, % from the deformation gradient, 
thus obtaining the \emph{symmetrized} strain tensor,
$$%\begin{equation}
\varepsilon_{\alpha \beta} = \frac{1}{2} \left( u_{\alpha,\beta} + u_{\beta,\alpha} \right).
$$%\end{equation}
$\varepsilon_{\alpha \beta}$ is a convenient measure of local strain,
as it only depends on \emph{relative} displacements of two adjacent material points,
and not on their absolute translation or rotation with respect to some reference 
configuration.

In this work we shall be primarily concerned with the effects of a 
spatially inhomogeneous strain. The third-rank \emph{strain gradient} tensor 
can be defined in two different ways, both important for the derivations that follow.
The first (\emph{type-I}) form consists in the gradient of the 
\emph{unsymmetrized} strain,
\begin{equation}
\eta_{\alpha, \beta \gamma}({\bf r}) = 
\frac{ \partial \tilde{\varepsilon}_{\alpha \beta}({\bf r})}{\partial r_\gamma} = 
\frac{\partial^2 u_\alpha({\bf r})}{\partial r_\beta \partial r_\gamma}.
\label{eqeta1}
\end{equation}
Note that $\eta_{\alpha, \beta \gamma}$, manifestly invariant upon $\beta \leftrightarrow \gamma$ exchange,
corresponds to the $\nu_{\alpha \beta \gamma }$ tensor of 
Ref.~\onlinecite{Hong-11}, and to the symbol $\partial \epsilon_{ \alpha \beta} / \partial r_\gamma$
of Ref.~\onlinecite{Tagantsev}.
Alternatively, the strain gradient tensor can be defined (\emph{type-II}) as the gradient of 
the \emph{symmetric} strain, $\varepsilon_{\alpha \beta}$,
$$%\begin{equation}
\varepsilon_{\alpha \beta, \gamma}({\bf r}) = \frac{\partial \varepsilon_{\alpha \beta} ({\bf r})}{\partial r_\gamma},
$$%\end{equation}
invariant upon $\alpha \leftrightarrow \beta$ exchange.
It is straightforward to verify that the two tensors contain exactly the
same number of independent entries, and that a one-to-one relationship can
be established to express the former as a function of the latter and viceversa.
For example,
\begin{equation}
\eta_{\alpha, \beta \gamma} = \varepsilon_{\alpha  \beta, \gamma} + 
\varepsilon_{ \gamma \alpha, \beta} - \varepsilon_{ \beta \gamma,\alpha}.
\label{eqeta2}
\end{equation}

In Fig.~\ref{fig1} we illustrate the three independent components of the 
$\eta_{\alpha, \beta \gamma}$ and $\varepsilon_{\alpha  \beta,\gamma}$
tensors on a square two-dimensional (2D) lattice, evidencing analogies 
and differences.
It is clear from the figure that the \emph{longitudinal} and \emph{shear}
components are elementary objects in both type-I and type-II forms.
The main difference between the two representations concerns the third
independent component, which assumes the form of a flat displacement
pattern in the type-I form, and has the more intuitive interpretation
of a pure \emph{bending} (one can show that $\varepsilon_{11,2}= 1/R$,
where $R$ is the curvature radius) in the type-II form; the latter will 
be indicated as \emph{transverse} strain gradient henceforth. 
In fact, these three components of the strain gradient tensor are, by
symmetry, the only types of independent perturbations in a cubic
material, and are therefore very important in the context of flexoelectricity.

\subsection{Long-wavelength acoustic phonons}

A macroscopic strain gradient breaks the translational
symmetry of the crystal lattice. For this reason, the response to 
such a perturbation cannot be straightforwardly represented in periodic 
boundary conditions. This makes the theoretical study of flexoelectricity
more challenging than other forms of electromechanical couplings, e.g.
piezoelectricity.
To circumvent this difficulty, we shall base our analysis on 
the study of long-wavelength acoustic phonons.
These perturbations, while generally incommensurate with the crystal lattice,
can be conveniently described~\cite{Baroni/deGironcoli/DalCorso:2001},
in terms of functions that are lattice-periodic, and therefore are formally and 
computationally very advantageous.

\begin{figure}
\begin{center}
%\begin{tabular}{c @{\hspace{30pt}} c}
\includegraphics[width=2.5in]{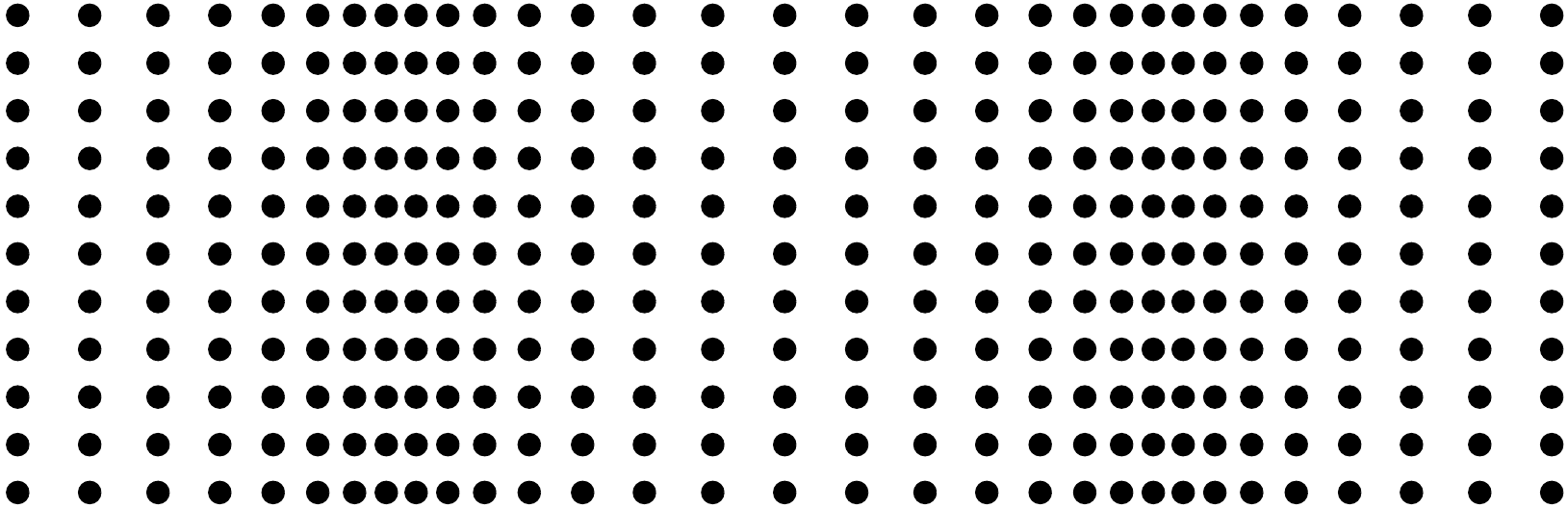} \\
\vspace{20pt}
\includegraphics[width=2.5in]{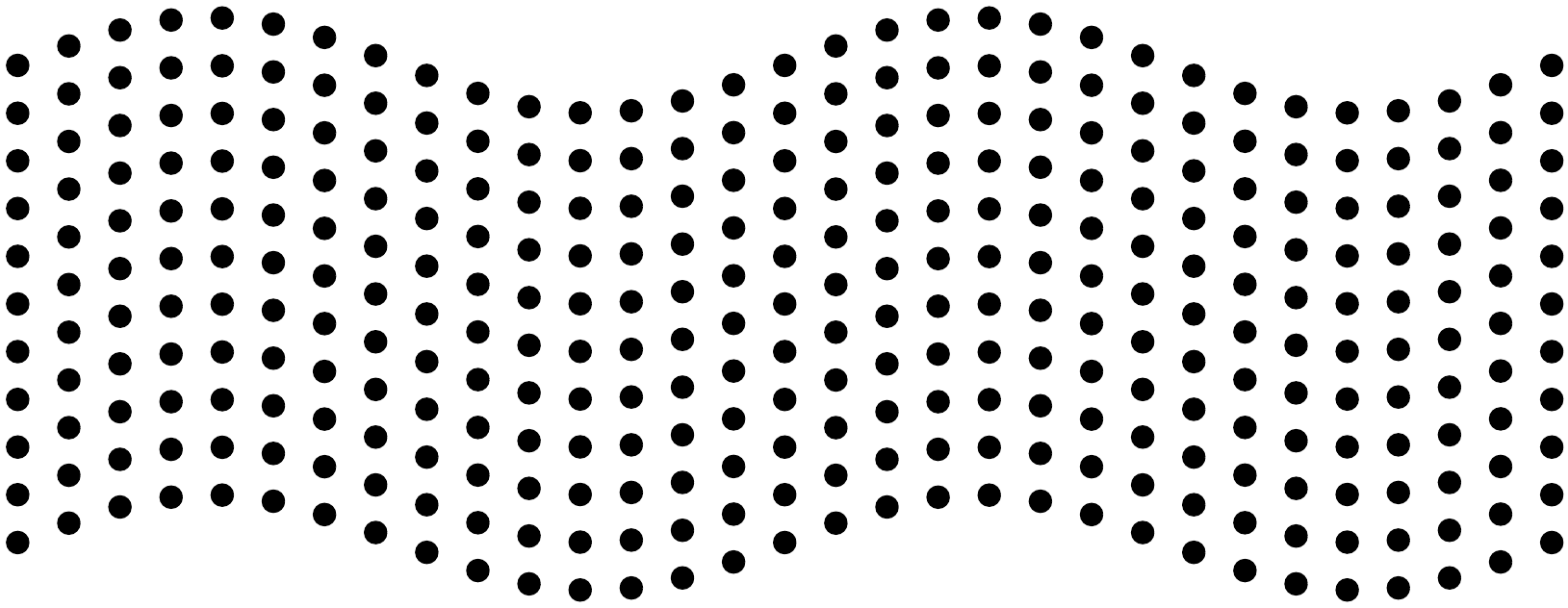} 
%$\eta_{1,11}$ & $\eta_{2,11}$ \\
%$\varepsilon_{11,1}$ & $\varepsilon_{12,1} = \varepsilon_{21,1}$ \\[20pt]
%\end{tabular}
%\vspace{20pt}
%\begin{tabular}{c @{\hspace{30pt}} c}
%\vspace{20pt}
%\includegraphics[width=1in]{eta3.pdf} &
%\includegraphics[width=1in]{eta333a.pdf} \\
%$\eta_{1,12} = \eta_{1,21}$ &
%$\varepsilon_{11,2}$ \\
%\end{tabular}
\end{center}
\caption{ Displacement fields in longitudinal (top) and transversal
(bottom) sound waves. 
%One can easily recognize the longitudinal and transversal
%strain gradient patterns, respectively. 
\label{fig2} }
\end{figure}

The direct relationship between an acoustic phonon and a mechanical deformation
is clear from Fig.~\ref{fig2}: in the longitudinal and transversal waves 
one can visually identify regions of negative and positive strain gradients, 
respectively of the longitudinal and shear type.
Mathematically, this observation can be formalized by writing (at the lowest order in the 
wavevector ${\bf q}$) an acoustic phonon as a homogeneous displacement of every 
material point of the type
$$%\begin{equation}
u_\beta({\bf r},t)  = U_\beta  \, e^{i {\bf q} \cdot {\bf r} - i \omega t},
$$%\end{equation}
where $U_\beta$ is the displacement amplitude and $\omega$ the frequency.
Consider now the microscopic polarization currents (these are due to 
the displacements of the charged particles, electron and nuclei, from their
equilibrium positions) induced by the phonon at the linear-response level,~\footnote{
  We use a tilde symbol in the following equations to indicate the total relaxed-ion 
  polarization.
  This includes the effects of the internal strains that are dynamically 
  produced by the deformation field, $u_\beta({\bf r},t)$. % electronic and
  In the remainder of this work we shall use $P^{\bf q}$, without tilde, to 
  indicate the elementary polarization response function, i.e. in absence of 
  internal strains.
  }
$$%\begin{equation}
\tilde{P}_\alpha ({\bf r}, t) = U_\beta \tilde{P}^{\bf q}_{\alpha \beta}({\bf r})  \, e^{i {\bf q} \cdot {\bf r} - i \omega t}.
$$%\end{equation}
Assuming for the moment that $\tilde{P}^{\bf q}$ is an analytic function of 
${\bf q}$ at the $\Gamma$ point, in a neighborhood of $\Gamma$ (i.e. in the 
long-wavelength regime) we can replace it with its second-order Taylor expansion,
\begin{equation}
\tilde{P}^{\bf q}_{\alpha \beta} \simeq \tilde{P}^{(0)}_{\alpha \beta} -iq_\gamma
\tilde{P}^{(1,\gamma)}_{\alpha \beta} - \frac{q_\gamma q_\lambda}{2} \tilde{P}^{(2,\gamma \lambda)}_{\alpha \beta}.
\label{ptot}
\end{equation}
Now, by applying Eq.~(\ref{defgrad}) and Eq.~(\ref{eqeta1}), we can compute
the local deformation gradient and strain gradient that are associated
with the acoustic phonon,
\begin{eqnarray}
\tilde{\varepsilon}_{\beta \gamma}({\bf r},t) &=& \frac{\partial u_\beta}{\partial r_\gamma} = 
i U_\beta q_\gamma \, e^{i {\bf q} \cdot {\bf r} - i \omega t}, 
\label{nonsymm} \\
\eta_{ \beta, \gamma \lambda}({\bf r},t) &=& 
\frac{\partial^2 u_\beta}{\partial r_\gamma \partial r_\lambda} = 
-U_\beta q_\gamma q_\lambda \, e^{i {\bf q} \cdot {\bf r} - i \omega t}.
\label{eqeta}
\end{eqnarray}
A comparison of Eqs.~(\ref{ptot}), (\ref{nonsymm}) and (\ref{eqeta}) suggests that the polarization 
response to a uniform strain (piezoelectricity) and to a strain-gradient (flexoelectricity) are, 
respectively, related to the first- and second-order Taylor expansion in powers of ${\bf q}$ of the
polarization field produced by a sound wave.
The latter can be computed by working with lattice-periodic functions only,
implying that all the theoretical and computational weaponry developed
so far within periodic boundary conditions (e.g. Bloch theorem, plane wave
basis set, pseudopotentials, etc.) can be proficiently applied to the flexoelectric problem.
This is precisely the approach that we shall take in the remainder of this work.

While this appears conceptually simple, there are a number of important issues that 
need to be addressed before one can establish the formal link between the 
%long-wave analysis of acoustic phonons 
electrical properties of sound waves
and the macroscopic electromechanical tensors.
%a 
%number of important issues need to be addressed.
%
First of all, it is not clear \emph{a priori} whether the above strategy
is even applicable. Long-wavelength phonons are generally accompanied by macroscopic 
electric fields. This is a substantial complication from the operational point
of view: the longitudinal character of the electrostatic screening causes a 
nonanalytic behavior of most response functions (e.g. the atomic eigendisplacements and the
electronic polarization, see Section~\ref{cochran} for a detailed discussion) 
in a vicinity of $\Gamma$, thwarting their expansion in powers of ${\bf q}$.
Whether (and how) these nonanalyticities can be tamed will need to 
be assessed prior to starting the actual derivations.
Second, phonon eigenmodes also contain, in addition to genuine 
macroscopic deformations, translations and rotations 
%[Eq.~(\ref{nonsymm})
%is written in terms of the \emph{unsymmetrized} strain tensor, and the
%induced polarization contains a translational 
%contribution $\tilde{P}^{(0)}_{\alpha \beta}$] 
of a given 
crystal cell with respect to its reference configuration at rest. It will be 
therefore necessary to show that such rototranslations do not contribute to the 
macroscopic electrical response.
Third, a phonon is inherently a dynamic perturbation, and whether the
effects derived for a sound wave are equally applicable to a static
deformation will need to be carefully demonstrated.
In the following Sections we shall first introduce the basic ingredients that
we need in order to derive the total polarization response, Eq.~(\ref{ptot});
next, we shall proceed to the formal derivation of the electromechanical
tensors, and to their validation in relation to the aforementioned 
sources of concern.

\section{Density-functional perturbation theory}

\label{dfpt}

This Section will provide a brief introduction
to the DFPT formalism. This is mainly aimed at specifying the general 
context of our derivations, as well as at pointing out the 
key modifications to the standard approach~\cite{Baroni/deGironcoli/DalCorso:2001}
that are necessary in the context of this work. 
In particular, we shall put the emphasis on the following three
technical points: the treatment of the macroscopic fields;
the definition of the microscopic polarization response;
the practical calculation of the relevant response functions 
by means of publicly available codes.
% (especially regarding
%the proper treatment of the phase factors in the perturbing 
%potential, and the differentiation in ${\bf q}$-space).
%
%; the treatment of the macroscopic fields;
%the 

\subsection{Linear response to monochromatic perturbations}

Our starting point is an insulating crystal, whose equilibrium configuration is described by
the three primitive translation vectors, ${\bf a}_{1,2,3}$, and by a 
basis of $N$ atoms located at positions $\bm{\tau}_\kappa$ ($\kappa=1,\ldots,N$) 
within the primitive unit cell.
Within density-functional theory, the electronic ground state can be written in terms of 
the self-consistent (SCF) Kohn-Sham equation,
$$%\begin{equation}
\hat{H}_{\bf k} |\phi_{n{\bf k}} \rangle = \epsilon_{n{\bf k}} |\phi_{n{\bf k}} \rangle,
$$%\end{equation}
where $\hat{H}_{\bf k}$ is the SCF Hamiltonian at the point ${\bf k}$ in the Brillouin zone, 
and $|\phi_{n{\bf k}} \rangle$ and $\epsilon_{n{\bf k}}$ are respectively the ground-state 
Bloch orbitals and eigenvalues.
In full generality, the Hamiltonian  
$$%\begin{equation}
\hat{H}_{\bf k} = \hat{T}_{\bf k} + \hat{V}^{\rm ext}_{\bf k} + \hat{V}^{\rm Hxc} 
$$%\end{equation}
contains a single-particle kinetic energy operator, $\hat{T}_{\bf k}$, the external
potential of the nuclei, $\hat{V}^{\rm ext}_{\bf k}$ and the Hartree and exchange
and correlation potential, the latter depending self-consistently on the electronic charge 
density $\rho^{\rm el}$,
\begin{equation}
\rho^{\rm el}({\bf r}) =  
-s \frac{\Omega}{(2\pi)^3} \sum_n \int_{\rm BZ} d^3 k \, \phi^*_{n{\bf k}} ({\bf r}) \phi_{n{\bf k}} ({\bf r}).
\end{equation}
($s$ is the occupation of the orbital, equal to 2 if spin pairing is assumed.)
The total charge density, $\rho({\bf r}) = \rho^{\rm el}({\bf r}) + \rho^{\rm ion}({\bf r})$, 
includes the contribution of the nuclear point charges, 
\begin{equation}
\rho^{\rm ion}({\bf r}) = \sum_{l\kappa} Z_\kappa \delta({\bf r-R}_{l\kappa}),
\end{equation}
where $Z_\kappa$ is the bare pseudopotential charge (or the atomic number in
the case of an all-electron description), and $\delta({\bf r-R}_{l\kappa})$ is a
Dirac delta function. 
[Note that ${\bf R}_{l\kappa} = {\bf R}_l + \bm{\tau}_\kappa$ 
is the equilibrium (unperturbed) atomic position in the crystal, and $l$ is a cell 
index.]
% Note that $n({\bf r})$ is a particle density, while $\rho({\bf r})$
%is a charge density, hence the negative sign of $n$ in the above expression.

Consider now a monochromatic perturbation, where the atoms in the
sublattice $\kappa$ undergo a small displacement along $\beta$ of 
the type
\begin{equation}
u_{\kappa \beta}^{l}  = \lambda e^{i {\bf q} \cdot {\bf R}_{l\kappa}}.
\label{elementary}
\end{equation}
%
%~\footnote{
%  Note that the quantities $U_{\kappa \beta}$ are independent of ${\bf q}$, and 
%  therefore represent the basis vectors of the dynamical matrix, 
%  \emph{not} its eigenvectors (phonons).}
The linear response of the crystal to such a perturbation can be readily 
computed in the framework of density-functional perturbation theory~\cite{Baroni/deGironcoli/DalCorso:2001},
by solving the following Sternheimer equation,
%$$
%\left( \hat{H}_{\bf k+q} + \alpha \hat{P}_{\bf k+q} - \epsilon_{n{\bf k}} \right) |\Delta \phi^{{\bf q},\kappa \beta}_{n{\bf k}} \rangle =
%$$
%\begin{equation}
%-\hat{Q}_{\bf k+q} \, \Delta \hat{V}^{{\rm SCF},\kappa \beta}_{\bf k+q,k}  |\phi_{n{\bf k}} \rangle.
%\label{stern}
%\end{equation}
\begin{align}
\left( \hat{H}_{\bf k+q} + \alpha \hat{P}_{\bf k+q} - \epsilon_{n{\bf k}} \right) &
|\Delta \phi^{{\bf q},\kappa \beta}_{n{\bf k}} \rangle = \nonumber \\
-\hat{Q}_{\bf k+q} \, \Delta \hat{V}^{{\rm SCF},\kappa \beta}_{\bf k+q,k} & |\phi_{n{\bf k}} \rangle.
\label{stern}
\end{align}
Here $|\Delta \phi^{{\bf q},\kappa \beta}_{n{\bf k}} \rangle$ are the desired first-order wavefunctions,
$\hat{P}_{\bf k} = \sum_n |\phi_{n{\bf k}} \rangle \langle \phi_{n{\bf k}} |$ and
$\hat{Q}_{\bf k} = \hat{1} -  \hat{P}_{\bf k}$ are
the projection operators on the valence and conduction subspaces, 
and $\Delta \hat{V}^{{\rm SCF},\kappa \beta}_{\bf k+q,k}$ is the sum of the external 
perturbing potential (due to the atomic displacements) and the linear variation 
in the Hxc potential due to the rearrangement of the electron cloud. 
The arbitrary parameter $\alpha$ guarantees 
orthogonality between $|\phi_{n{\bf k+q}} \rangle$ and $|\Delta \phi^{{\bf q},\kappa \beta}_{n{\bf k}} \rangle$
and is otherwise irrelevant.
Note that Eq.~(\ref{stern}) involves lattice-periodic functions only, and thus provides
a convenient route to accessing the relevant response functions at an arbitrary
wavevector ${\bf q}$.

%In order to derive the electrical response to a macroscopic deformation,
In the context of the present work, we need to focus on three basic response 
functions, all of which are linear in the perturbation amplitude $\lambda$.
(To avoid overburdening the notation, from now on we shall omit the ``$\Delta$'' 
prefix whenever the linearity of a given response function with respect to 
$\lambda$ is obvious from the context.)
%~\footnote{We implicitly assume that
%  the trivial displacement of the nuclear point charge is \emph{included}
%  in all the electronic response functions. The only effect on the 
%  macroscopic quantities discussed in this work is the obvious inclusion 
%  of the bare nuclear charge in the Born dynamical tensors.} 
%
The first quantity is the variation of the total charge density, 
$$%\begin{equation}
\frac{d \rho ({\bf r})}{d \lambda} = \rho_{\kappa \beta}^{\bf q} ({\bf r}) e^{i {\bf q \cdot r}}.
$$%\end{equation}
%(We adopt from now on the usual summation convention on repeated indices.)
%
Similar to $\rho({\bf r})$, the cell-periodic function 
$\rho_{\kappa \beta}^{\bf q} ({\bf r}) = 
\rho_{\kappa \beta}^{{\rm el},{\bf q}} ({\bf r}) +
\rho_{\kappa \beta}^{{\rm ion},{\bf q}} ({\bf r})$ can be also 
decomposed into an electronic and a (trivial) ionic contribution, 
%which 
%is readily available as an output of 
\begin{align}
\rho_{\kappa \beta}^{{\rm el},{\bf q}} ({\bf r}) =& - 2s \frac{\Omega}{(2\pi)^3} 
\sum_n \int_{\rm BZ} d^3 k \, \phi^*_{n{\bf k}} ({\bf r}) \, 
\Delta \phi^{{\bf q},\kappa \beta}_{n{\bf k}} ({\bf r}) \nonumber \\ 
\rho_{\kappa \beta}^{{\rm ion},{\bf q}} ({\bf r}) =&  Z_\kappa \sum_{l} 
\left[ -\frac{\partial \delta({\bf r - R}_{l\kappa})}{\partial r_\beta } 
        - i q_\beta \delta({\bf r - R}_{l\kappa}) \right], \nonumber
%
%e^{ i {\bf q} \cdot ({\bf R}_{l\kappa}-{\bf r})}. \nonumber
\end{align}
where $\Delta \phi^{{\bf q},\kappa \beta}_{n{\bf k}}$ is the solution
of Eq.~(\ref{stern}).

%The second term is due to the displacement of the nuclear point charges, and
%
%concerns the ion cores (which are described by point charges) -- its
%cell integral yields $Z_\kappa \delta_{\beta \gamma}$.

%) that can be represented on a 
%reciprocal-space grid,
%\begin{eqnarray}
%\rho_{\kappa \beta}^{\bf q} ({\bf r}) &=& \sum_{\bf G} \tilde{\rho}_{\kappa \beta}^{\bf q} ({\bf G}) e^{i {\bf G \cdot r}}, \\
%\tilde{\rho}_{\kappa \beta}^{\bf q} ({\bf G}) &=& 
%\frac{1}{\Omega} \int_{\rm cell} \rho_{\kappa \beta}^{\bf q} ({\bf r}) e^{-i {\bf G \cdot r}} d^3 r.
%\end{eqnarray}
The second quantity is the \emph{microscopic} polarization response, defined as 
the current density, ${\bf j}({\bf r},t)$, that is linearly induced when the perturbation
is adiabatically switched on via a \emph{time-dependent} parameter $\lambda$,
$$%\begin{equation}
\lambda \rightarrow  \lambda(t), \qquad  j_\alpha ({\bf r},t) = 
\frac{d P_\alpha ({\bf r})}{d \lambda} \dot{\lambda}(t).
$$%\end{equation}
The variation of $P_\alpha ({\bf r})$ can also be written as a cell-periodic
part multiplied by a phase,
$$%\begin{equation}
\frac{d P_\alpha ({\bf r})}{d \lambda} = % U_{\kappa \beta} 
 P_{\alpha \, \kappa \beta}^{\bf q}({\bf r}) e^{i {\bf q \cdot r}},
$$%\end{equation}
and decomposed into an electronic and ionic part, $P_{\alpha \, \kappa \beta}^{\bf q}
= P_{\alpha \, \kappa \beta}^{{\rm el},{\bf q}} + P_{\alpha \, \kappa \beta}^{{\rm ion},{\bf q}}$. 
The ionic contribution has again a simple expression,
\begin{equation}
P_{\alpha \, \kappa \beta}^{{\rm ion},{\bf q}}({\bf r})  = Z_\kappa \delta_{\alpha \beta} 
\sum_l \delta({\bf r - R}_{l\kappa}),
\end{equation}
independent of ${\bf q}$. It is easy to verify that 
$$\nabla \cdot \left( {\bf P}_{\kappa \beta}^{{\rm ion},{\bf q}} e^{i {\bf q \cdot r}} \right) =
- \rho^{{\rm ion},{\bf q}} e^{i {\bf q \cdot r}}.$$
The electronic contribution, $P_{\alpha \, \kappa \beta}^{{\rm el},{\bf q}}$, is a new 
quantity that is not part of currently available DFPT implementations. Further details on
how it can be calculated in practice are provided in Section~\ref{polarization}.
%we describe a viable route to compute it in practice.
%
%Even if the microscopic polarization $P_{\alpha \, \kappa \beta}^{\bf q}({\bf r})$ has seldom 
%been considered in the first-principles literature to date (Ref.~\onlinecite{puma} is
%one exception) this quantity is physically well defined (the probability current is a 
%fundamental quantum-mechanical observable). In Section~\ref{polarization} we shall 
%outline a general procedure to compute $P_{\alpha \, \kappa \beta}^{\bf q}({\bf r})$ 
%based on the current density operator.
%, to the author's knowledge . Further details about how this quantity can 
%be calculated in practice will be provided in a forthcoming publication.
%

The third and last basic response function that we shall consider in this work 
is the force induced on the atom $l\kappa'$ along $\alpha$, 
$d f^l_{\kappa' \alpha} / d \lambda$, whose cell-periodic part is 
the ${\bf q}$-space force constant matrix, $\Phi$,
\begin{equation}
\frac{d f^l_{\kappa' \alpha}}{d \lambda} = 
-\Phi_{\kappa' \alpha \, \kappa \beta}^{\bf q}  e^{i {\bf q} \cdot {\bf R}_{l\kappa'}}.
\label{force}
\end{equation}
This is, of course, a central quantity in DFPT, and can be readily 
computed following the prescriptions of Refs.~\onlinecite{Baroni/deGironcoli/DalCorso:2001},
\onlinecite{Gonze} and \onlinecite{Gonze/Lee}. (With respect to
the procedure described in these works, note that there is an important 
subtlety related to the \emph{phase} of the perturbing potentials and 
response functions, which we shall discuss in Section~\ref{secphase}.)

\subsection{Taylor expansion in a vicinity of $\Gamma$}

As we shall see in Section~\ref{longw}, in order to obtain 
the long-wave limit of the polarization response to a phonon, 
Eq.~(\ref{ptot}), one needs to evaluate a number of 
intermediate quantities.
%To calculate the response to a macroscopic deformation, we 
%will need to taking the long-wave limit of an acoustic phonon.
%
%The long-wave limit, in turn, consists in 
These are the lowest terms of the Taylor expansion (in ${\bf q}$ space) 
of the fundamental response functions 
$\Phi^{\bf q}$, $\rho^{\bf q}$ and ${\bf P}^{\bf q}$ introduced earlier
in this Section.
%
%
%
%with a 
%$\kappa$-independent displacement amplitude, $U_{\kappa \beta} = U_\beta$.
%
%It is expedient to introduce additional symbols for the charge and
%polarization response to such a perturbation,
%\begin{equation}
%P^{\bf q}_{\alpha \beta}({\bf r}) = \sum_\kappa P^{\bf q}_{\alpha \, \kappa \beta}({\bf r}), 
%\qquad \rho^{\bf q}_\beta({\bf r}) = \sum_\kappa \rho^{\bf q}_{\kappa \beta}({\bf r}).
%\end{equation}
%
%
%
Unfortunately, these functions are plagued by a nonanalytic behavior
at $\Gamma$, %(see the Appendix for a detailed discussion), 
which implies that their direct Taylor expansion is not feasible.
The nonanaliticity is related to
%Before processing $\rho_{\beta}^{\bf q} ({\bf r})$ any further, we need to 
%deal with 
the macroscopic electric fields that occur in response to
the perturbation.
% a 
%long-wavelength perturbation at a given ${\bf q}$.
%
To clarify this point, it is useful to write the induced electric field as
$$%\begin{equation}
\frac{d \mathcal{E}_\alpha ({\bf r})}{d \lambda} = 
\mathcal{E}_{\alpha \, \kappa \beta}^{\bf q}({\bf r}) \, e^{i {\bf q \cdot r}}.
$$%\end{equation}
After expanding the cell-periodic part, $\mathcal{E}_{\alpha \, \kappa \beta}^{\bf q}({\bf r})$ 
into its reciprocal-space coefficients (indexed by the reciprocal-lattice vectors ${\bf G}$), 
it becomes apparent that the ${\bf G}=0$ term (indicated by a wide bar symbol),
$$%\begin{equation}
 \overline{\mathcal{E}}_{\alpha,\kappa \beta}^{\bf q} = 
\frac{1}{\Omega} \int_{\rm cell} \mathcal{E}_{\alpha, \kappa \beta}^{\bf q}({\bf r}) \, d^3 r,
$$%\end{equation}
which is purely longitudinal, is problematic for ${\bf q} \rightarrow 0$.
In fact, one can show (a rigorous derivation is provided in Section~\ref{secphon}) 
that, at order zero in ${\bf q}$, $\overline{\mathcal{E}}_{\alpha, \kappa \beta}^{\bf q}$
is a direction-dependent constant,
\begin{equation}
\overline{\mathcal{E}}_{\alpha, \kappa \beta}^{{\bf q} \rightarrow 0} \sim
-\frac{4 \pi}{\Omega} \, \hat{q}_\alpha \frac{ ( \hat{\bf q} \cdot {\bf Z}_\kappa )_\beta} %\hat{q}_\alpha Z^*_{\kappa,\alpha \beta}} 
{\hat{\bf q} \cdot \bm{\epsilon} \cdot \hat{\bf q}},
\label{nonanalytic}
\end{equation}
where $\bm{\epsilon}$ is the macroscopic dielectric tensor, 
$Z^*_{\kappa,\alpha \beta}$ is the Born dynamical charge tensor,
and $\hat{\bf q} = {\bf q} / q$.
Such a nonanalytic behavior of 
$\overline{\mathcal{E}}_{\alpha,\kappa \beta}^{{\bf q} \rightarrow 0}$
propagates to the charge, polarization and lattice 
responses, thwarting their Taylor expansion at $\Gamma$. %$q \rightarrow 0$.
%
%Another way to see this is to observe that the long-range character
%of the electrostatic interactions in real space leads to dipolar
%fields decaying as $r^{-3}$ -- the Fourier transform of such fields
%is discontinuous at $\Gamma$, consistent with Eq.~(\ref{nonanalytic}).
%
%Note that, in addition to dynamical diploes, higher-order multipolar 
%fields are generally present. These are responsible for discontinuities
%at the first (quadrupole) and second (octupole) order in $q$. As 
%the octupolar term is always present regardless of symmetry, the 
%long-wave analysis of the charge and polarization response functions 
%is problematic \emph{even in elemental solids}, where the Born dynamical 
%charge tensors are identically zero.

We shall circumvent this difficulty by \emph{removing} the macroscopic
electrostatic component (corresponding to the ${\bf G}=0$ vector
of the reciprocal lattice) from the self-consistent electrostatic potential.
This prescription has the effect of screening the longitudinal fields
associated with the long-wavelength phonon. Therefore, it corresponds 
to adopting \emph{short-circuit} electrical boundary conditions in 
the calculation of the response functions, which is indeed the standard 
convention in the definition of the electromechanical coupling coefficients.
This way, we have solved two problems at once: i) all the response
functions become analytic at $\Gamma$ and their polynomial expansion 
is, in principle, well defined at any order in ${\bf q}$; ii) we have specified 
once and for all that the response functions are calculated with 
a macroscopic electric field kept constant and equal to zero,
i.e. in short circuit.
A formal demonstration of these claims, based on the
dielectric matrix approach,~\cite{rmm_thesis}
%an analysis of the
%nonanalytic properties of the dielectric matrix,
%description of crystalline insulators, 
is provided in Section~\ref{cochran}.

By using the aforementioned precautions, it is now formally possible to
perform the Tayor expansion of the charge density response (we shall assume
from now on that repeated indices are implicitly summed over),
%Similarly to the polarization, the microscopic charge density 
%response can be expanded in powers of ${\bf q}$ in a vicinity of 
%the $\Gamma$ point,
%To second order, for example, we have 
\begin{equation}
\rho_{\kappa \beta}^{\bf q}({\bf r}) \simeq \rho_{\kappa \beta}^{(0)}({\bf r}) -i
q_\gamma  \rho_{\kappa \beta}^{(1, \gamma)}({\bf r}) - \frac{q_\gamma q_\lambda}{2} \rho_{\kappa \beta}^{(2, \gamma \lambda)}({\bf r}), %+\mathcal{O}(q^3), 
%\\
%P_{\alpha \beta}^{\bf q}({\bf r}) &\simeq& P_{\alpha \beta}^{(0)}({\bf r}) 
%-iq_\gamma  P_{\alpha \beta, \gamma}^{(1)}({\bf r}) - \frac{q_\gamma q_\lambda}{2} P_{\alpha \beta, \gamma \lambda}^{(2)}({\bf r}). \quad \quad
\label{rhoq}
\end{equation}
the microscopic polarization,
\begin{equation}
P_{\alpha , \kappa \beta}^{\bf q}({\bf r}) \simeq P_{\alpha , \kappa \beta}^{(0)}({\bf r}) 
 - iq_\gamma  P_{\alpha, \kappa \beta}^{(1,\gamma)}({\bf r}) - 
 \frac{q_\gamma q_\lambda}{2} P_{\alpha, \kappa \beta}^{(2,\gamma \lambda)}({\bf r}),
\label{pq}
\end{equation}
and the force-constant matrix,
\begin{equation}
\Phi_{\kappa \alpha , \kappa' \beta}^{\bf q} \simeq 
\Phi^{(0)}_{\kappa \alpha , \kappa' \beta} -
iq_\gamma \Phi^{(1,\gamma)}_{\kappa \alpha , \kappa' \beta} -
\frac{q_\gamma q_\lambda}{2} \Phi^{(2,\gamma \lambda)}_{\kappa \alpha , \kappa' \beta}. %\qquad
\label{phiq}
\end{equation}
Note the choice of the prefactors, which is motivated by the relationship to
the localized real-space representation (see Section~\ref{elec}).
In practice, at an arbitrary order and for a given response function $g^{\bf q}_{\kappa \beta}({\bf r})$,
we define
\begin{equation}
g^{(n,\gamma_1 \ldots \gamma_n)}_{\kappa \beta}({\bf r}) = i^n \,
 \frac{ \partial^n g^{{\bf q}}_{\kappa \beta}({\bf r})} {\partial q_{\gamma_1} \ldots \partial q_{\gamma_n}} \Big|_{{\bf q}=0}.
\label{gn}
\end{equation}
This prescription also guarantees that the functions $g^{(n,\gamma_1 \ldots \gamma_n)}_{\kappa \beta}({\bf r})$
are always real.

\subsection{Practical considerations}

%\subsection{Practical calculation}

\subsubsection{Phase factors}

\label{secphase}

Our definition of the elementary monochromatic perturbations, Eq.~(\ref{elementary}),
differs from that used by Gonze~\cite{Gonze} and Gonze and Lee~\cite{Gonze/Lee} (GL),
\begin{displaymath}
u_{\kappa \beta}^{{\rm GL},l}  = \lambda e^{i {\bf q} \cdot {\bf R}_{l}},
\end{displaymath}
by a sublattice-dependent (but cell-independent) phase factor,
$$
u_{\kappa \beta}^{l}  = u_{\kappa \beta}^{{\rm GL},l} \, e^{i {\bf q} \cdot \bm{\tau}_{\kappa}}.
$$
Such a modification is irrelevant in the calculation of phonon dispersion 
curves, but is crucial in the context of the long-wave expansion performed 
here.
In fact, it guarantees that the acoustic phonon eigenmodes 
%reduce to
%rigid shifts of the crystal lattice in a vicinity of $\Gamma$ (zero 
%order in $q$). 
\emph{do not} depend on the (arbitrary) assignment of each basis 
atom to a given cell in the crystal, and therefore we regard it 
as a very natural choice on general physical grounds.
%only depend on an arbitrary choice of the origin
%of the whole crystal lattice coordinates, and \emph{not}
%on the 
%This is what one expects on general physical grounds, 
%and enormously facilitates the analysis of the higher orders in $q$.

From the point of view of practical calculations, it should be kept in 
mind that all the response functions discussed in this work generally 
differ from the quantities that are computed within the publicly available 
implementations of DFPT.
Given that the modification consists in a trivial phase, however, it 
is easy to write the correspondence between the response functions 
defined by GL and those considered here.
For example, concerning the charge density response, we have (by 
using the linearity of the response functions in the perturbation)
$$
\rho^{\bf q}_{\kappa \beta}({\bf r}) = \rho^{{\rm GL},{\bf q}}_{\kappa \beta}({\bf r}) \, 
e^{i {\bf q} \cdot \bm{\tau}_{\kappa}}.
$$
In the case of the force-constant matrix, there is an additional phase factor 
coming from the factorization Eq.~(\ref{force}), which leads to the following 
correspondence,
$$
\Phi_{\kappa \alpha \, \kappa' \beta}^{\bf q} = \widetilde{C}^{\rm GL}_{\kappa \alpha \, \kappa' \beta} ({\bf q}) \,
e^{i {\bf q} \cdot (\bm{\tau}_{\kappa'} - \bm{\tau}_{\kappa})}.
$$
Of course, the real-space force constants (i.e. the second derivative
of the total energy with respect to the displacements of individual 
atoms) must be consistent with the definition given by GL, 
$$
\Phi_{\kappa \alpha \, \kappa' \beta}^l = C^{\rm GL}_{\kappa \alpha \, \kappa' \beta}(0,l).
$$
Therefore, our modification essentially concerns the definition of the
Fourier transform that is used to move between direct and reciprocal space.
Here we have [compare with Eq.~(10) of Ref.~\onlinecite{Gonze/Lee}]
$$
\Phi_{\kappa \alpha \, \kappa' \beta}^{\bf q} = \sum_l \Phi_{\kappa \alpha \, \kappa' \beta}^l 
e^{i {\bf q} \cdot ({\bf R}_l + \bm{\tau}_{\kappa'} - \bm{\tau}_{\kappa})}.
$$

\subsubsection{Differentiation in ${\bf q}$-space}

To calculate the Taylor expansion of the response functions one can
follow two different routes. Ideally, it would be desirable to take
the analytical gradients of the Sternheimer equation, Eq.~(\ref{stern}),
in ${\bf q}$-space and solve directly for the perturbed wavefunctions at
a given order in ${\bf q}$,
$$%\begin{equation}
|\phi^{\bf q}_{n{\bf k}} \rangle \simeq |\phi^{(0)}_{n{\bf k}} \rangle
- i q_\gamma |\phi^{(1,\gamma)}_{n{\bf k}} \rangle - 
\frac{q_\gamma q_\lambda}{2} |\phi^{(2,\gamma \lambda)}_{n{\bf k}} \rangle.
$$%\end{equation}
(The dependence on the sublattice index $\kappa$ and the
displacement direction $\beta$ has been kept implicit to avoid overburdening the
notation.)
Then the response functions could be simply calculated from the orbitals at the 
desired order in ${\bf q}$ by exploiting the linearity of the respective Taylor 
expansions. For example, the charge density at linear order in ${\bf q}$ would read
\begin{align}
\rho_{\kappa \beta}^{(1,\gamma)} ({\bf r}) =& -2f \frac{\Omega}{(2\pi)^3} \sum_n 
\int_{\rm BZ} d^3 k \, \phi^*_{n{\bf k}} ({\bf r}) \, \phi^{(1,\gamma)}_{n{\bf k}} ({\bf r}) \nonumber \\
& + Z_\kappa \delta_{\beta \gamma} \sum_{l} \delta({\bf r - R}_{l\kappa}).
%
%
%Z_\kappa \, \frac{\partial \delta({\bf r - R}_{l\kappa})}
%{\partial r_\beta }(r - R_{l\kappa})_\gamma. \nonumber
\end{align}
%
%
%The minus sign of the first term on the right-hand side comes from the negative 
%electron charge. The second term is due to $\rho^{\rm ext}$, and
%concerns the ion cores (which are described by point charges) -- its
%cell integral yields $Z_\kappa \delta_{\beta \gamma}$.
%
%
%Performing the analytic long-wave expansion of the Sternheimer equation goes beyond the 
%scopes of this work, and
We have not implemented the analytic long-wave expansion of the Sternheimer equation
here. (The explicit derivation is under way and will be the subject of a future
communication.) Instead we propose, for the time being, to extract the
needed Taylor-expanded response functions by using a finite-difference approach
in ${\bf q}$-space.~\footnote{
  Note that the quantities that undergo the differentiation in ${\bf q}$ are
  first-order response functions in the perturbation amplitude, $\lambda$,
  to be calculated within DFPT.}
This has the advantage of allowing the calculation of the flexoelectric
tensor in arbitrary solids by means of the existing implementations of 
DFPT. 
%
%Also, we expect this procedure to be computationally efficient, as a small 
%number of ${\bf q}$ points surrounding $\Gamma$ need to be explicitly 
%calculated.
%
In practice, it suffices to discretize Eq.~(\ref{gn}) (replace $g$ with
the desired response function), by using an appropriate grid of ${\bf q}$ 
points surrounding $\Gamma$.
%An easier alternative to 
%is to extract the 
%
This procedure is good for performing the long-wave analysis of the charge 
density and the force-constant matrix, as both quantities are fully 
implemented in publicly available codes.
The calculation of the polarization response deserves a separate comment,
as currently available implementations of DFPT provide access only to the macroscopic 
(cell-averaged) part, and not to the full microscopic current density.
In the following Section we shall outline a viable procedure to access
the latter quantity.

\subsection{Microscopic polarization response}

\label{polarization}

%As mentioned earlier in this Section, the microscopic polarization response 
%to an arbitrary perturbation can be obtained via adiabatic perturbation 
%theory. 
%
%First of all, note that the polarization, like the charge density, has
%two contributions: one coming from the electrons and the other coming from
%the displacement of the ionic charges,
%\begin{align}
%P^{\bf q}_{\alpha,\kappa \beta}({\bf r}) &= P^{{\rm el},{\bf q}}_{\alpha,\kappa \beta}({\bf r}) +
% P^{{\rm ion},{\bf q}}_{\alpha,\kappa \beta}({\bf r}).
%\end{align}
%
%The ionic contribution is trivial, and can be written as
%\begin{equation}
%P^{\bf q}_{\alpha,\kappa \beta}({\bf r}) = \delta_{\alpha \beta} Z_\kappa \sum_l 
%\delta({\bf r}-{\bf R}_{l\kappa}), %e^{ i {\bf q} \cdot ({\bf R}_{l\kappa}-{\bf r})}.
%\end{equation}
%independent of ${\bf q}$.

To derive the electronic contribution to the microscopic polarization response, we shall 
work in reciprocal space, and write $P^{{\rm el},{\bf q}}_{\alpha,\kappa \beta}$ 
in terms of its Fourier coefficients (we shall omit the superscript ``el'' in the remainder
of this section, as the absence of the ionic contribution is obvious from the context),
%it is obvious from the context that we are discussing the electronic
%contribution only),
\begin{equation}
P^{{\bf q}}_{\alpha,\kappa \beta}({\bf G}) = \frac{1}{\Omega} 
\int_{\rm cell} d^3 r P^{{\bf q}}_{\alpha,\kappa \beta}({\bf r}) e^{-i{\bf G\cdot r}}.
\end{equation}
We seek an (unknown) operator $\hat{P}_{\alpha,{\bf q + G}}$ such that
\begin{equation}
P^{\bf q}_{\alpha,\kappa \beta}({\bf G}) = 2 \frac{ s \Omega}{(2\pi)^3} \sum_n 
\int_{\rm BZ} d^3 k \, 
\langle \psi_{n{\bf k}} |\hat{P}_{\alpha,{\bf q + G}} | \Delta \psi^{{\bf q},\kappa \beta}_{n{\bf k}} \rangle,
\end{equation}
where 
%we have made the dependence of the first-order wavefunction, $\Delta \psi$, on $\kappa$ and $\beta$
%explicit, and 
$\psi_{n{\bf k}} ({\bf r}) = \phi_{n{\bf k}} ({\bf r}) e^{i {\bf k \cdot r}}$ and 
$\Delta \psi^{{\bf q},\kappa \beta}_{n{\bf k}}({\bf r})  = 
\Delta \phi^{{\bf q},\kappa \beta}_{n{\bf k}}({\bf r}) e^{i {\bf (k+q) \cdot r}}  $. 
(Strictly speaking, only the ${\bf G}=0$ component of the polarization response, corresponding
to $\overline{P}^{\bf q}_{\alpha,\kappa \beta}$, is sufficient for the scopes of
the present work; we keep the ${\bf G}$-dependence for the sake of generality.)
%such that the desired reciprocal
% the first-order wavefunctions at a given ${\bf q}$ are known,
%
It is convenient to simplify the notation, and write
\begin{equation}
P^{\bf q}_{\alpha,\kappa \beta}({\bf G}) = 2 \sum_v
\langle \psi_v |\hat{P}_{\alpha,{\bf q + G}} | \Delta \psi_v^{{\bf q},\kappa \beta} \rangle.
\end{equation}
where the index $v$ runs over valence (occupied) wavefunctions.
As the first-order wavefunctions belong, by construction, to the conduction 
manifold, one can insert a projector $\hat{Q}=\sum_c |\psi_c \rangle \langle \psi_c |$,
$$
P^{\bf q}_{\alpha,\kappa \beta}({\bf G}) = 2 \sum_{v,c}
\langle \psi_v |\hat{P}_{\alpha,{\bf q + G}} |\psi_c \rangle 
\langle \psi_c | \Delta \psi_v^{{\bf q},\kappa \beta} \rangle,
$$
where $c$ runs over the unoccupied orbitals.
Since both $|\psi_c \rangle$ and $|\psi_v \rangle$ are eigenstates of the
unperturbed Hamiltonian, $\hat{H}$, one can readily write 
\begin{equation}
P^{\bf q}_{\alpha,\kappa \beta}({\bf G}) = 2 \sum_{v,c}
\frac{\langle \psi_v | \left[ \hat{P}_{\alpha,{\bf q + G}}, \hat{H} \right] |\psi_c \rangle}
{\epsilon_c - \epsilon_v} 
\langle \psi_c | \Delta \psi_v^{{\bf q},\kappa \beta} \rangle.
\label{eqeq}
\end{equation}
Now, assuming that $\hat{P}_{\alpha,{\bf q + G}}$ does not depend explicitly on time
we have, from Ehrenfest theorem,
%Now recall Ehrenfest theorem,
\begin{equation}
\frac{d}{dt} \langle \hat{A} \rangle = -i\langle [ \hat{A}, \hat{H} ] \rangle,
\end{equation}
where $\langle \hat{O} \rangle$ stands for the expectation value of the operator
$\hat{O}$.
Since in a nonmagnetic insulator 
${\bf j}({\bf r},t) = d {\bf P} ({\bf r},t) / dt$,
it follows that the commutator in Eq.~\eqref{eqeq} must correspond to 
the \emph{current density operator},
\begin{equation}
\hat{j}_{\alpha,{\bf q + G}} = -i\left[ \hat{P}_{\alpha,{\bf q + G}}, \hat{H} \right].
\end{equation}
Hence, we have
\begin{equation}
P^{\bf q}_{\alpha,\kappa \beta}({\bf G}) = -2i \sum_{v}
\langle \tilde{\psi}^{\alpha,{\bf q + G}}_v | \Delta \psi_v^{{\bf q},\kappa \beta} \rangle,
%\label{eqeq}
\end{equation}
where
\begin{equation}
|\tilde{\psi}^{\alpha,{\bf q + G}}_v \rangle = 
-\sum_c |\psi_c \rangle \frac{ \langle \psi_c |\hat{j}_{\alpha,{\bf q + G}} |\psi_v \rangle}
{\epsilon_c - \epsilon_v}.
\end{equation}
are the first-order orbitals induced by $\hat{j}_{\alpha,{\bf q + G}}$ as a perturbing operator.
These can be conveniently obtained by solving the nonselfconsistent Sternheimer equation,
\begin{equation}
( \hat{H} - \epsilon_v ) \, |\tilde{\psi}^{\alpha,{\bf q + G}}_v \rangle =
- \hat{Q} \, \hat{j}_{\alpha,{\bf q + G}} \, |\psi_v \rangle.
\end{equation}
This result allows us to calculate the cross-gap matrix elements of the unknown
microscopic polarization operator, $\hat{P}_{\alpha,{\bf q + G}}$, by means of
the more familiar current density operator. The probability current is a 
fundamental quantum-mechanical observable, and implementing it in an existing
DFPT code should not present major conceptual obstacles; such a task will
be the topic of a future communication.
In this context, it is worth mentioning the work of Umari, Dal Corso and 
Resta~\cite{puma}, where the microscopic polarization response to a uniform electric
field perturbation was derived and computed; the authors used an approach that is 
closely related to the one presented here.

\section{Long-wave analysis}

\label{longw}

In the following, we shall use the long-wave method~\cite{Born/Huang} 
to derive the electromechanical response (either piezoelectric
or flexoelectric) of the crystal in terms of the elementary 
ingredients introduced above.
We shall first focus on the atomic displacements induced by a
``short-circuited'' (in the sense specified in the previous Section) 
acoustic phonon, and later compute the polarization field 
associated with the deformation.

\subsection{Internal strains}

\label{intstrains} 

%(This paragraph loosely follows the lines taken in 
%Refs.~\onlinecite{Born/Huang} and~\onlinecite{Tagantsev}.)
%
Consider the real-space atomic equation of motion, 
$$%\begin{equation}
m_\kappa \ddot u_{\kappa \alpha}^{0}(t) = -\Phi^l_{\kappa \alpha \kappa' \beta} u_{\kappa' \beta}^{l}(t),
$$%\end{equation}
where $u$ are the displacements, $\Phi$ is the real-space force-constant matrix and $m_\kappa$ 
is the mass of the specie $\kappa$. ($l$ indexes the lattice cell where the atom $\kappa'$ is
located; the atom $\kappa$ is located at the $l=0$ cell; $\alpha$ and $\beta$ refer to Cartesian 
directions.)
We seek solutions of the type
$$%\begin{equation}
u_{\kappa \beta}^{l}(t)  = U_{\kappa \beta}^{\bf q} e^{i {\bf q} \cdot {\bf R}_{l\kappa} - i\omega t}, %\qquad U_{\kappa \beta}^{q \rightarrow 0} = U_\beta
$$%\end{equation}
These are given by the eigenvalue problem
\begin{equation}
m_\kappa \omega^2 U_{\kappa \alpha}^{\bf q} = \Phi_{\kappa \alpha \, \kappa' \beta}^{\bf q} U_{\kappa' \beta}^{\bf q}.
\label{eom}
\end{equation}
We solve Eq.~(\ref{eom}) perturbatively~\cite{Tagantsev,Born/Huang} 
in a vicinity of ${\bf q}=0$ by writing the wavevector as~\cite{Born/Huang} 
$\epsilon {\bf q}$, where $\epsilon$ is a dimensionless perturbation 
parameter.
%expanding $U_{\kappa \alpha}^{\bf q}$ in powers of $q$ along a direction 
%$\hat{\bf q}$.
%\begin{equation}
%U_{\kappa \alpha}^{\bf q} = U^{(0,\hat{\bf q})}_{\kappa \alpha} + iq U^{(1,\hat{\bf q})}_{\kappa \alpha} +
%q^2 U^{(2,\hat{\bf q})}_{\kappa \alpha} +  \mathcal{O}(q^3).
%\end{equation}
For an acoustic branch, $\omega$ and $U_{\kappa \alpha}^{\bf q}$ can be
expanded as follows,~\cite{Born/Huang} 
\begin{align}
\label{omega}
\omega(\epsilon {\bf q}) &= \epsilon \omega^{(1,{\bf q})} + \epsilon^2 \omega^{(2,{\bf q})} + \ldots, \\
U^{\bf \epsilon {\bf q}}_{\kappa \alpha} &= 
U^{(0,{\bf q})}_{\kappa \alpha} + i \epsilon U^{(1,{\bf q})}_{\kappa \alpha} + 
\epsilon^2 U^{(2,{\bf q})}_{\kappa \alpha} + \ldots.
\label{uq}
\end{align}
(The expansion of $\omega$ starts with the linear term, as for  
acoustic waves the frequency approaches zero as $\epsilon {\bf q} \rightarrow 0$.)
%
%Note that in Eq.~\eqref{omega} and Eq.~\eqref{uq} we have temporarily abandoned the usual 
%convention of the $(-i)^n/n!$ prefactors in the $n$-th order term; both $\omega(\epsilon {\bf q})$
%and $U^{\bf \epsilon {\bf q}}_{\kappa \alpha}$ are intermediate quantities, and such prefactors 
%would have added an unneeded burden in the notation.
%
We shall now proceed to calculating the induced displacements by plugging 
Eq.~\eqref{omega}, Eq.~\eqref{uq} and the Taylor expansion of the force-constant
matrix, Eq.~\eqref{phiq}, into Eq.~(\ref{eom}), and by grouping 
the different terms according to their perturbative order.

%An acoustic phonon is described by the solutions of  Eq.~(\ref{eom}) with $\omega=0$ 
%for $q \rightarrow 0$. 
At order zero in $\epsilon$ we have
\begin{equation}
\Phi^{(0)}_{\kappa \alpha , \kappa' \beta} U^{(0,{\bf q})}_{\kappa \beta} = 0 \quad \rightarrow \quad 
U^{(0)}_{\kappa \beta} = U_{\beta},
\label{ozero}
\end{equation}
i.e. the phonon eigenvector must be independent of $\kappa$.
In fact, the matrix $\Phi^{(0)}$ is the zone-center dynamical matrix of the
crystal, and has three zero-frequency eigenmodes corresponding to the rigid
translations of the whole lattice along each Cartesian direction.
This means that Eq.~(\ref{ozero}) is identically verified by any real-space
vector ${\bf U}$. 
%which is therefore an independent parameter of the problem.

At first order in $\epsilon$ we have
\begin{equation}
\Phi^{(0)}_{\kappa \alpha , \kappa' \beta} U^{(1,{\bf q})}_{\kappa' \beta} %q_\lambda \Lambda^{\kappa'}_{\beta \gamma \lambda} 
 -  U_\beta q_\lambda \sum_{\kappa'} \Phi^{(1,\lambda)}_{\kappa \alpha , \kappa' \beta} = 0.
\end{equation}
%(Summation is implicit over $\kappa'$ even when this index is not
%repeated.)
%
Solvability requires that
$$%\begin{equation}
\sum_{\kappa \kappa'} \Phi^{(1,\lambda)}_{\kappa \alpha , \kappa' \beta} = 0
$$%\end{equation}
be identically satisfied for any $\alpha,\beta$, which is indeed the case~\cite{Born/Huang}.
The explicit solution can be written as
\begin{eqnarray}
U^{(1,{\bf q})}_{\kappa \alpha} &=& \Gamma^\kappa_{\alpha \beta \gamma} U_\beta q_\gamma, \\
\Gamma^\kappa_{\alpha \beta \gamma} &=& \widetilde{\Phi}^{(0)}_{\kappa \alpha , \kappa' \lambda}
\Lambda^{\kappa'}_{\lambda \beta \gamma}
%\sum_{\kappa'} \Phi^{(1,\lambda)}_{\kappa'' \gamma , \kappa' \beta}.
\end{eqnarray} 
Here we have introduced $\widetilde{\Phi}^{(0)}_{\kappa \alpha \, \kappa' \beta}$ as the 
pseudoinverse~\cite{Wu/Vanderbilt/Hamann:2005} of the singular matrix 
$\Phi^{(0)}_{\kappa \alpha \, \kappa' \beta}$
(the zero eigenvalues of $\bm{\Phi}$, corresponding to rigid translations,
are mapped into zero eigenvalues of $\widetilde{\bm{\Phi}}$, while the 
nonsingular remainder of the matrix is inverted), and
\begin{equation}
\Lambda^\kappa_{\alpha \beta \gamma} = \sum_{\kappa'} \Phi^{(1,\gamma)}_{\kappa \alpha \, \kappa' \beta},
\end{equation}
is the piezoelectric \emph{force-response} tensor (following the
notation of Ref.~\onlinecite{Wu/Vanderbilt/Hamann:2005}).
$\Lambda^\kappa_{\alpha \beta \gamma}$ describes the force induced on the 
sublattice $\kappa$ along $\alpha$ when the crystal undergoes a homogeneous strain 
deformation,~\footnote{
  Strictly speaking, $\Lambda^\kappa_{\alpha \beta \gamma}$ is defined as 
  the response to an \emph{unsymmetrized} strain, $\tilde{\varepsilon}_{\beta \gamma}$.
  However, due to the invariance of $\Lambda^\kappa_{\alpha \beta \gamma}$ with respect
  to $\beta \leftrightarrow \gamma$ exchange, we can readily identify it as the 
  force response to a \emph{symmetrized} strain, $\varepsilon_{\beta \gamma}$.}
$\varepsilon_{\beta \gamma}$,
and is symmetric with respect to $\beta \leftrightarrow \gamma$ exchange.~\cite{Born/Huang}
The internal-strain tensor~\cite{Martin}, $\Gamma^\kappa_{\alpha \beta \gamma}$,  describes the 
atomic relaxations induced by $\varepsilon_{\beta \gamma}$, and inherits the $\beta \leftrightarrow \gamma$ 
invariance from $\Lambda$.
Note that $\Gamma^\kappa_{\alpha \beta \gamma}$ is specified only modulo an 
arbitrary $\kappa$-independent (but possibly $\alpha \beta \gamma$-dependent) 
constant, which physically corresponds to a rigid shift of the whole lattice.

At second order in $\epsilon$ we obtain
\begin{eqnarray}
\Phi^{(0)}_{\kappa \alpha  \,\kappa'' \mu } U^{(2,{\bf q})}_{\kappa'' \mu} &=&
 m_\kappa [\omega^{(1,{\bf q})}]^2  U_{\alpha} 
- q_\gamma q_\lambda T^\kappa_{\alpha \beta, \gamma \lambda}  
U_{\beta},
\label{u2-a}
\end{eqnarray}
where $m_\kappa$ are atomic masses, and we have introduced the 
\emph{type-I flexoelectric force-response tensor} ${\bf T}$ as follows,
\begin{equation}
T^\kappa_{\alpha \beta, \gamma \lambda} = \left[ \alpha \beta, \gamma \delta \right]^\kappa +
\frac{1}{2} \Big[ \left( \alpha \gamma,\beta \lambda \right)^\kappa + \left( \alpha \lambda,\beta \gamma \right)^\kappa\Big].
\label{tk} 
\end{equation}
The square brackets and round brackets are defined (in loose analogy with
the notation of Ref.~\onlinecite{Born/Huang}) as
\begin{eqnarray}
\left[ \alpha \beta, \gamma \delta \right]^\kappa &=&
-\frac{1}{2} \sum_{\kappa'} \Phi^{(2,\gamma \delta)}_{\alpha \kappa \, \beta \kappa'},
\label{squarek} \\
%\end{eqnarray}
%\begin{eqnarray}
\left( \alpha \lambda,\beta \gamma \right)^\kappa &=&
\Phi^{(1,\lambda)}_{\kappa \alpha  \, \kappa' \rho} 
\Gamma^{\kappa'}_{\rho \beta \gamma}.
%\widetilde{\Phi}^{(0)}_{\kappa' \rho \, \kappa'' \mu} 
%\sum_{\kappa'''} \Phi^{(1, \lambda)}_{\kappa'' \mu \, \kappa''' \beta}.
\label{roundk}
\end{eqnarray}
%
%The former, Eq.~\eqref{squarek}, 
$\left[ \alpha \beta, \gamma \delta \right]^\kappa$ describes the force 
induced along $\alpha$ on a given atomic sublattice $\kappa$ by a ``frozen-ion'' 
(in the sense specified in Ref.~\onlinecite{Hong-11}) strain gradient 
$\eta_{\beta,\gamma \delta}$ (in type-I form).
%
%The latter, Eq.~\eqref{roundk}, 
$\left( \alpha \lambda,\beta \gamma \right)^\kappa$ describes the additional 
force produced by the atomic relaxations that are first-order in $\epsilon$ and is, 
therefore, only relevant to crystals that have one or more free Wyckoff parameters.
%
%Recall that every atom undergoes an internal $\epsilon_{\beta \gamma}$,
%relaxation along $\alpha$ equal to $\Gamma^{\kappa}_{\alpha \beta \gamma}$.
%
%A macroscopic strain gradient produces a gradient in the
Note that the round bracket is a type-II object (i.e. it relates the force along
$\alpha$ to the type-II strain gradient component $\varepsilon_{\beta \gamma,\lambda}$), 
hence the symmetrization in Eq.~\ref{tk} (the ${\bf T}$ tensor is a type-I object).

The linear problem, Eq.~(\ref{u2-a}), admits solution if and only if
the following condition on $\omega^{(1,{\bf q})}$ is satisfied~\cite{Born/Huang,Tagantsev},
\begin{equation}
\left(  M [\omega^{(1,{\bf q})}]^2 \delta_{\alpha \beta} - 
q_\gamma q_\lambda T_{\alpha \beta, \gamma \lambda} \right) U_\beta = 0,
\label{acoustic}
\end{equation}
where $M=\sum_\kappa m_\kappa$ is the total mass of the
primitive cell, and $T_{\alpha \beta, \gamma \lambda} = 
\sum_\kappa  T^\kappa_{\alpha \beta, \gamma \lambda}.$
We shall see in Section~\ref{secelastic} that the quantity 
$T_{\alpha \beta, \gamma \lambda} / \Omega$ can be considered 
a ``type-I'' representation of the macroscopic elastic tensor, $\bm{\mathcal{C}}$.~\footnote{
  Note that in piezoelectrically active materials the long-range 
  electrostatic fields contribute to sound propagation via a nonanalytic
  term in the elastic tensor. Our neglect of the macroscopic ${\bf G}=0$ 
  term in the electrostatics implies that such contribution is not
  accounted for in Eq.~(\ref{acoustic}).}
One will then easily recognize Eq.~(\ref{acoustic}) as the sound
wave equation.~\cite{Born/Huang} Its solutions depend only on 
$\bm{\mathcal{C}}$, on the mass density $M/\Omega$ and on the
propagation direction $\hat{\bf q}={\bf q}/q$, and are therefore
characterized by a linear dispersion relation
along any given $\hat{\bf q}$.
%[In fact, it coincides with the sound wave 
%equation in nonpiezoelectric crystals. 
%In piezoelectrically active materials the long-range
%electrostatic fields contribute to sound propagation -- 
%]
By combining Eq.~(\ref{acoustic}) with Eq.~(\ref{u2-a}), we readily obtain
\begin{eqnarray}
U^{(2,{\bf q})}_{\kappa \alpha} &=& -U_\beta q_\gamma q_\lambda N^\kappa_{\alpha \beta \gamma \lambda}, \\
N^\kappa_{\alpha \beta \gamma \lambda} &=& \widetilde{\Phi}^{(0)}_{\kappa \alpha \kappa' \rho} \hat{T}^{\kappa'}_{\rho \beta, \gamma \lambda},
\label{u2-b}
\end{eqnarray}
where $N^\kappa_{\alpha \beta \gamma \lambda}$ is the \emph{type-I flexoelectric 
internal-strain tensor}, and we have also introduced the \emph{mass-compensated} 
force-response tensor,
\begin{eqnarray}
\hat{T}^\kappa_{\alpha \beta, \gamma \lambda} &=& T^\kappa_{\alpha \beta, \gamma \lambda} - 
\frac{m_\kappa}{M} T_{\alpha \beta, \gamma \lambda}.
\label{htk}
\end{eqnarray}
Note that the sum over $\kappa$ of the $\hat{\bf T}$ tensor identically vanishes
by construction; this is a necessary condition for the linear problem Eq.~(\ref{u2-b})
to be solvable, proving that our derivations are internally consistent.
 
In summary, the lattice response to a (short-circuited) long-wavelength acoustic 
phonon can be written as
\begin{equation}
U_{\kappa \alpha}^{\bf q} = U_\beta \left[ \delta_{\alpha \beta} +i q_\gamma \Gamma^\kappa_{\alpha \beta \gamma} 
- q_\gamma q_\lambda N^\kappa_{\alpha \beta \gamma \lambda} \right] + \mathcal{O}(q^3) ,
\end{equation}
where $\Gamma^\kappa_{\alpha \beta \gamma}$ and $N^\kappa_{\alpha \beta \gamma \lambda}$
are the desired internal-strain tensors.

Before closing this part, it is useful to briefly comment on the
relationship between our derivation and Tagantsev's. 
%which at first 
%sight looks essentially identical. 
%
Our approach accurately follows the formalism of Ref.~\onlinecite{Tagantsev},
except for the procedure to extract the relevant physical quantities from the
force constants of the crystal.~\footnote{The force-response tensor 
  $\hat{T}^\kappa_{\alpha \beta, \gamma \lambda}$ of this work is closely related 
  to the $T^{jl}_{ip,i'p'}$ of Ref.~\onlinecite{Tagantsev}, with the exception that
  the latter is not summed over one of the sublattice indices. Note that our Eq.~\eqref{tk}
  differs from the unnumbered equation after Eq.~(A7) of Ref.~\onlinecite{Tagantsev}:
  in the former, the contribution that depends on $\Gamma$ (round brackets) is correctly 
  symmetrized over the $\gamma \lambda$ indices, while in the latter it is not.}
Tagantsev wrote the moments of the force constant matrix 
as lattice sums in real space, whose convergence is not guaranteed unless a 
specific prescription for dealing with macroscopic electrostatics is formulated.
A heuristic treatment of the macroscopic fields might be possible in 
atomistic models, where the charge response to individual atomic 
displacements is trivially simple. 
In the present quantum-mechanical context,
the problem is complicated by the presence of higher-order multipolar
interactions~\cite{Resta-10,Hong-11}, whose impact on lattice dynamics might
be cumbersome to keep track of.
Here we solve this issue by working in ${\bf q}$ space, where a rigorous 
strategy to suppress the problematic electric fields in the long-wavelength 
limit is easy to implement once and for all, and does not require any special
effort.

%Here $\Lambda$ and $N$ are internal-strain tensors,
%which can be readily computed by using the atomic masses $M_\kappa$
%and the first three terms of the small-$q$ expansion of the force
%constant matrix, 
%\begin{eqnarray}
%\Phi_{\kappa \alpha , \kappa' \beta}^{\bf q} &\simeq& 
%\Phi^{(0)}_{\kappa \alpha , \kappa' \beta} + 
%iq_\gamma \Phi^{(1,\gamma)}_{\kappa \alpha , \kappa' \beta} +
%\frac{q_\gamma q_\lambda}{2} \Phi^{(2,\gamma \lambda)}_{\kappa \alpha , \kappa' \beta}. \qquad
%\end{eqnarray}
%
%(The detailed derivations closely follow the approach of 
%Refs.~\onlinecite{Born/Huang} and~\onlinecite{Tagantsev}
%and will not be reported here.)

\subsection{Macroscopic flexoelectric coefficients in type-I form}

%To compute the electrical response of the acoustic phonon 
%described above we need, in addition to the internal
%strain tensors, the small-$q$ expansion of the polarization
%response to atomic displacements,
%\begin{eqnarray}
%P_{\alpha , \kappa' \beta}^{\bf q}({\bf r}) &\simeq& P_{\alpha , \kappa' \beta}^{(0)}({\bf r}) 
% + iq_\gamma  P_{\alpha, \kappa' \beta}^{(1,\gamma)}({\bf r}) + \frac{q_\gamma q_\lambda}{2} P_{\alpha, \kappa' \beta}^{(2,\gamma \lambda)}({\bf r}).
%\end{eqnarray}
%
In order to write the total polarization response to the long-wavelength phonon (and
hence to an arbitrary mechanical deformation), we need to combine the eigendisplacements 
derived in the previous section with the small-${\bf q}$ expansion of the induced 
polarization, Eq.~(\ref{pq}).
As in this work we are only concerned with the macroscopic response, we shall work on the cell-averaged 
counterparts of the polarization-response functions, which we indicate by an overline symbol,
\begin{equation}
\overline{P}^{(n,\gamma_1, \ldots, \gamma_n)}_{\alpha, \kappa \beta} = 
\frac{1}{\Omega} \int_{\rm cell} d^3 r \, P^{(n,\gamma_1, \ldots, \gamma_n)}_{\alpha, \kappa \beta} ({\bf r}).
\end{equation}
By construction, the zero-order term is proportional to the \emph{Born effective charge} tensor
of the specie $\kappa$,
\begin{equation}
\Omega \overline{P}^{(0)}_{\alpha \, \kappa \beta} = Z^*_{\kappa, \alpha \beta}.
\end{equation}
It follows that, due to the acoustic sum rule~\cite{rmm_thesis}, the sum over $\kappa$ of
the $\overline{P}^{(0)}$ tensor identically vanishes.
%The physical nature of the higher-order terms will be further clarified in 
%Section~\ref{richard}.
%
Thus, the contribution of the rigid translation to the macroscopic
$\overline{\bf P}$ vanishes as well (as anticipated above), leaving us 
with the terms that are linear and quadratic in ${\bf q}$,
\begin{equation}
\overline{P}_\alpha({\bf r},t) = \left( i U_\beta q_\gamma e_{\alpha \beta \gamma} -
U_\beta q_\gamma q_\lambda \mu^{\rm I}_{\alpha \beta, \gamma \lambda} \right) \, e^{i {\bf q} \cdot {\bf r} - i \omega t},
\label{pmac1}
\end{equation}
where the relaxed-ion response tensors are given by
\begin{eqnarray}
e_{\alpha \beta \gamma} &=& \bar{e}_{\alpha \beta \gamma} + \frac{1}{\Omega} Z^*_{\kappa, \alpha \rho} \Gamma^\kappa_{\rho \beta \lambda}, \\
\mu^{\rm I}_{\alpha \beta, \gamma \lambda} &=& \bar{\mu}^{\rm I}_{\alpha \beta, \gamma \lambda} - 
\frac{1}{2} \left( \Gamma^\kappa_{\rho \beta \gamma} P_{\alpha, \kappa \rho}^{(1,\lambda)} + 
\Gamma^\kappa_{\rho \beta \lambda} P_{\alpha, \kappa \rho}^{(1,\gamma)} \right) + \nonumber \\
&& + \frac{1}{\Omega} Z^*_{\kappa, \alpha \rho}  N^\kappa_{\rho \beta \gamma \lambda} 
\label{mui}
\end{eqnarray}
In the above expressions we have used the bar symbol to indicate, following the
notation of Ref.~\onlinecite{Wu/Vanderbilt/Hamann:2005}, the frozen-ion counterparts
of the tensors,
\begin{eqnarray}
\bar{e}_{\alpha \beta \gamma} &=& -\sum_\kappa \overline{P}_{\alpha, \kappa \beta}^{(1,\gamma)}, \\
\bar{\mu}^{\rm I}_{\alpha \beta, \gamma \lambda} &=& \frac{1}{2} \sum_\kappa \overline{P}_{\alpha, \kappa \beta}^{(2,\gamma \lambda)}.
\end{eqnarray}
The tensors $e$ and $\bar{e}$ correspond to the well-known relaxed-ion and
frozen-ion piezoelectric coefficients, and are both symmetric with respect to 
$\beta \leftrightarrow \gamma$ exchange. 
(This property of the latter tensor, as well as its relationship to 
Martin's theory of piezoelectricity~\cite{Martin}, will be rigorously demonstrated 
in Sections~\ref{secsymm} and~\ref{secmartin}, respectively.)
Hence, the unsymmetrized stress tensor, 
$\tilde{\varepsilon}_{\beta \gamma}({\bf r},t)=
i U_\beta q_\gamma e^{i {\bf q} \cdot {\bf r} - i \omega t}$,
can be replaced with its symmetric counterpart in Eq.~(\ref{pmac1}), leading to an expression that 
is fully invariant with respect to either translations or rotations of the original reference,
\begin{equation}
\overline{P}_\alpha ({\bf r},t) =  \varepsilon_{\beta \gamma} ({\bf r},t) \, e_{\alpha \beta \gamma} +
\eta_{\beta, \gamma \lambda} ({\bf r},t) \, \mu^{\rm I}_{\alpha \beta, \gamma \lambda}.
\end{equation}
This formula, which is a central result of this work, allows us to 
identify $\mu^{\rm I}$ and $\bar{\mu}^{\rm I}$ with the sought-after relaxed-ion and frozen-ion 
flexoelectric tensors, respectively, i.e.
\begin{equation}
\mu^{\rm I}_{\alpha \beta, \gamma \lambda} = \frac{\partial \overline{P}_\alpha}{\partial \eta_{\beta, \gamma \lambda}}
\end{equation}
%
%In the following, we shall specify the meaning of the I superscript on the flexoelectric
%tensors, and establish a formal link between flexoelectricity and elasticity.
%
The superscript I indicates that $\mu^{\rm I}$ and $\bar{\mu}^{\rm I}$ are ``type-I'' 
objects, i.e. they describe the response to a type-I strain gradient tensor 
$\eta_{\beta, \gamma \lambda}$.
One can, of course, write the flexoelectric tensor, $\mu$, in type-II form --
we shall do this explicitly hereafter, as such a conversion is important
for later derivations (and, in particular, for tracing the important 
link to elasticity that we anticipated in Section~\ref{intstrains}).

\subsection{Type-II form and ``elastic sum rule''}

\label{secelastic}
From the definition of the type-II strain gradient tensor, $\varepsilon_{\beta \gamma, \lambda}$,
it follows that
\begin{equation}
\mu^{\rm II}_{\alpha \lambda, \beta \gamma} = \frac{\partial \overline{P}_\alpha}{\partial \varepsilon_{\beta \gamma, \lambda}},
\end{equation}
where the type-II flexoelectric tensor is related to $\bm{\mu}^{\rm I}$ via 
a cyclic permutation of 
the last three indices,
%the usual operation on the last three indices,
\begin{equation}
\mu^{\rm II}_{\alpha \lambda, \beta \gamma} = \mu^{\rm I}_{\alpha \beta, \gamma \lambda} + 
\mu^{\rm I}_{\alpha \gamma, \lambda \beta} - \mu^{\rm I}_{\alpha \lambda, \beta \gamma}.
\label{muii-a}
\end{equation}
Note that $\mu^{\rm II}_{\alpha \lambda, \beta \gamma}$ is invariant upon exchange of the 
last two indices, consistent with the analogous symmetry of the type-II strain gradient tensor.
By combining Eq.~(\ref{muii-a}) with Eq.~(\ref{mui}) we have
\begin{equation}
\mu^{\rm II}_{\alpha \lambda, \beta \gamma} = \bar{\mu}^{\rm II}_{\alpha \lambda, \beta \gamma} -
\overline{P}_{\alpha, \kappa \rho}^{(1,\lambda)} \Gamma^\kappa_{\rho \beta \gamma}  + 
 \frac{1}{\Omega} Z^*_{\kappa, \alpha \rho}  L^\kappa_{\rho \lambda, \beta \gamma} 
\label{muii-b}
\end{equation}
The tensor $\bar{\mu}^{\rm II}$ is defined from $\bar{\mu}^{\rm I}$ via the 
symmetrization Eq.~(\ref{muii-a}). The internal-strain tensor $L^\kappa$
follows from $N^\kappa$ via an analogous operation on the indices $\beta \gamma \lambda$, 
which can be ultimately traced back to a redefinition of the force-response tensor,
\begin{eqnarray}
L^\kappa_{\rho \lambda, \beta \gamma} &=& \widetilde{\Phi}^{(0)}_{\kappa \rho \kappa' \alpha}  \hat{C}^{\kappa'}_{\alpha \lambda, \beta \gamma}, \\
\hat{C}^\kappa_{\alpha \lambda, \beta \gamma} &=& \hat{T}^\kappa_{\alpha \beta, \gamma \lambda} + 
\hat{T}^\kappa_{\alpha \gamma, \lambda \beta} - \hat{T}^\kappa_{\alpha \lambda, \beta \gamma}.
\end{eqnarray}
In turn, the tensor $\hat{C}^\kappa_{\alpha \lambda, \beta \gamma}$
can be written explicitly in terms of the type-II flexoelectric force-response 
tensor,
\begin{equation}
C^\kappa_{\alpha \lambda, \beta \gamma} = [\alpha \beta, \gamma \lambda]^\kappa + [\alpha \gamma, \lambda \beta ]^\kappa 
- [\alpha \lambda, \beta \gamma]^\kappa + (\alpha \lambda, \beta \gamma)^\kappa,
\label{ccc}
\end{equation}
after separating the mass-dependent part,
\begin{eqnarray}
\hat{C}^\kappa_{\alpha \lambda, \beta \gamma} &=& C^\kappa_{\alpha \lambda, \beta \gamma } - 
\frac{m_\kappa}{M} \Omega \mathcal{C}_{\alpha \lambda, \beta \gamma}.
\label{hck}
\end{eqnarray}
Under the assumption that the crystal at rest is free of stresses (see Section 28 of Ref.~\onlinecite{Born/Huang}),
\begin{equation}
\mathcal{C}_{\alpha \lambda, \beta  \gamma} = \frac{1}{\Omega} \sum_\kappa C^\kappa_{\alpha \lambda, \beta \gamma }
\label{elastic}
\end{equation}
is the macroscopic \emph{elastic tensor} calculated in short-circuit boundary 
conditions (zero macroscopic electric field).
This is another key result of this work, which we shall indicate as
``elastic sum rule'' henceforth.~\footnote{
  Apart from its formal appeal, from the point of view of 
  first-principles applications this result can turn out to be very handy 
  as a cross-check for the numerical accuracy of the calculated values.}

The detailed proof that Eq.~\eqref{elastic} indeed is the elastic tensor
%In retrospect, the existence of such a relationship between the flexoelectric 
%tensor and the elastic tensor could have already been anticipated from the expression
%we obtained for the sound-wave equation Eq.~(\ref{acoustic}), and is fully 
%consistent with the analysis provided 
can be found in the Born and Huang (BH) book.~\cite{Born/Huang} 
In fact, our choice of notation for some intermediate lattice-dynamical quantities
was motivated by their direct relationship to the $[\alpha \beta, \gamma \lambda]$
and $(\alpha \beta, \gamma \lambda)$ of BH, %Ref.~\onlinecite{Born/Huang} (indicated
%by BH),
\begin{eqnarray}
[\alpha \beta, \gamma \lambda]^{\rm BH} &=& \frac{1}{\Omega} \sum_\kappa [\alpha \beta, \gamma \lambda]^\kappa, \nonumber\\
(\alpha \beta, \gamma \lambda)^{\rm BH} &=& \frac{1}{\Omega} \sum_\kappa (\alpha \beta, \gamma \lambda)^\kappa. \nonumber
\end{eqnarray}
By combining these definitions with Eq.~(\ref{ccc}) and Eq.~(\ref{elastic}) we 
recover the BH formula for the elastic tensor [Eq.~(27.26) therein],
\begin{displaymath}
\mathcal{C}_{\alpha \lambda, \beta  \gamma} = [\alpha \beta, \gamma \lambda] + [\alpha \gamma, \lambda \beta ] 
- [\alpha \lambda, \beta \gamma] + (\alpha \lambda, \beta \gamma).
\end{displaymath}
As mentioned in Section~\ref{intstrains}, the square brackets have to do with the frozen-ion
deformation of the lattice, and are the only contribution in high-symmetry crystals,
while the round brackets have to do with the internal degrees of freedom that respond 
to a uniform strain %(and thus contribute to the elastic constant) 
in lower-symmetry crystals.
To further illustrate the implications of these statements in the context of elasticity,
it is useful to write $C^{\kappa}_{\alpha \lambda, \beta \gamma} = 
\bar{C}^{\kappa}_{\alpha \lambda, \beta \gamma} +
(\alpha \lambda, \beta \gamma)^{\kappa}$, 
%We shall come back to this point in Section~\ref{seclattice}, by 
where we have introduced the auxiliary quantity
\begin{equation}
\bar{C}^{\kappa}_{\alpha \lambda, \beta \gamma} = [\alpha \beta, \gamma \lambda]^\kappa + [\alpha \gamma, \lambda \beta ]^\kappa
- [\alpha \lambda, \beta \gamma]^\kappa.
\end{equation}
The bar symbol was motivated by the direct relationship between 
$\bar{C}^{\kappa}_{\alpha \lambda, \beta \gamma}$ and the 
\emph{frozen-ion} elastic tensor~\cite{Wu/Vanderbilt/Hamann:2005},
$\bar{\mathcal{C}}_{\alpha \lambda, \beta \gamma}$, 
\begin{equation}
\sum_\kappa \bar{C}^{\kappa}_{\alpha \lambda, \beta \gamma} =
\Omega \bar{\mathcal{C}}_{\alpha \lambda, \beta \gamma}.
\end{equation}
Thus, the quantity $(\alpha \lambda, \beta \gamma)$ is simply the additional 
contribution to the elastic tensor, $\mathcal{C}_{\alpha \lambda, \beta \gamma}=
\bar{\mathcal{C}}_{\alpha \lambda, \beta \gamma} + (\alpha \lambda, \beta \gamma)$, 
that is associated with the relaxation of the internal degrees of freedom of the cell.

In a hand-waving way, %one can say that the computation of the lattice-mediated 
one can say that the type-II flexoelectric force-response tensor is
%flexoelectric response simply requires computing 
a ``sublattice-resolved'' version of the macroscopic elastic coefficients.
This statement can be substantiated by invoking a general result of continuum
mechanics, relating the divergence of the stress field to the net force, $f_\alpha$,
acting on a volume, $\Omega$, of the material,
\begin{equation}
f_\alpha = \int_\Omega d^3 r \, \nabla_\beta \sigma_{\alpha \beta} ({\bf r}),
\end{equation}
Recall the definition of the stress tensor in a linear material,
\begin{equation}
\sigma_{\alpha \beta} ({\bf r}) = \mathcal{C}_{\alpha \beta \gamma \lambda} \varepsilon_{\gamma \lambda} ({\bf r}).
\end{equation}
Assuming a bulk crystal, the elastic tensor is independent of position;
therefore, for a unit cell of the crystal we immediately have 
\begin{equation}
\sum_\kappa f^\kappa_\alpha = \Omega \mathcal{C}_{\alpha \beta \gamma \lambda} \varepsilon_{\gamma \lambda, \beta} ({\bf r}).
\label{totfor}
\end{equation}
This conclusively proves our claim: the macroscopic elastic tensor can be 
interpreted as a net force acting on the primitive cell in response to a 
strain gradient. This must correspond to the basis sum of the 
force induced \emph{on individual atoms} (again in response a strain gradient), which
is nothing but the flexoelectric force-response tensor, ${\bf C}$.

\subsection{Dynamic and static flexoelectricity}

It is apparent from the above derivations that the flexoelectric internal-strain
tensors (either type-I, ${\bf N}$, or type-II, ${\bf L}$) directly depend 
on the atomic masses via $\hat{\bf T}$ or $\hat{\bf C}$ [Eq.~(\ref{htk})].
This poses a conceptual problem, as many experiments involve an external load
that is \emph{statically} applied to a sample. If the flexoelectric tensor
is an intrinsically \emph{dynamic} quantity, as one would conclude based on
its mass dependence, is there a hope that our theory might be able to 
interpret such data? Tagantsev~\cite{Tagantsev} argued that one must consider
two distinct versions of the flexoelectric tensor, a static and a dynamic 
one. 
In the following we shall discuss this point in detail, in light of the 
results presented so far.

%Before 
%The physical justification of why the flexoelectric coefficient \emph{must} 
%depend on masses rest on the analysis of the last few paragraphs, 
To start with, it is useful to analyze the physical origin of the aforementioned
mass dependence in the dynamical context of a long-wavelength acoustic phonon.
Suppose that the perturbed crystal is characterized by a macroscopic strain
gradient $\varepsilon_{\beta \gamma,\lambda}({\bf r},t)$ at a given
position ${\bf r}$ and time $t$.
%Suppose that at a given position ${\bf r}$ and time $t$ the perturbed crystal is
%characterized by a macroscopic strain gradient $\varepsilon_{\beta \gamma,\lambda}$.
% 
According to Eq.~\eqref{totfor}, the unit cell at (${\bf r}$, $t$) feels 
a net force that depends on $\varepsilon_{\beta \gamma,\lambda}$,
on the macroscopic elastic tensor of the material and 
on its mass density. Such force produces, in turn, an acceleration ${\bf a}$
equal to
$$a_\alpha = \frac{\Omega \mathcal{C}_{\alpha \lambda, \beta \gamma}}{M} 
\varepsilon_{\beta \gamma,\lambda}.$$ 
%
%undergoes an 
%acceleration that is equal to 
Then, in the moving frame of such material point, each individual atom
must experience, in addition to the force induced by the strain gradient
in the laboratory frame, a fictitious force equal to $-m_\kappa a_\alpha$,
$$\tilde{f}^\kappa_\alpha = -m_\kappa \frac{\Omega \mathcal{C}_{\alpha \lambda, \beta \gamma}}{M} 
 \varepsilon_{\beta \gamma,\lambda}.$$
Such fictitious force indeed coincides with the mass-dependent contribution to the 
\emph{compensated} flexoelectric tensor, $\hat{C}^\kappa_{\alpha \lambda,\beta \gamma}$.
The fact that the sublattice sum of $\hat{C}^\kappa_{\alpha \lambda,\beta \gamma}$
vanishes identically is consistent with the obvious fact that, in its own accelerated frame, 
the material point does not move by definition, so the total force acting on it must vanish.
%
%While the above arguments clearly establish the dynamical nature of the flexoelectric
%tensor, 
%According to the equivalence principle of general relativity, 

The above arguments clearly establish  the dynamical nature of the flexoelectric 
internal-strain tensor that we derived in Section~\ref{intstrains}. 
Nevertheless, it is straightforward to show that the same tensor (${\bf L}$ or ${\bf N}$)
is equally valid to describing the static response of the system to a uniform gravitational 
field.
This can be demonstrated, for example, by assuming that an external force, proportional
to the mass $m_\kappa$, is applied to every atom of the crystal, and by performing 
the explicit derivation all over again.
Alternatively, and more simply, one could invoke the 
equivalence principle of general relativity: the fictitious forces occurring
in the accelerated frame described above must be analogous to those occurring in 
an inertial frame under the action of a gravitational field. 
%The proof then immediately follows from 
%the discussion of the above paragraphs.
%
%(That gravity can produce a strain gradient is intuitively easy to understand.
%Consider for example a thick slab of a deformable solid sitting on a hard
%support; the vertical compressive force acting on the top layers is different than 
%that acting on the bottom layers, resulting in a longitudinal strain gradient.)
%Are there any other physically meaningful solutions?
%Strictly speaking, an alternative (\emph{albeit} unrealistic) 
%way of eliminating the problematic net force would consist in applying a strong 
%gravitational field to the sample. This would lead, again, to a mass-dependent
%flexoelectric response that is exactly equivalent to that of a sound wave,
%although gravitation is way too weak to induce any measurable effect
%in a nanoscale object.

%In absence of either inertial or strong enough gravitational fields 
%(i.e. in a static experiment performed on Earth) 
Assuming a static regime and that the effects of gravitation
are neglibly small on the experimentally relevant scale, the 
following condition for mechanical equilibrium must hold at every
point in the sample,
%In absence of gravitational fields, and assuming that the
%sample is at rest (i.e. in a typical static experiment), 
%each material point located in the \emph{interior} of the sample must be at mechanical 
%equilibrium, which means that the net force acting on it must vanish.
%
%This leads to an explicit condition on the strain gradient field,
\begin{equation}
\sum_{\beta \gamma \lambda} \mathcal{C}_{\alpha \lambda, \beta \gamma} \, \varepsilon_{\beta \gamma, \lambda} ({\bf r}) = 0.
\label{static}
\end{equation}
Eq.~\eqref{static} implies that an individual component of the strain 
gradient tensor, $\varepsilon_{\beta \gamma, \lambda}$ cannot be 
sustained statically at any point in a material unless 
$\mathcal{C}_{\alpha \lambda, \beta \gamma}=0$ for all $\alpha$.
Thus, in a static deformation field two (or more) 
inequivalent strain gradient components generally coexist, in such a 
way that their respective net force mutually cancels.
%e only way of compensating the 
%net force statically is by superimposing 
%
Let's see the consequences of this observation on the flexoelectric
response.
The lattice-mediated contribution (note that the other contributions to 
$\Delta \overline{\bf P}$ are independent of masses and therefore not a 
concern here) to the flexoelectric polarization is 
\begin{displaymath}
\Delta \overline{P}^{\rm latt}_\alpha = \frac{Z^*_{\kappa, \alpha \rho}}{2\Omega}  \widetilde{\Phi}^{(0)}_{\kappa \rho \kappa' \chi}  \left( 
C^{\kappa'}_{\chi \lambda, \beta \gamma} - \frac{\Omega m_\kappa}{M} \mathcal{C}_{\chi \lambda, \beta \gamma} \right)
\varepsilon_{\beta \gamma, \lambda}.
\end{displaymath}
%L^\kappa_{\rho \beta \gamma \lambda} 
By summing up all the components of $\varepsilon_{\beta \gamma, \lambda}$ and by
imposing the equilibrium condition Eq.~(\ref{static}), the mass-dependent part
disappears, and we have 
\begin{equation}
\Delta \overline{P}^{\rm latt}_\alpha = \frac{1}{2\Omega} Z^*_{\kappa, \alpha \rho} \widetilde{\Phi}^{(0)}_{\kappa \rho \kappa' \chi}  
C^{\kappa'}_{\chi \lambda, \beta \gamma}  
\varepsilon_{\beta \gamma, \lambda}({\bf r}).
\end{equation}
This expression depends, as it should, only on static properties of the 
material, i.e. the interatomic force constants and the linear response of the
electron cloud to a displacement of the nuclei.
%\begin{equation}
%
This result resolves the paradox that we formulated at the beginning of this 
subsection, and provides us with the conforting proof that our theory is
indeed applicable to both static and dynamic phenomena alike. 
In particular, the above derivations confirm that \emph{the flexoelectric 
tensor is a genuine dynamical quantity, but is readily applicable
to static regimes, as the troublesome mass dependence disappears in such
cases.}

There is, therefore, no need to seek the definition of a distinct ``static'' 
tensor~\cite{Tagantsev}, and in fact such a quest would be thwarted by the 
inherent indeterminacy of the problem.
Suppose we have found some definition of the (lattice-mediated) tensor that reproduces
the results of static measurements, and call it $\mu^{\rm II,a}_{\alpha \lambda, \beta \gamma}$. 
It is straightforward to see that any of the infinite variants of 
$\bm{\mu}^{\rm II,a}$ that can be obtained by writing
$$%\begin{equation}
\mu^{\rm II,b}_{\alpha \lambda, \beta \gamma} = \mu^{\rm II,a}_{\alpha \lambda, \beta \gamma} + 
Z^*_{\kappa, \alpha \rho} \widetilde{\Phi}^{(0)}_{\kappa \rho \kappa' \chi} \lambda_{\kappa'} \mathcal{C}_{\chi \lambda, \beta \gamma},
$$%\end{equation}
where $\lambda_{\kappa'}$ is a set of completely arbitrary values (apart
that their sum over $\kappa'$ must vanish), 
describes the \emph{static} behavior of the material equally well.
[Experimentally this fact translates in the formal impossibility (already 
pointed out in Ref.~\onlinecite{Pavlo}) of
measuring the individual components of the flexoelectric tensor 
by static means, even by combining the results of a vast number
of geometries and configurations.]
Of these infinite variants it suffices, of course, to choose one -- 
few will disagree on the mass-compensated $\bm{\mu}$ being the most sensible choice,
as it is good for \emph{both} static and dynamic phenomena.

As an academic excercise, it is interesting to briefly comment on other
conceivable ways (not necessarily realistic) of compensating 
the net flexoelectric force on the unit cell, which might make
sense in the context of a computational or \emph{Gedanken-} experiment.
This can be done by treating the masses $m_\kappa$ in Eq.~(\ref{hck}) 
as free parameters, and by setting them by hand to some arbitrary value.
Interestingly, one can show that by setting all masses to the same 
value we would recover Tagantsev's definition of the ``static'' 
flexoelectric tensor (whose appellation as \emph{static} appears 
therefore questionable).
In turn, by setting all masses to zero except one single atom (say, A) 
in the basis, we would allow only atoms A to feel inertia,
while the other species would be, at any given time, fully relaxed 
in the deformation field generated by the A sublattice.
This is reminiscent of the computational strategy used 
by Hong {\em et al.}~\cite{Hong-10}, of ``freezing in'' the displacements
of one sublattice while letting the others relax.
We stress that these ``alternative'' definition of the flexoelectric
tensor do not correspond to any physically measurable quantity,
and therefore their use appears of little interest, except as a 
conceptual aid to check the internal consistency of the theory.

%In Section~\ref{secmass} we shall discuss this point in detail. We
%shall argue that there is no such thing as a ``static'' tensor (nor an actual need 
%thereof), and that the mass dependence cancels out once the sample reaches
%static equilibrium with the applied loads.

%\subsubsection{Physical interpretation of the flexoelectric effect}

\section{The electronic response functions}

\label{elec}

%So far we have almost exclusively dealt with the polarization response to 
%atomic displacements. Earlier treatments, however, have almost systematically
%dealt with the \emph{charge-density} response to perturbations. 
The scope of this Section is to derive a number of useful properties
of the electronic response functions, i.e. the charge density and 
polarization.
We shall focus on their mutual relationship, on their
symmetry properties, and on their representation in
terms of localized functions.
%first to focus our attention on the charge density, 
%by establishing a number of formal links to the perturbative expansion of
%the polarization. 
%
%Next, we shall use these relationships to investigate more closely
%the physical nature and the symmetry properties of the electronic tensors
%(polarization and charge).
%
%Finally, we shall illustrate the soundness of our formalism by
%providing alternative derivations of well-known results, such 
%as Martin's theory of piezoelectricity~\cite{Martin}, and the theory of the 
%longitudinal flexoelectric response developed by Resta~\cite{Resta-10} and Hong 
%and Vanderbilt~\cite{Hong-11}

\subsection{Charge versus polarization response}

By using the fundamental relationship 
$$%\begin{equation}
\nabla \cdot {\bf P}({\bf r}) = -\rho({\bf r})
$$%\end{equation} 
one can verify that (recall the ${\bf q}$-dependent phase factor in both $\rho$ and ${\bf P}$; 
in the following equations we omit the dependence on ${\bf r}$ and use
$\partial / \partial r_\alpha \rightarrow \partial_\alpha$ in order to lighten 
the notation)
\begin{eqnarray}
\rho_{\kappa \beta}^{(0)} &=& -\partial_\alpha P_{\alpha, \kappa \beta}^{(0)}, \nonumber \\
\rho_{\kappa \beta}^{(1,\gamma)} &=& P_{\gamma, \kappa \beta}^{(0)} -
\partial_\alpha P_{\alpha, \kappa \beta}^{(1,\gamma)}, \nonumber \\
\rho_{\beta}^{(2, \gamma_1 \gamma_2)} &=& 
P_{ \gamma_1, \kappa \beta }^{ (1, \gamma_2 ) } + 
P_{ \gamma_2, \kappa \beta }^{ (1, \gamma_1 ) } 
- \partial_\alpha P_{\alpha, \kappa \beta}^{(2, \gamma_1 \gamma_2)}, \label{order2} \nonumber \\
\rho_{\kappa \beta}^{(3, \gamma_1 \gamma_2 \gamma_3)} &=& 
P_{ \gamma_1, \kappa \beta }^{ (2, \gamma_2 \gamma_3 ) } + 
P_{ \gamma_2, \kappa \beta }^{ (2, \gamma_3 \gamma_1 ) } +
P_{ \gamma_3, \kappa \beta }^{ (2, \gamma_1 \gamma_2 ) } \nonumber \\
 && - \partial_\alpha P_{\alpha, \kappa \beta}^{(3, \gamma_1 \gamma_2 \gamma_3)}, \label{order3} \nonumber \\
 & \ldots & \nonumber
%\rho_{\beta}^{(n, \alpha_1 \ldots \alpha_n)}({\bf r}) &=& 
%\sum_p P_{ \alpha_{p(1)}, \kappa \beta }^{ ( n-1, \alpha_{p(2)} \ldots \alpha_{p(n)} ) }({\bf r}) + \nonumber \\
% && - \partial_\gamma P_{\gamma, \kappa \beta}^{(n, \alpha_1 \ldots \alpha_n)}({\bf r}), \label{ordern}
\end{eqnarray}
where the rule to extend this to an arbitrary order is
self-explanatory.
It is interesting to look at the cell averages of the above expressions, 
as these are immediately relevant for the calculation of the
macroscopic electromechanical coefficients.
The cell average of the divergence of a periodic function is zero, 
and therefore we have
%\begin{eqnarray}
\begin{subequations}
\begin{align}
\overline{\rho}_{\kappa \beta}^{(0)} &= 0, \label{barrho0} \\
\overline{\rho}_{\kappa \beta}^{(1,\gamma)} &= \overline{P}_{\gamma, \kappa \beta}^{(0)}, \label{barrho1} \\
\overline{\rho}_{\kappa \beta}^{(2, \gamma_1 \gamma_2)} &= \overline{P}_{ \gamma_1, \kappa \beta }^{ (1, \gamma_2 ) } + 
\overline{P}_{ \gamma_2, \kappa \beta }^{ (1, \gamma_1 ) }, \label{barrho2} \\
\overline{\rho}_{\kappa \beta}^{(3, \gamma_1 \gamma_2 \gamma_3)} &= 
\overline{P}_{ \gamma_1, \kappa \beta }^{ (2, \gamma_2 \gamma_3 ) } + 
\overline{P}_{ \gamma_2, \kappa \beta }^{ (2, \gamma_3 \gamma_1 ) } +
\overline{P}_{ \gamma_3, \kappa \beta }^{ (2, \gamma_1 \gamma_2 ) }. \label{barrho3}
\end{align}
\end{subequations}
%\end{eqnarray}
%
As a first observation, note that the average of the induced charge upon rigid 
displacement of the sublattice $\kappa$ must be zero, to preserve
neutrality. 
Second, it is clear that the electronic polarization-response tensor 
at the order $n-1$ contains sufficient information (in general more 
than necessary) to fully determine the charge-response tensor at the order $n$.
In the $n=1$ case such a relationship is one-to-one -- both $\overline{\rho}^{(1)}$ and
$\overline{P}^{(0)}$ tensors correspond to the Born effective charge tensor (apart
from a trivial factor of volume).
The relationship at $n=2,3$ will be clarified in the following subsections.

\subsection{Extended and localized representations}

Several authors (starting from Martin~\cite{Martin}) have based their treatment of 
electromechanical effects on the charge-density response to the displacement
of an \emph{isolated atom}, rather than to extended collective modes as we
have done so far in this work.
In the following we shall establish the rigorous link between these two 
perspectives on the same problem, thus putting our own approach on firmer 
theoretical grounds.
To that end, we need to move to a \emph{localized} representation of
the response functions at a given order in ${\bf q}$. 
In close analogy to the theory of Wannier functions~\cite{Marzari/Vanderbilt:1997,Stengel/Spaldin:2005,RMP_Wannier}, 
we can achieve this via a Fourier transform in ${\bf q}$-space. 
For example, in the case of the charge density we have
\begin{equation}
f_{\kappa \beta} ({\bf r}- {\bf R}_{l\kappa}) = \frac{\Omega}{(2\pi)^3}
\int_{\rm BZ} d^3 q \, \rho_{\kappa \beta}^{\bf q} ({\bf r}) 
e^{-i {\bf q} \cdot ({\bf R}_{l\kappa}-{\bf r})} ,
\label{fk}
\end{equation}
where $f_{\kappa \beta} ({\bf r})$ is (analogously to Ref.~\onlinecite{Martin}) the
response to the displacement of an \emph{isolated} atom at the lattice site ${\bf R}_{l\kappa}$.
%and sublattice index $\kappa$. (At difference with the case of Wannier functions,
%the present theory is not affected by any phase indeterminacy, and the functions $f_{\kappa \beta}$ 
%correspond to physically observable quantities.~\cite{Martin}) 
%
Since we are considering a periodic bulk crystal, of course $f_{\kappa \beta}$ is 
independent of the cell index $l$.
Similarly, for the polarization we can readily extract the $\alpha$ component
of the polarization field, $\mathcal{P}_{\alpha, \kappa \beta}$, 
induced by a small displacement of the atom $l\kappa$ along $\beta$,
\begin{equation}
\bm{\mathcal{P}}_{\kappa \beta} ({\bf r}- {\bf R}_{l\kappa}) = \frac{\Omega}{(2\pi)^3}
\int_{\rm BZ} d^3 q \, {\bf P}_{\kappa \beta}^{\bf q} ({\bf r}) 
e^{-i {\bf q} \cdot ({\bf R}_{l\kappa}-{\bf r})} .
\label{pk}
\end{equation}
Before proceeding any further, however, we need to stop for a moment and
make a parenthetical digression.
%make an important consideration about macroscopic electrostatics.
%
In fact, the forthcoming derivations will heavily rely on the decay 
properties of the $\bm{\mathcal{P}}$ and $f$ functions in real space.
Such decay is fast enough only if the dependence of the extended 
response functions, ${\bf P}^{\bf q}$ and $\rho^{\bf q}$, on ${\bf q}$
is analytic across the full Brillouin zone, which brings us back to
macroscopic electrostatics.

%\subsubsection{Electrostatic screening}

Recall that in Section~\ref{dfpt} we discussed a procedure 
to ``cure'' the nonanalytic behavior of the electric 
fields near $\Gamma$, by simply removing the ${\bf G}=0$ term
from the self-consistent electrostatic potential. 
Such a prescription is, however, well-defined \emph{only in the
context of a Taylor expansion around} $\Gamma$, and is therefore 
unsuitable to the present purposes -- the above Fourier 
transforms are integrals over the full Brillouin zone.
To have a truly localized representation of the charge-density 
(and polarization) response to the displacement of an isolated
atom, we need to devise a strategy that ensures: (i) the analyticity 
of ${\bf P}^{\bf q}$ and $\rho^{\bf q}$ over all reciprocal space,
and (ii) their periodicity in ${\bf q}$, i.e. 
$\rho^{\bf q+G}({\bf r}) e^{i{\bf G \cdot r}} = \rho^{\bf q}({\bf r})$. 
Clearly, the ``${\bf G}\neq 0$'' prescription cannot satisfy both
requirements, hence the need for an alternative approach.

To address this issue, we shall follow the strategy proposed 
by Martin~\cite{Martin} of assuming that a very low-density gas of mobile
carriers is superimposed to the insulating crystal lattice.
As demonstrated in Section~\ref{secmobile}, this assumption modifies 
the Coulomb kernel as follows,
\begin{equation}
\frac{4\pi}{q^2} \rightarrow \frac{4\pi}{q^2 + k_{\rm TF}^2}, % \left( 1-e^{-q^2 \sigma^2 / 4} \right).
\label{kq}
\end{equation}
where $k_{\rm TF}$ is the inverse of the Thomas-Fermi screening length.
As the Coulomb kernel is now analytic over all reciprocal space,
we can now safely include the ${\bf G}=0$ component of the electrostatic
potential, and therefore fullfill both (i) and (ii).
Of course, the response functions that result from the former (``${\bf G}\neq0$'')
and the latter (``Thomas-Fermi'' or ``TF'') procedures generally differ (and
further depend on $k_{\rm TF}$ in the latter case).
One can show (see Section~\ref{secmobile}), however, that the lowest orders
in their Taylor expansion around $\Gamma$ --and this includes all quantities 
that enter the piezoelectric and flexoelectric tensor-- are unsensitive to
whether one uses ``${\bf G}\neq0$'' or ``TF''.
% (and are, therefore, independent of
%$k_{\rm TF}$ in the latter case).

Based on this result, we shall implicitly assume from now on that
all response functions are defined within the ``TF'' model,
%
%We can thus 
and proceed to deriving the relationship between the 
${\bf q}$-expansion of their extended representation ($\rho^{\bf q}$
and ${\bf P}^{\bf q}$) and the real-space moments of their
localized representation ($f$ and $\bm{\mathcal{P}}$).
%
%polarization and charge-density response and
%their localized representations.
%
For both physical quantities, the converse Fourier transforms can be written as
\begin{eqnarray}
\rho_{\kappa \beta}^{\bf q} ({\bf r}) &=& \sum_l f_{\kappa \beta} ({\bf r}- {\bf R}_{l\kappa}) e^{i {\bf q} \cdot ({\bf R}_{l\kappa}-{\bf r})} \label{converserho} \\
\bm{\mathcal{P}}_{\kappa \beta}^{\bf q} ({\bf r}) &=& \sum_l {\bf P}_{\kappa \beta} ({\bf r}- {\bf R}_{l\kappa}) e^{i {\bf q} \cdot ({\bf R}_{l\kappa}-{\bf r})}
\label{conversep}
\end{eqnarray}
By differentiating Eq.~(\ref{converserho}) in $q$-space, we readily obtain 
%the desired formulas,
%formal relationship between the extended functions $\rho^{(0,1,2)}$ 
%and the localized functions $f$,
%\begin{eqnarray}
\begin{subequations}
\begin{align}
\rho_{\kappa \beta}^{(0)}({\bf r}) &= \sum_{l} f_{\kappa \beta} ({\bf r - R}_{l\kappa}), \label{rho0} \\
\rho_{\kappa \beta}^{(1,\gamma)}({\bf r}) &=  \sum_{l} f_{\kappa \beta} ({\bf r - R}_{l\kappa}) (r-R_{l\kappa})_\gamma, \label{rho1} \\
\rho_{\kappa \beta}^{(2,\gamma \lambda)} ({\bf r})&= \sum_{l}f_{\kappa \beta} ({\bf r - R}_{l\kappa}) (r-R_{l\kappa})_\gamma(r-R_{l\kappa})_\lambda. \quad
\label{rho2}
\end{align}
\end{subequations}
%\end{eqnarray}
%
Analogous formulas
link the extended $P^{(0,1,2)}$ to the localized $\mathcal{P}$. (These
are simply obtained by replacing $\rho \rightarrow P_{\alpha}$ and
$f \rightarrow \mathcal{P}_\alpha$ in the above expressions.)
It is useful to introduce the moments of the localized functions,
by means of the following integrals over all space,
\begin{eqnarray}
Q^{(n,\gamma_1 \ldots \gamma_n)}_{\kappa \beta} &=& \int d^3 r \, f_{\kappa \beta} ({\bf r}) r_{\gamma_1} \ldots r_{\gamma_n}, \\
J^{(n,\gamma_1 \ldots \gamma_n)}_{\alpha, \kappa \beta} &=& \int d^3 r \, \mathcal{P}_{\alpha, \kappa \beta} ({\bf r}) r_{\gamma_1} \ldots r_{\gamma_n}.
\end{eqnarray}
$Q^{(0)}_{\kappa \beta}$ vanishes because of charge neutrality, 
while $Q^{(n=1,2,3)}$ are, respectively, the dipolar ($n=1$), 
quadrupolar~\cite{Martin} ($n=2$) and octupolar~\cite{Resta-10,Hong-11} 
($n=3$) moments of the induced charge distribution $f$.
Following Eqs.~(\ref{rho0}), (\ref{rho1}) and (\ref{rho2}), such moments are 
trivially related to the cell-average of the extended functions that we have used 
throughout this work,
\begin{eqnarray}
Q^{(n,\gamma_1 \ldots \gamma_n)}_{\kappa \beta} &=& \Omega \overline{\rho}^{(n,\gamma_1 \ldots \gamma_n)}_{\kappa \beta}, \\
J^{(n,\gamma_1 \ldots \gamma_n)}_{\alpha, \kappa \beta} &=& \Omega \overline{P}^{(n,\gamma_1 \ldots \gamma_n)}_{\alpha,\kappa \beta}.
\end{eqnarray}
Later in this Section we shall use these results to demonstrate the 
consistency of the present theory with earlier works on the subject.
Before doing that, we need to demonstrate a number of key symmetry 
properties of the response functions, which we shall discuss in the following.

\subsection{Symmetry properties}

\label{secsymm}

The symmetry properties of the charge response were
established by Martin~\cite{Martin}; we shall translate
his results to our notation, and later extend these ideas to
the polarization response.
We shall be concerned with the \emph{basis sums} of the electronic 
response functions, 
\begin{eqnarray}
\rho_{\beta}^{(n,\ldots)}({\bf r}) &=&  \sum_\kappa \rho_{\kappa \beta}^{(n,\ldots)}({\bf r}), \\
P_{\alpha \beta}^{(n,\ldots)}({\bf r}) &=&  \sum_\kappa P_{\alpha, \kappa \beta}^{(n,\ldots)}({\bf r}),
\label{rhotrans}
\end{eqnarray}
which are relevant for the frozen-ion contribution
to the electromechanical tensors.
Translation invariance requires~\cite{Martin} that
\begin{equation}
\rho_\beta^{(0)}({\bf r}) = -\frac{\partial \rho({\bf r})}{\partial r_\beta},
\end{equation}
where $\rho({\bf r})$ is the ground-state charge density of the crystal at rest.
The polarization counterpart of this property reads
\begin{equation}
P_{\alpha \beta}^{(0)}({\bf r}) = \delta_{\alpha \beta} \rho({\bf r}).
\label{ptrans}
\end{equation}
The physics behind Eq.~(\ref{ptrans}) is clear: upon rigid translation of
the crystal, the induced current density must be proportional to the charge density at ${\bf r}$,
and directed along the translation direction (i.e. the electron cloud must undergo the same rigid 
shift as the nuclei).

Rotation invariance states that the electronic charge density must 
accompany a rigid \emph{rotation} of the atomic lattice. This leads~\cite{Martin} 
immediately to the invariance of $\rho_\beta^{(1,\gamma)}$ with respect
to $\beta \leftrightarrow \gamma$ exchange,
\begin{equation}
\rho_\beta^{(1,\gamma)}({\bf r}) = \rho_\gamma^{(1,\beta)}({\bf r}).
\end{equation}
[This result follows from Eq.~(18c) of Ref.~\onlinecite{Martin}, by using 
Eq.~(\ref{rhotrans}), (\ref{rho0}) and (\ref{rho1}).]
In the context of the polarization, the same invariance property holds,
\begin{equation}
P_{\alpha \beta}^{(1,\gamma)}({\bf r}) = P_{\alpha \gamma}^{(1,\beta)}({\bf r}).
\end{equation}
To see this, describe the rotation with a displacement ${\bf u}$ of a point ${\bf r}$ 
in the crystal as
\begin{equation}
u_\alpha = \epsilon^{\alpha \beta \gamma} \hat{\theta}_\beta r_\gamma,
\end{equation}
where $\hat{\bm{\theta}}$ is an axial vector and $\epsilon^{\alpha \beta \gamma}$
is the antisymmetric Levi-Civita symbol.
We impose as above that the charge density transforms the same way as the atomic lattice,
$$
\sum_{l\kappa} \sum_{\beta \gamma \lambda} \epsilon^{\beta  \gamma \lambda} 
\mathcal{P}_{\alpha,\kappa \beta} ({\bf r - R}_{l\kappa})  \hat{\theta}_\gamma ({\bf R}_{l\kappa})_\lambda = 
$$
\begin{equation}
= \sum_{\gamma \lambda} \epsilon^{\alpha  \gamma \lambda} \rho({\bf r}) \hat{\theta}_\gamma r_\lambda,
\end{equation}
where we have written the sums over the Cartesian indices explicitly for clarity.
By using Eq.~(\ref{ptrans}) we can write 
$$
\rho({\bf r}) = \sum_\beta P_{\alpha \beta}^{(0)}({\bf r}) = \sum_{l \kappa \beta} \mathcal{P}_{\alpha,\kappa \beta} ({\bf r - R}_{l\kappa}).
$$
This leads immediately to
\begin{equation}
\sum_{l\kappa} \sum_{\beta \gamma \lambda} \epsilon^{\beta  \gamma \lambda} \hat{\theta}_\gamma P_{\alpha \beta}^{(1,\lambda)}({\bf r}) = 0,
\end{equation}
which must be satisfied for any $\hat{\bm{\theta}}$, thus completing the proof.
We have accumulated enough results now to compare our theory to 
earlier treatments of the electromechanical problem, respectively
piezoelectricity and flexoelectricity.

\subsection{Martin's theory of piezoelectricity}

\label{secmartin}

Based on the theory developed in this work, we can write the polarization
response to a deformation (to first order in ${\bf q}$, which includes the
relevant terms for piezoelectricity) as
\begin{eqnarray}
%\begin{equation}
\overline{P}_\alpha ({\bf r}) &=&  \varepsilon_{\beta \gamma} ({\bf r}) \, 
\left[ \overline{P}^{(0)}_{\alpha, \kappa \rho} \Gamma^\kappa_{\rho \beta \gamma}
-  \overline{P}^{(1,\gamma)}_{\alpha \beta} \right]. 
\label{piezo}
\end{eqnarray}
As before, $\overline{P}^{(0)}_{\alpha, \kappa \beta} =  Z^*_{\kappa,\alpha \beta} / \Omega$ 
is the Born effective charge tensor divided by the volume. 
Concerning $\overline{P}^{(1,\gamma)}_{\alpha \beta}$, recall the relationship 
between the polarization and the charge response functions, Eq.~(\ref{barrho2}),
\begin{equation}
\overline{P}^{(1,\gamma)}_{\alpha \beta} + \overline{P}^{(1,\alpha)}_{\gamma \beta} = \overline{\rho}^{(2,\alpha \gamma)}_\beta.
\end{equation}
Next, observe that $\overline{P}^{(1,\gamma)}_{\alpha \beta}$ is symmetric with respect
to $\gamma \leftrightarrow \beta$, while $\overline{\rho}^{(2,\alpha \gamma)}_\beta$ is
symmetric with respect to $\gamma \leftrightarrow \alpha$. This means that the
two tensors have the same number of independent entries (eighteen); hence,
the above relationship can be readily inverted to yield the polarization tensor 
components as a unique function of the charge response tensor,
%. As the two tensors have the same number of independent entries (eighteen),
%the solution is unique and reads
\begin{equation}
\overline{P}^{(1,\gamma)}_{\alpha \beta} = \frac{1}{2 \Omega} \left[ Q^{(2,\alpha \gamma)}_\beta + 
Q^{(2, \gamma \beta)}_\alpha -  Q^{(2,\alpha \beta)}_\gamma \right],
\label{rich}
\end{equation}
where we have expressed the latter in terms of the induced quadrupolar 
moments. By inserting Eq.~(\ref{rich}) into Eq.~(\ref{piezo}) we recover Eq.~(26) 
of Ref.~\onlinecite{Martin}.

The same result could be deduced, via a somewhat clumsier algebra, from the 
total charge density response to a deformation. At first order in ${\bf q}$
the net induced charge is zero; therefore, we need to push our expansion
to the second order in ${\bf q}$, i.e. to the strain-gradient term
\begin{eqnarray}
\overline{\rho} ({\bf r}) &=& 
\eta_{\beta, \gamma \lambda} ({\bf r}) \, 
\Big[ \overline{\rho}^{(2,\gamma \lambda)}_\beta - \nonumber \\
&& - \frac{1}{2} \Gamma^\kappa_{\rho \beta \gamma} \overline{\rho}_{\kappa \rho}^{(1,\lambda)} - 
\frac{1}{2} \Gamma^\kappa_{\rho \beta \lambda} \overline{\rho}_{\kappa \rho}^{(1,\gamma)} \Big]
%+ \nonumber \\
%&& + \frac{1}{2\Omega} Z^*_{\kappa, \alpha \rho}  N^\kappa_{\rho \beta \gamma \lambda} 
%
% \mu^{\rm I}_{\alpha \beta, \gamma \lambda}.
%\end{equation}
\end{eqnarray}
[This equation was obtained by replacing the polarization response tensors
in Eq.~(\ref{mui}) with the corresponding charge density tensors, and by 
eliminating the vanishing terms.]
By rewriting the same expression in terms of the type-II strain gradient tensor, $\varepsilon_{\beta \gamma, \lambda}$,
we readily obtain
\begin{equation}
\overline{\rho} ({\bf r}) = \varepsilon_{\beta \gamma, \lambda} ({\bf r})  \, \Big[
\overline{P}^{(1,\gamma)}_{\lambda \beta} -
\frac{Z^*_{\kappa,\lambda \rho}}{\Omega} \Gamma^\kappa_{\rho \beta \gamma} \Big].
\label{piezo2}
\end{equation}
[We have used Eq.~(\ref{rich}) and $\Omega \overline{\rho}_{\kappa \rho}^{(1,\lambda)} = Z^*_{\kappa,\lambda \rho}$.]
This is very similar to Eq.~(\ref{piezo}), except that here we have the charge 
instead of the polarization, the type-II strain gradient instead of the strain tensor,
and a minus sign.
To demonstrate that Eq.~(\ref{piezo}) and (\ref{piezo2}) are, in fact, 
equivalent one just needs to write Eq.~(\ref{piezo}) in a 
macroscopic strain gradient,
\begin{equation}
\overline{P}_\alpha ({\bf r}) =  \varepsilon_{\beta \gamma,\lambda} ( r_\lambda e_{\alpha \beta \gamma} + 
\mu^{\rm II}_{\alpha \lambda,\beta \gamma} ),
\end{equation}
where $e_{\alpha \beta \gamma}$ is the piezoelectric tensor, and the flexoelectric
contribution is included for completeness.
By applying $\nabla \cdot {\bf P} = -\rho$ one readily recovers Eq.~(\ref{piezo2}),
thus completing the proof.

This derivation tells us that, in order to extract the piezoelectric tensor, one 
can look indifferently at the polarization induced by a strain or at
the \emph{net charge} associated with a strain gradient. 
In the latter case, the purely electronic (frozen-ion) contribution
is written in terms of the quadrupolar response tensor, consistent 
with Martin's arguments.
Interestingly, our derivation also shows that the peculiar symmetrization of the 
quadrupolar tensor indices, Eq.~(\ref{rich}), which was inferred by Martin 
via symmetry arguments, is intimately related to the 
analogous relationship, Eq.~(\ref{eqeta2}), between the type-I and 
type-II \emph{strain gradient} tensors. 
In particular, the tensor $Q^{(2,\gamma \lambda)}_\beta / (2 \Omega)$ describes 
the macroscopic charge density response to a frozen-ion strain gradient of type I, 
$\eta_{\beta, \gamma \lambda}$; this can be recast into type-II form via the 
symmetrization Eq.~\eqref{rich}, yielding the frozen-ion piezoelectric tensor.

As a closing remark, note that the present theory is consistent with
the definition of the frozen-ion piezoelectric coefficient proposed 
by Hong and Vanderbilt~\cite{Hong-11}. In our notation, Eq.~(13) of
Ref.~\onlinecite{Hong-11} reads
\begin{equation}
\bar{e}_{\alpha, \beta \gamma} = -\frac{1 }{\Omega} J^{(1,\gamma)}_{\alpha \beta} = -\overline{P}^{(1,\gamma)}_{\alpha \beta}.
\label{hong}
\end{equation}

\subsection{Earlier treatments of the
flexoelectric problem}
%Resta and Hong/Vanderbilt 
%theories of flexoelectricity}

We shall discuss the literature works that are most relevant 
to the present theory, and that have most directly contributed to
its conceptual foundation.
The connection to Tagantsev's theory~\cite{Tagantsev} of the 
lattice-mediated response has been extensively discussed in 
Section~\ref{longw}.
There we have pointed out the crucial necessity for an adequate 
treatment of the nonanalyticities due to the long-range electrostatic
forces, which we implemented by suppressing the ${\bf G}=0$ term
in the self-consistent electrostatic potential.
%but apart from this issue, our derivations of the internal strain
%response essentially follow Tagantsev's approach.
%
Interestingly, by using the ``screened'' Coulomb kernel described
in Section~\ref{elec} we can draw an even closer link
to Ref.~\onlinecite{Tagantsev}, i.e. express the $\Phi^{(0,1,2)}$
matrices that we used in this work in terms of the moments of the real-space 
force constants.
Define, following the prescriptions of Section~\ref{dfpt}, the real-space 
force constant matrix as a Fourier transform of the reciprocal-space one,
\begin{equation}
\Phi_{\kappa \alpha,\kappa' \beta}^l  = \frac{\Omega}{(2\pi)^3} \int_{\rm BZ} d^3 q \, \Phi_{\kappa \alpha,\kappa' \beta}^{\bf q}
e^{-i {\bf q} \cdot ({\bf R}_{l}+\bm{\tau}_{\kappa'}-\bm{\tau}_\kappa)}.
\label{dirphi}
\end{equation}
(Note the phase factor dependent on the relative sublattice positions,
$\bm{\tau}_{\kappa'}-\bm{\tau}_\kappa$; see Section~\ref{dfpt}
for an explanation.)
Thanks to the screening of the long-range Coulomb forces, the real-space
$\Phi$ decays exponentially as a function of the interatomic distance,
${\bf R}_{l}+\bm{\tau}_{\kappa'}-\bm{\tau}_\kappa$.
This means that the moments of $\Phi^l$ are well defined up to any order.
Consider now the converse transform,
\begin{equation}
\Phi_{\kappa \alpha,\kappa' \beta}^{\bf q} = \sum_l \Phi_{\kappa \alpha,\kappa' \beta}^l 
e^{i {\bf q} \cdot ({\bf R}_{l}+\bm{\tau}_{\kappa'}-\bm{\tau}_\kappa)}.
\label{convphi}
\end{equation}
By differentiating with respect to the components of ${\bf q}$ we readily obtain
the desired link between the small-${\bf q}$ Taylor expansion of $\Phi^{\bf q}$ and
the moments of $\Phi^l$.

We shall now focus more specifically on the theory of the
electronic flexoelectric response that was proposed by Resta~\cite{Resta-10} (RR)
and Hong and Vanderbilt~\cite{Hong-11} (HV).
RR demonstrated that the (purely electronic) \emph{longitudinal} 
flexoelectric response in elemental crystals (this statement was later 
generalized to all insulating crystals by HV) is uniquely determined by the 
basis sum of the induced octupolar moments, $Q^{(3,\alpha \gamma \lambda)}_\beta$ 
in our notation.
To see this, recall our result for the flexoelectric polarization 
in a ``frozen-ion'' sound wave under fixed-$\mathcal{E}$ electrical
boundary conditions (short-circuit, SC),
%~\footnote{Strictly speaking, all the following formulas  
%  should have a factor of ${\bf q} \cdot \bm{\epsilon} \cdot {\bf q}$ at the denominator, 
%  where $\bm{\epsilon}$ is the high-frequency macroscopic dielectric tensor. 
%  Such a factor, originating from the ,
%  have been omitted to avoid overburdening the notation.}
\begin{equation}
\overline{P}^{\rm SC}_\alpha ({\bf r}) = 
-\frac{U_\beta q_\gamma q_\lambda}{2} \, 
\overline{P}^{(2,\gamma \lambda)}_{\alpha \beta} e^{i {\bf q\cdot r}},
\end{equation}
Now assume that the wave is purely longitudinal ($U_\beta = U \hat{q}_\beta$), 
and consider the longitudinal component of the \emph{dielectrically screened} 
(i.e. we assume a phonon propagating in an ideal insulator, in absence
of mobile carriers) polarization by 
projecting it over $\hat{\bf q}$,
\begin{equation}
\overline{P}_{\hat{\bf q}} ({\bf r}) = 
-\frac{U q^2}{2 \, \hat{\bf q} \cdot \bm{\epsilon} \cdot \hat{\bf q}} 
\hat{q}_\alpha \hat{q}_\beta \hat{q}_\gamma \hat{q}_\lambda
\, \overline{P}^{(2,\gamma \lambda)}_{\alpha \beta} e^{i {\bf q\cdot r}},
\end{equation}
where we have inserted a factor of $\hat{\bf q} \cdot \bm{\epsilon} \cdot \hat{\bf q}$
($\bm{\epsilon}$ is the high-frequency dielectric tensor) to account for the fixed-$D$
electrical boundary conditions that characterize a long-wave phonon along the propagating direction.
Note that the amplitude of the longitudinal strain gradient tensor along ${\bf q}$ is
$\eta_{\hat{\bf q}} = -U q^2 e^{i {\bf q\cdot r}}$; also, we use $\overline P^{(2)} = J^{(2)}/\Omega$ 
and Eq.~(\ref{barrho3}),
%given that summation
%is performed on all indices of the flexoelectric tensor, we can use 
\begin{equation}
J^{(2, \gamma \lambda)}_{\alpha \beta} +
J^{(2, \alpha \gamma)}_{\lambda \beta} +
J^{(2, \lambda \alpha)}_{\gamma \beta} = 
Q^{(3, \alpha \gamma \lambda)}_{\beta}.
\label{jjj}
\end{equation}
After a cyclic permutation of the indices we readily obtain
\begin{equation}
\frac{\partial \overline{P}_{\hat{\bf q}} ({\bf r})}{\partial \eta_{\hat{\bf q}}} =  
\frac{1}{6 \Omega}
\frac{\hat{q}_\alpha \hat{q}_\beta \hat{q}_\gamma \hat{q}_\lambda Q^{(3, \alpha \gamma \lambda)}_{\beta} }{ \hat{\bf q} \cdot \bm{\epsilon} \cdot \hat{\bf q}},
\end{equation}
where the numerator of the fraction is the longitudinal component of the 
octupolar tensor along $\hat{\bf q}$, consistent with RR and HV.
(The dielectric screening factor $\hat{\bf q} \cdot \bm{\epsilon} \cdot \hat{\bf q}$
is implicit in the multipolar moments defined by HV and RR, see Section~\ref{ebcq} for a 
detailed derivation.) 
%$Q^{(3)}_{\beta, \alpha \gamma \lambda}$ tensor defined in this 
%work and the octupolar moments 

Unlike the piezoelectric case, Eq.~(\ref{jjj}) cannot be inverted to 
express the $J^{(2)}$ tensor as a function of $Q^{(3)}$ -- the octupolar 
charge response contains enough information to describe the (purely
electronic) longitudinal flexoelectric effect, but additional 
data, contained in $J^{(2)}$, is necessary to describe the transversal
response~\cite{Hong-11}.
To address the latter, HV proposed to write [Eq.~(13) therein,
expressed in our notation]
\begin{equation}
\bar{\mu}^{\rm I}_{\alpha  \beta,\gamma\lambda} = \frac{1}{2\Omega} J^{(2,\gamma \lambda)}_{\alpha \beta}.
\end{equation}
This formula is, again, fully consistent with the results derived here.

\section{Nonanalytic behavior of the response functions at $\Gamma$}

\label{cochran}

%\subsection{Introduction}

In several part of this work we have stressed that the long-range electrostatic
interactions are responsible for a nonanalytic behavior of the response functions
at the $\Gamma$ point of the Brillouin zone.
We have also argued that, for a correct derivation of the electromechanical 
coupling tensors, these macroscopic fields need to be suppressed, either by 
removing the ${\bf G}=0$ term in the self-consistent electrostatic potential
(Section~\ref{longw}) or by appropriately screening the Coulomb kernel 
by means of a low-density gas of mobile charges (Section~\ref{elec}).
In this Section we shall provide a formal justification for these two
apparently dissimilar prescriptions, and show that they are indeed equivalent
in the context of the piezoelectric or flexoelectric response of a generic
insulator.
We shall also clarify the physical nature of the aforementioned nonanalyticity, 
and its impact on the main response functions considered in this work, the 
charge density and the interatomic force-constant matrix.

That the force-constant matrix is nonanalytic at $\Gamma$ has been well known 
since the early days of lattice-dynamics theory~\cite{Born/Huang}. 
Macroscopic electric fields have a dramatic impact on the propagation of
optical phonons in a neigborhood of the zone center, as they are responsible
for the frequency splitting between longitudinal and transverse modes (Lyddane-Sachs-Teller
relationship~\cite{lst}, LST henceforth).
Cochran and Cowley~\cite{Cochran/Cowley} showed, based on phenomenological
arguments, that the LST relationship holds in a generic crystalline insulator.
Later, the microscopic expressions for the interatomic force constants
(together with the Cochran-Cowley formula) were rigorously derived, based on 
a fundamental quantum-mechanical framework, by Pick, Cohen and Martin.~\cite{rmm_thesis}
More recently the LST relationship was revisited by Resta~\cite{Resta_LST} in
the context of magnetoelectric materials, where both electric and magnetic
fields were found to be important in the long-wavelength limit.

Unfortunately, all the aforementioned works have limited their analysis to
the nonanalyticities of the force-constant matrix at the lowest 
(zero) order in the phonon wavevector ${\bf q}$.
This is by far the most important term in the context of lattice 
dynamics, but it is insufficient to the scopes of the present study, 
where an expansion of the response functions up to (and including) order 
$\mathcal{O}(q^2)$ is needed to access the relevant electromechanical 
tensors. 
%
%Of course, there is no need to worry about these issues at the
%macroscopic level, where a knowledge of the electromechanical
%tensors calculated at zero $\mathcal{E}$ suffices for a complete 
%solution of the Poisson equation of electrostatics in a given 
%deformation field.
%
%
%Detailed information about the behavior of the force-constant 
%matrix at small $q$ is, however, important in the context of atomistic
%modeling, especially if the parameters of the force field are 
%to be derived from first-principles. 
%
%In particular, pushing the accuracy of the 
%model to order $q$ and $q^2$ is mandatory to correctly describe the 
%piezoelectric and flexoelectric properties of the material, respectively.
%
%xplicit calculation of the above expressions (with their
%full $\hat{\bf q}$-dependence) might be therefore useful for achieving an 
%accurate sampling of the phonon dispersion curves at small $q$, while keeping
%the computational workload to a moderate level. 
%
%For example, we expect that in piezoelectric materials the correct sound velocity may be 
%difficult to describe unless $\Phi_{\kappa \alpha, \kappa' \beta}^{{\rm mac},{(1)}}$ is explicitly
%accounted for.
%
%
In the following we shall address this point by extending the Cochran-Cowley 
formula to $\mathcal{O}(q^2)$, showing that the quadrupolar and octupolar 
moments of the charge response enter naturally as higher-order counterparts 
of the dynamical Born charge tensors. 
In particular, they mediate long-range 
interatomic force constants that decay as $r^{-4}$ and $r^{-5}$, respectively,
as one would expect for  classical dipole-quadrupole and dipole-octupole /
quadrupole-quadrupole terms.
Interestingly, we also find an additional $r^{-5}$ term, of less obvious 
physical interpretation, that is related to the ${\bf q}$-dispersion of the macroscopic
dielectric tensor.

\subsection{The dielectric matrix approach}

We shall frame the following discussions with an exact all-electron
description of the periodic solid in mind. This means that the 
nuclei are described by $\delta$ functions, each carrying a
positive charge that corresponds to its atomic number.
All the physical properties of relevance to the present work
can then be expressed in terms of the \emph{dielectric matrix}
(or the closely related polarizability matrix), which describes
the response of the screened electrostatic potential to an
external perturbation. %(Unless otherwise specified, we assume
%that the screening is operated by bound charges.)
%
%The dielectric properties of a crystalline insulator can be 
%conveniently describes in terms of the response of the electron
%density (and, in turn, of the electrostatic potential) to an
%external perturbation.
%
The fundamental law relating the total (screened) potential $V$ 
to the external perturbing potential $V^{\rm ext}$ and the induced 
potential $V^{\rm ind}$ (related to the rearrangement of the
electron cloud that follows the perturbation) is, in full generality,
\begin{equation}
V({\bf r}) =  V^{\rm ext} ({\bf r}) + V^{\rm ind}({\bf r}).
\end{equation}
It is most practical to exploit the periodicity of the 
system and work in reciprocal space, where the above 
relationship reads
\begin{equation}
V_{\bf G}({\bf q}) =  V^{\rm ext}_{\bf G} ({\bf q}) + V^{\rm ind}_{\bf G}({\bf q}).
\label{vg}
\end{equation}
(We assume a monochromatic external perturbation of wavevector ${\bf q}$,
and we expand all quantities on the usual reciprocal-lattice grid,
indexed by ${\bf G}$.)
The induced potential $V^{\rm ind}$ is due to the electrostatic 
perturbation produced by the induced charge,
\begin{equation}
V^{\rm ind}_{\bf G}({\bf q}) = \frac{4 \pi}{|{\bf q + G}|^2} \rho^{\rm ind}_{\bf G}({\bf q}).
\end{equation}
In turn, the induced charge can be written in terms of the total (screened)
potential $V$ by introducing the polarizability matrix
\begin{equation}
\rho^{\rm ind}_{\bf G}({\bf q}) = \sum_{{\bf G}'} \Pi_{\bf GG'}({\bf q}) V_{\bf G'}({\bf q}).
\label{indpol}
\end{equation}
[Note that the matrix $\Pi_{\bf GG'}({\bf q})$ defined by Eq.~(\ref{indpol})
corresponds to the symbol $\pi ({\bf q+G, q+G'})$ of Ref.~\onlinecite{rmm_thesis}.]
By combining the above, Eq.~(\ref{vg}) can be written as
\begin{equation}
%\left[ \delta_{\rm GG'} - \frac{4 \pi}{|{\bf q + G}|^2} \Pi_{\bf GG'}({\bf q}) \right] 
\sum_{\bf G'} \epsilon ({\bf q+G, q+G'}) V_{\bf G'}({\bf q})
= V^{\rm ext}_{\bf G} ({\bf q}),
\label{vvv}
\end{equation}
where $\epsilon$ is the static dielectric matrix,~\cite{rmm_thesis}
\begin{equation}
\epsilon ({\bf q+G, q+G'}) = \delta_{\rm GG'} - \frac{4 \pi}{|{\bf q + G}|^2} \Pi_{\bf GG'}({\bf q}).
\label{epsq}
\end{equation}
In the context of the present work, we find it convenient to study the
potential response to an external \emph{charge} perturbation (e.g. produced
by the displacement of a nucleus from its equilibrium lattice position), 
rather than a to potential,
\begin{equation}
%\left[ \frac{|{\bf q + G}|^2}{4 \pi} \delta_{\rm GG'} - \Pi_{\bf GG'}({\bf q}) \right]
\sum_{\bf G'} \Xi_{\bf GG'}({\bf q}) V_{\bf G'}({\bf q})
= \rho^{\rm ext}_{\bf G} ({\bf q}),
\label{Poisson}
\end{equation}
where we have introduced a new matrix
\begin{equation}
\Xi_{\bf GG'}({\bf q}) = \frac{|{\bf q + G}|^2}{4 \pi} \delta_{\rm GG'} - \Pi_{\bf GG'}({\bf q}).
%\label{vvv}
\end{equation}
$\Xi$ is related to the dielectric matrix, Eq.~(\ref{epsq}) by $\Xi = K \epsilon$, 
where $K_{\bf GG'}({\bf q}) = \delta_{\bf GG'}{|{\bf q + G}|^2}/{4 \pi}$ is 
diagonal.
The matrix $\Xi_{\bf GG'}({\bf q})$ is Hermitian and analytic at all ${\bf q}$, 
since $\Pi$ enjoys both properties.~\cite{rmm_thesis}
%
%In the framework of an exact all-electron treatment of the many-electron 
%problem,

To obtain the electrostatic potential response to $\rho^{\rm ext}$ it 
suffices to invert $\Xi$,
\begin{equation}
V_{\bf G}({\bf q}) = \sum_{\bf G'} \Xi^{-1}_{\bf GG'}({\bf q}) \, \rho^{\rm ext}_{\bf G'} ({\bf q}).
\end{equation}
%
%As $\Xi$ is the product of a diagonal matrix and a Hermitian matrix, its
%inverse is 
Since $\Xi = K \epsilon$, one immediately has % $\Xi^{-1} = \epsilon^{-1} K^{-1}$,
the following relationship,
\begin{equation}
\Xi^{-1}_{\bf GG'}({\bf q}) = \epsilon^{-1}({\bf q+G,q+G'}) \frac{4 \pi}{|{\bf q + G'}|^2}.
\end{equation}
where $\epsilon^{-1}$ is the \emph{inverse dielectric matrix}. (Most 
literature works have used $\epsilon^{-1}$ as the fundamental dielectric
function; hereafter we shall instead work with the closely related
quantity $\Xi^{-1}$.)
Note that, unlike $\Xi$, $\Xi^{-1}$ is generally nonanalytic in ${\bf q}$. 
To see this, it is instrumental to divide the two matrices into four blocks,
\begin{equation}
\Xi = \left( \begin{array}{c c}
  A   & B \\
  B^* & C \end{array} \right), \qquad
  \Xi^{-1} = \left( \begin{array}{c c}
  P   & Q \\
  Q^* & S \end{array} \right). 
\end{equation}
The ``heads'', $A$ and $P$, are $1 \times 1$ matrices, corresponding to the
${\bf G=G'=}0$ elements; the ``wings'' $B$ and $Q$ are one-dimensional vectors, 
while $C$ and $S$ are square Hermitian matrices.
It is easy to show that
\begin{eqnarray}
Q_i &=& -P \, (B \cdot C^{-1})_i \label{eqq} \\
P &=& (A - B \cdot C^{-1} \cdot B^*)^{-1} \label{eqp} \\
S_{ij} &=& C^{-1}_{ij} +  (C^{-1} \cdot B^*)_i \, P \, (B \cdot C^{-1})_{j},
\label{eqs}
\end{eqnarray}
where the indices $i$ and $j$ stand for reciprocal lattice
vectors excluding ${\bf G} = 0$.
From the above relationships, it is manifest that all the elements in the 
$\Xi^{-1}$ matrix display a nonanalytic behavior. This is due to the quantity
$P({\bf q})$, which constitutes the head of the matrix, and also appears in
$Q_i$ and $S_{ij}$. 
It can be shown that, for a generic insulator,
$P({\bf q})$ diverges as $\sim q^{-2}$ in a vicinity of $\Gamma$.
%
%Such a divergence, in turn, can be traced back to the behavior 
%(details of the
%derivation can be found in Ref.~\onlinecite{rmm_thesis}) 
%of $A$ and $B$ near $\Gamma$.
%
Indeed, at the leading order one has~\cite{rmm_thesis}
\begin{eqnarray}
A({\bf q}) &\sim& {\bf q} \cdot \mathcal{A} \cdot {\bf q}, \\
B_i({\bf q}) &\sim& {\bf q} \cdot \mathcal{B}_i,
\end{eqnarray}
where $\mathcal{A}$ and $\mathcal{B}_i$ are, respectively, a $3 \times 3$
tensor and 3-vectors, both independent of ${\bf q}$. (The matrix elements
of $C({\bf q})$, on the other hand, tend to a finite constant.)
The $\mathcal{O}(q^{-2})$ divergence of $P({\bf q})$ then follows from
Eq.~(\ref{eqp}).

\subsection{Application to the phonon problem}

\label{secphon}

The $\Xi^{-1}$ matrix is a fundamental property of a crystalline solid 
within the adiabatic approximation. To appreciate the physical meaning of
this quantity, 
%
%
%To understand the physical implications of the above formulas, 
consider 
the ``bare'' charge perturbation induced by a collective displacement of the
sublattice $\kappa$ along the Cartesian direction $\alpha$,
\begin{equation}
\rho^{\rm ext}_{\kappa \alpha {\bf G}}({\bf q}) = 
-\frac{i Z_\kappa}{\Omega} 
  (q_\alpha + G_\alpha) 
  e^{ -i {\bf G} \cdot \bm{\tau}_\kappa},
\end{equation}
where we have made the dependence on $\kappa$ and $\alpha$ explicit.
(The above formula describes the nucleus as a $\delta$-function of 
charge $Z_\kappa$, consistent with the exact all-electron treatment
that is assumed in this Section.)
The induced charge response to such a perturbation reads
\begin{equation}
\tilde{\rho}^{\rm ind}_{\kappa \alpha {\bf G}}({\bf q}) =
\Pi_{\bf G \cdot}({\bf q}) \cdot \Xi^{-1}({\bf q}) \cdot \rho^{\rm ext}_{\kappa \alpha}({\bf q}).
\label{tilderho}
\end{equation}
where the scalar products indicate summation over repeated reciprocal lattice 
indices (${\bf G'}$, ${\bf G''}$, etc.).
The force-constant matrix has also a simple expression in terms of $\Xi^{-1}$,
%
%he above derivations allow for a natural separation of the force-constant 
%matrix into an analytic and a nonanalytic part.
%
%We can write the full force-constant matrix as
\begin{equation}
\tilde{\Phi}_{\kappa \alpha, \kappa' \beta}^{\bf q} = 
\Omega \, \left[ \rho^{\rm ext}_{\kappa \alpha}({\bf q}) \right]^* \cdot \Xi^{-1}({\bf q}) \cdot
\rho^{\rm ext}_{\kappa' \beta}({\bf q})
\label{tildephi}
\end{equation}
The tilde sign emphasizes that the response functions of Eq.~(\ref{tilderho}) and
Eq.~(\ref{tildephi}) are \emph{inclusive} of the macroscopic fields. 
These should be, therefore, distinguished from the closely related quantities,
$\rho^{\bf q}$ and $\Phi^{\bf q}$, which have been introduced in the early 
sections of this work. In fact, the latter were defined by prescribing that 
the macroscopic ${\bf G}=0$ term in the electrostatics should be switched off,
while we didn't take such a precaution in the derivation of $\tilde{\rho}^{\rm ind}$
and $\tilde{\Phi}$.

To make the link with $\rho^{\bf q}$ and $\Phi^{\bf q}$, we proceed
to solving again the Poisson equation, Eq.~(\ref{Poisson}), this time by 
imposing that the ${\bf G}=0$ component of the screened potential vanishes,
\begin{eqnarray}
\Xi_{\bf GG'}({\bf q}) \tilde{V}_{\bf G'}({\bf q})
&=& \rho^{\rm ext}_{\bf G} ({\bf q}), \qquad {\bf G, G'}\neq 0 \\
\tilde{V}_{{\bf G}}({\bf q}) &=& 0 \qquad \qquad \qquad {\bf G}= 0. 
\end{eqnarray}
We obtain analogous expressions for the density- and force-response 
functions,
%
%If we calculate $\Phi$ while suppressing the macroscopic fields, as above,
%we obtain
\begin{eqnarray}
\tilde{\rho}^{\rm ind}_{\bf G}({\bf q}) &=& 
\Pi_{\bf G \cdot}({\bf q}) \cdot C^{-1}({\bf q}) \cdot \rho^{\rm ext}({\bf q}), \\
\Phi_{\kappa \alpha, \kappa' \beta}^{{\bf q}} &=& 
\Omega \, \left[ \rho^{\rm ext}_{\kappa \alpha}({\bf q}) \right]^* \cdot C^{-1}({\bf q}) \cdot
\rho^{\rm ext}_{\kappa' \beta}({\bf q}).
\end{eqnarray}
(Note that the scalar products now run over the ${\bf G} \neq 0$ 
components, consistent with the array dimensions of $C^{-1}$.) 
$\Phi_{\kappa \alpha, \kappa' \beta}^{{\bf q}}$ coincides with the force-constant
matrix (indicated by the same symbol) that we have used in the remainder of this work,
and whose Taylor expansion yields the electromechanical internal-strain tensors. 
On the other hand, the \emph{total}
density response, inclusive of the external perturbing function,
\begin{equation}
\rho^{\bf q}_{\kappa \alpha} ({\bf G}) = \rho^{\rm ext}_{\kappa \alpha {\bf G}} ({\bf q})
+ \rho^{\rm ind}_{\kappa \alpha {\bf G}} ({\bf q}).
\end{equation}
corresponds to the function $\rho_{\kappa \beta}^{\bf q} ({\bf G})$ 
[the Fourier transform of $\rho_{\kappa \beta}^{\bf q} ({\bf r})$] that we
have extensively used in the previous Sections.
The above derivations provide the rigorous proof that both functions are
indeed analytic, and thus their expansions in powers of
${\bf q}$ is formally justified.

So what is it, physically, that causes the nonanaliticity of the ``full'' 
(tilded) response functions?
The answer is well known, and lies in the long-ranged character of the 
electrostatic interactions; this makes the behavior of the macroscopic 
fields (by ``macroscopic'' here we really mean the ${\bf G}=0$ component)
nonanalytic in a vicinity of $\Gamma$.
To see this, it is useful to calculate the macroscopic electrostatic 
potential resulting from the macroscopic induced charge,
$\overline{\rho}^{\bf q}_{\kappa \alpha} = \rho^{\bf q}_{\kappa \alpha} ({\bf G}=0)$,
\begin{equation}
\overline{V}^{\bf q}_{\kappa \alpha} = \Xi^{-1}_{00}({\bf q}) \overline{\rho}^{\bf q}_{\kappa \alpha}.
\end{equation}
We can express this more conveniently as
\begin{equation}
\overline{\bm{\mathcal{E}}}^{\bf q}_{\kappa \alpha} = -i \, 4\pi {\bf q} \, 
\frac{ \overline{\rho}^{\bf q}_{\kappa \alpha} } {\xi({\bf q})},
\label{efieldq}
\end{equation}
where we have used the fact that the electric field is minus the gradient of
the potential, and we have replaced $\Xi^{-1}_{00}({\bf q})$ with a new symbol,
\begin{equation}
\xi({\bf q}) = \frac{1}{4\pi P({\bf q})} = \frac{1}{4 \pi \Xi^{-1}_{00}({\bf q})} = \frac{q^2}{\epsilon^{-1}({\bf q,q})}.
\end{equation}
%
%Both $\overline{\rho}^{\bf q}_{\kappa \alpha}$ and 
$\xi({\bf q})$ is an analytic function of ${\bf q}$ 
[this property is obvious from the expression of $P$ 
given in Eq.~(\ref{eqp})].
However, the fact that  $\xi({\bf q})$ appears at the denominator
in Eq.~(\ref{efieldq}) makes the macroscopic electric field 
strongly nonanalytic at $\Gamma$.
To see this, it is helpful to replace both the numerator and the 
denominator with the leading order term in their respective ${\bf q}$-expansion,
\begin{eqnarray}
\Omega \overline{\rho}^{\bf q}_{\kappa \alpha} &\sim& 
-i q_\beta Z^*_{\kappa, \alpha \beta}, \\
\xi({\bf q}) &\sim& q_\alpha \epsilon_{\alpha \beta} q_\beta,
%+ q_\alpha  q_\beta q_\gamma q_\lambda
%\epsilon^{(4)}_{\alpha \beta \gamma \lambda} + \mathcal{O}(q^6),
\end{eqnarray}
where $Z^*_{\kappa, \alpha \beta}$ is the Born dynamical charge tensor 
(the star here does not indicate complex conjugation; this
is the commonly used notation to distinguish the Born tensor 
from the bare nuclear charge $Z_\kappa$),
and $\epsilon_{\alpha \beta}$ is the macroscopic high-frequency dielectric 
tensor.
One readily obtains
\begin{equation}
\overline{\bm{\mathcal{E}}}^{{\bf q}\rightarrow 0}_{\kappa \alpha} = - \frac{4\pi}{\Omega} {\bf q} \, 
\frac{({\bf q} \cdot {\bf Z}^*_{\kappa})_\alpha} {{\bf q} \cdot \bm{\epsilon} \cdot {\bf q}},
\end{equation}
i.e. for small values of $q$ the macroscopic field tends to a direction-dependent 
constant, and is therefore discontinuous at $\Gamma$.
Such a nonanalytic behavior propagates to both the charge density and
force-constant response functions, causing them to be nonanalytic as well.

\subsection{Higher-order generalization of the Cochran-Cowley formula}

It is interesting to work out the example of the force-constant 
matrix explicitly, to make contact with the existing knowledge
on its nonanalytic behavior near $\Gamma$.
%and analyze the consequences of the macroscopic
%fields at higher orders in ${\bf q}$.
%
%For example, after observing that $\overline{\rho}^{\bf q}_{\kappa \alpha} = 
%\rho^{\bf q}_{\kappa \alpha} ({\bf G}=0)$, one can define
To that end, consider the function
\begin{equation}
%
%
%is an important quantity because it coincides with the charge-density response
%that we considered in the main part of this work,
%\begin{eqnarray}
%\tilde{\rho}^{\rm tot}_{0} ({\bf q}) &=& \overline{\rho}_{\kappa \beta}^{\bf q}, \\
%\tilde{\rho}^{\rm tot}_{\bf G} ({\bf q}) &=& \rho_{\kappa \beta}^{\bf q} ({\bf G}).
%\end{eqnarray}
%
%
%If we define
%\begin{equation}
\Phi_{\kappa \alpha, \kappa' \beta}^{{\bf q},{\rm NA}} = 
\Omega \, \left[ \overline{\rho}^{\bf q}_{\kappa \alpha} \right]^*   P({\bf q}) \,
\overline{\rho}^{\bf q}_{\kappa' \beta},
\end{equation}
where the NA superscript indicates that this quantity is nonanalytic in
${\bf q}$, because of the factor of $P({\bf q})$.
It is a straightforward excercise to show that
\begin{equation}
\tilde{\Phi}_{\kappa \alpha, \kappa' \beta}^{\bf q} = 
\Phi_{\kappa \alpha, \kappa' \beta}^{{\bf q}} +\Phi_{\kappa \alpha, \kappa' \beta}^{{\bf q},{\rm NA}}.
\end{equation}
Such a partition of the force-constant matrix into an analytic and a nonanalytic
part corresponds precisely to that of Ref.~\onlinecite{rmm_thesis}. 
[$\Phi_{\kappa \alpha, \kappa' \beta}^{{\bf q}}$ and 
$\Phi_{\kappa \alpha, \kappa' \beta}^{{\bf q},{\rm NA}}$ are, respectively,
$\bar{C}_{\kappa \kappa'}^{\alpha \beta}({\bf q},1)$ and
$\bar{C}_{\kappa \kappa'}^{\alpha \beta}({\bf q},2)$ of Ref.~\onlinecite{rmm_thesis}.]
Thus, our prescription of removing the ${\bf G}=0$ in the self-consistent 
electrostatic potential naturally yields the analytic part of the force-constant
matrix as defined by Pick, Cohen and Martin~\cite{rmm_thesis}.
%
%We are ready now to generalize the theory of Ref.~\onlinecite{rmm_thesis} to
%higher orders in ${\bf q}$, as required by the treatment of the flexoelectric
%problem.
%
The remainder, $\Phi_{\kappa \alpha, \kappa' \beta}^{{\bf q},{\rm NA}}$,
can be expressed more conveniently as
\begin{equation}
\Phi_{\kappa \alpha, \kappa' \beta}^{{\bf q},{\rm NA}} = 
4 \pi \Omega \frac{ \left[ \overline{\rho}^{\bf q}_{\kappa \alpha} \right]^* \, 
\overline{\rho}^{\bf q}_{\kappa' \beta} }{\xi({\bf q})},
\end{equation}
%where we have introduced a new symbol for the inverse of $P$,
%Eq.~\ref{eqp} states that $\xi({\bf q})$ is an analytic function, going to zero 
%quadratically in a vicinity of $\Gamma$.
%
Thus, similarly to the case of the electric field, the nonanalytic part of 
the force-constant matrix can be written, in full generality, as the ratio 
of two \emph{analytic} functions of ${\bf q}$, either of which can be expanded 
in a Taylor series.
We shall now push the Taylor expansion to higher orders in $q$, including
all terms that are potentially relevant in the present theory of the 
flexoelectric response,
\begin{align}
\Omega \overline{\rho}^{\bf q}_{\kappa \alpha} \sim & \,
-i q_\beta Q^{(1,\beta)}_{\kappa \alpha} - 
\frac{q_\beta q_\gamma}{2}  Q^{(2,\beta \gamma)}_{\kappa \alpha} \nonumber \\
& +i \frac{q_\beta q_\gamma q_\lambda}{6} Q^{(3,\beta \gamma \lambda)}_{\kappa \alpha}
+ \mathcal{O}(q^4), \\
\xi({\bf q}) \sim & \, q_\alpha q_\beta \epsilon_{\alpha \beta} + q_\alpha  q_\beta q_\gamma q_\lambda
\epsilon^{(4)}_{\alpha \beta \gamma \lambda} + \mathcal{O}(q^6),
\end{align}
where 
%$\epsilon^{(2)}_{\alpha \beta} = \epsilon_{\alpha \beta}$ is the macroscopic 
%high-frequency dielectric tensor, and 
$Q^{(1,\beta)}_{\kappa \alpha} = Z^*_{\kappa,\alpha \beta}$
is again the Born dynamical charge tensor. 
[In order to lighten the notation, we shall use the following conventions
henceforth,
\begin{align}
%({\bf q \cdot Z}^*_\kappa)_\alpha &= 
%  q_\beta Q^{(1,\beta)}_{\kappa \alpha}, \nonumber \\
({\bf q \, q \cdot Q}^*_\kappa)_\alpha &= 
  q_\beta q_\gamma Q^{(2,\beta \gamma)}_{\kappa \alpha} \nonumber \\
({\bf q \, q \, q \cdot O}^*_\kappa)_\alpha &= 
  q_\beta q_\gamma q_\lambda Q^{(3,\beta \gamma \lambda)}_{\kappa \alpha}, \nonumber \\
({\bf q \, q} \cdot \bm{\epsilon}^{(4)} \cdot {\bf q \, q}) &= 
q_\alpha  q_\beta q_\gamma q_\lambda
\epsilon^{(4)}_{\alpha \beta \gamma \lambda}, \nonumber
\end{align}
where the dynamic quadrupoles and octupoles are indicated as $Q^*$ and $O^*$,
respectively, in analogy with the dynamic dipoles $Z^*$.] 
Note the absence of the zero-th order term in the expansion of 
$\overline{\rho}^{\bf q}$ (because of the requirement of charge neutrality)
and the absence of the odd terms in the expansion of $\xi({\bf q})$ (because
of the requirement of time-reversal symmetry -- we assume that we are dealing
with a nonmagnetic insulator).
At the leading order, we recover the usual Cochran-Cowley formula,
\begin{equation}
\Phi_{\kappa \alpha, \kappa' \beta}^{{\bf q},{\rm DD}} = \frac{4 \pi}{\Omega}
\frac{({\bf q \cdot Z}^*_\kappa)_\alpha ({\bf q \cdot Z}^*_{\kappa'})_\beta }{ {\bf q} \cdot \bm{\epsilon} \cdot {\bf q}},
\end{equation}
which invloves the well-known dipole-dipole (DD) interactions.
 This term produces
a long-ranged contribution to the real-space interatomic force constants (IFC) that 
decays as $1/d^3$ (with the interatomic distance $d$).~\cite{Gonze/Lee}
The next order in the expansion,
\begin{eqnarray}
\Phi_{\kappa \alpha, \kappa' \beta}^{{\bf q},{\rm DQ}} &=& -i\frac{4 \pi}{2\Omega}
\frac{({\bf q \cdot Z}^*_\kappa)_\alpha ({\bf q \, q \cdot Q}^*_{\kappa'})_\beta }{ {\bf q} \cdot \bm{\epsilon} \cdot {\bf q}} \nonumber \\
&& +i\frac{4 \pi}{2\Omega}
\frac{({\bf q \, q \cdot Q}^*_\kappa)_\alpha ({\bf q \cdot Z}^*_{\kappa'})_\beta }{ {\bf q} \cdot \bm{\epsilon} \cdot {\bf q}},
\end{eqnarray}
contains dipole-quadrupole (DQ) interaction terms. It is easy to show that this contribution 
plays an important role in piezoelectric materials, where it is responsible for the 
boundary-dependent macroscopic electric fields that arise upon deformation.
%affects the sound velocity 
%(and its dependence on the electrical boundary conditions). 
Its contribution to the IFC 
decays as $1/d^4$.
Finally, we have three contributions, all at the same order in $q$. First,
the dipole-octupole (DO) term,
\begin{eqnarray}
\Phi_{\kappa \alpha, \kappa' \beta}^{{\bf q},{\rm DO}} &=& -\frac{4 \pi}{6\Omega}
\frac{({\bf q \cdot Z}^*_\kappa)_\alpha ({\bf q \, q \, q \cdot O}^*_{\kappa'})_\beta }{ {\bf q} \cdot \bm{\epsilon} \cdot {\bf q}} \nonumber \\
&& -\frac{4 \pi}{6\Omega}
\frac{({\bf q \, q \, q \cdot O}^*_\kappa)_\alpha ({\bf q \cdot Z}^*_{\kappa'})_\beta }{ {\bf q} \cdot \bm{\epsilon} \cdot {\bf q}},
\end{eqnarray}
which can be related to the purely electronic flexoelectric response (and, 
in particular, to the macroscopic electric fields generated by the latter 
under open-circuit boundary conditions).
The second is
a quadrupole-quadrupole interaction, 
\begin{equation}
\Phi_{\kappa \alpha, \kappa' \beta}^{{\bf q},{\rm QQ}} = \frac{4 \pi}{4\Omega}
\frac{({\bf q \, q \cdot Q}^*_\kappa)_\alpha ({\bf q \, q \cdot Q}^*_{\kappa'})_\beta }{ {\bf q} \cdot \bm{\epsilon} \cdot {\bf q}},
\end{equation}
which has an impact [via the square brackets Eq.~\eqref{squarek}]
on the elastic coefficients (and hence on sound velocity) in 
piezoelectric materials. 
%
%This term vanishes in nonpiezoelectric insulators (where the sublattice sum
%of the dynamic quadrupole tensor ${\bf Q}^*$ vanishes), and is therefore 
%irrelevant for the flexoelectric properties of a centrosymmetric solid.
%
The third term, of less obvious physical interpretation, is due to the 
$q$-dispersion of the macroscopic dielectric tensor, and reads
\begin{equation}
\Phi_{\kappa \alpha, \kappa' \beta}^{{\bf q},{\rm D\epsilon D}} = 
-\frac{4 \pi}{\Omega}
\frac{({\bf q \cdot Z}^*_\kappa)_\alpha \, ({\bf q \, q} \cdot \bm{\epsilon}^{(4)} \cdot {\bf q \, q}) \,
({\bf q \cdot Z}^*_{\kappa'})_\beta }{ ({\bf q} \cdot \bm{\epsilon} \cdot {\bf q})^2},
\end{equation}
Note that, in spite of being $\mathcal{O}(q^2)$, this term is
irrelevant for both flexoelectricity and elasticity, as 
it vanishes upon summation over one (or both) of the sublattice 
indices [as required, e.g. in Eq.~(\eqref{squarek})].
%does not contribute to the square brackets, Eq.~\eqref{squarek},
%as the latter involve a summation over one of the two sublattice indices;
%this summation vanishes because of the acoustic sum rule.
%
%(It does not enter the round brackets either, as the gradient of 
%$\Phi_{\kappa \alpha, \kappa' \beta}^{{\bf q},{\rm D\epsilon D}}$ vanishes
%identically at $\Gamma$.) 
%
In summary, we have
\begin{eqnarray}
\Phi_{\kappa \alpha, \kappa' \beta}^{{\bf q},{\rm NA}} &=&
\Phi_{\kappa \alpha, \kappa' \beta}^{{\bf q},{\rm DD}} +
\Phi_{\kappa \alpha, \kappa' \beta}^{{\bf q},{\rm DQ}} + 
 \Phi_{\kappa \alpha, \kappa' \beta}^{{\bf q},{\rm DO}} +\nonumber \\
&&   \Phi_{\kappa \alpha, \kappa' \beta}^{{\bf q},{\rm QQ}} +
   \Phi_{\kappa \alpha, \kappa' \beta}^{{\bf q},{\rm D\epsilon D}} + \mathcal{O}(q^3).
\end{eqnarray}
The DD and DQ terms are nonanalytic at zero-th and first-order in $q$,
respectively; DO, QQ and ${\rm D \epsilon D}$ are all nonanalytic at
the order $q^2$. This formula describes the long-range electrostatic 
interactions in an arbitrary insulator up to the order $1/d^5$ (included),
and constitutes therefore a higher-order generalization of the well-known
Cochran-Cowley formula (DD only, valid up to $1/d^3$).

This completes our discussion of the nonanalytic behavior of $\Phi$ in a vicinity of 
the $\Gamma$ point.
Apart from the direct interest to the study of electromechanical phenomena,
explicitly incorporating these terms in lattice-dynamical studies may
be instrumental to achieving an accurate sampling of the phonon dispersion curves 
(especially at small $q$), while keeping the computational workload to a moderate 
level. 
This might be done, for example, by using Ewald summation techniques similar
to those discussed in Ref.~\onlinecite{Gonze/Lee}.
%, only extended to higher orders 
%in the multipolar expansion. 

\subsection{Longitudinal versus transversal charge response}

\label{ebcq}

Hong and Vanderbilt~\cite{Hong-11,Hong-13}, building on the work 
of Resta~\cite{Resta-10} based their treatment of the flexoelectric
problem on the dipolar, quadrupolar and octupolar response to 
atomic displacements, in close analogy to the approach taken here.
However, at difference with the present work, Refs.~\onlinecite{Hong-11}
and~\onlinecite{Hong-13} defined the charge response functions under
longitudinal (fixed electric displacement) boundary conditions.
In order to trace a closer link to their approach, we shall briefly
discuss here the relationship between the transversal (fixed electric
field) quantities defined here and the longitudinal ones.

The charge response in longitudinal boundary conditions (which are the 
physically correct ones for the description of a phonon perturbation in
an insulating crystal) can be simply written by applying Gauss's law
to the nonanalytic macroscopic electric field,
\begin{equation}
\tilde{\rho}^{\bf q}_{\kappa \beta} = 
\frac{ i {\bf q} \cdot \overline{\bm{\mathcal{E}}}_{\kappa \beta}^{\bf q}}{4 \pi}.
\end{equation}
By using the formula for the electric field given in Eq.~(\ref{efieldq}), 
and by replacing the numerator and denominator
with their Taylor expansion in ${\bf q}$, we have
\begin{equation}
\tilde{\rho}^{\bf q}_{\kappa \beta} \sim \frac{q^2}{\Omega} 
\frac{-i ({\bf q \cdot Z}^*_\kappa)_\beta -\frac{({\bf q \, q \cdot Q}^*_\kappa)_\beta}{2}
+i\frac{({\bf q \, q \, q \cdot O}^*_{\kappa'})_\beta}{6} }
{{\bf q} \cdot \bm{\epsilon} \cdot {\bf q} + ({\bf q \, q} \cdot \bm{\epsilon}^{(4)} \cdot {\bf q \, q})}.
\end{equation}
This expression can be conveniently written, for a given direction $\hat{\bf q} = {\bf q}/q$, 
as
\begin{equation}
\tilde{\rho}^{\bf q}_{\kappa \beta} \sim -i q \tilde{\rho}^{(1,\hat{\bf q})}_{\kappa \beta}
- \frac{q^2}{2} \tilde{\rho}^{(2,\hat{\bf q})}_{\kappa \beta} 
+ i \frac{q^3}{6} \tilde{\rho}^{(3,\hat{\bf q})}_{\kappa \beta},
\end{equation} 
where $\tilde{\rho}^{(n,\hat{\bf q})}_{\kappa \beta}$ are direction-dependent 
constants. Their explicit formulas are
\begin{align}
\Omega \tilde{\rho}^{(1,\hat{\bf q})}_{\kappa \beta} &= 
\frac{(\hat{\bf q} \cdot {\bf Z}^*_\kappa)_\beta}{\hat{\bf q} \cdot \bm{\epsilon} \cdot \hat{\bf q}}, \\
\Omega \tilde{\rho}^{(2,\hat{\bf q})}_{\kappa \beta} &= 
\frac{(\hat{\bf q} \, \hat{\bf q} \cdot {\bf Q}^*_\kappa)_\beta}
{\hat{\bf q} \cdot \bm{\epsilon} \cdot \hat{\bf q}}, \\
\Omega \tilde{\rho}^{(3,\hat{\bf q})}_{\kappa \beta} &= 
\frac{(\hat{\bf q} \, \hat{\bf q} \, \hat{\bf q}\cdot {\bf O}^*_\kappa)_\beta}
{\hat{\bf q} \cdot \bm{\epsilon} \cdot \hat{\bf q}} + 
\frac{ (\hat{\bf q} \, \hat{\bf q} \cdot \bm{\epsilon}^{(4)} \cdot \hat{\bf q} \, \hat{\bf q})  
  (\hat{\bf q} \cdot {\bf Z}^*_\kappa)_\beta}{(\hat{\bf q} \cdot \bm{\epsilon} \cdot \hat{\bf q})^2}.
\end{align}
Note that the last equation simplifies upon summation over the sublattice index 
$\kappa$, as required by the formula for the flexoelectric tensor,
\begin{equation}
\Omega \sum_\kappa \tilde{\rho}^{(3,\hat{\bf q})}_{\kappa \beta} = 
\frac{(\hat{\bf q} \, \hat{\bf q} \, \hat{\bf q} \cdot \sum_\kappa {\bf O}^*_\kappa)_\beta}
{\hat{\bf q} \cdot \bm{\epsilon} \cdot \hat{\bf q}}.
\end{equation}
Thus, the two sets of quantities (longitudinal and transversal) are 
trivially related by a factor of $\hat{\bf q} \cdot \bm{\epsilon} \cdot \hat{\bf q}$,
which describes the macroscopic dielectric screening along the direction
$\hat{\bf q}$.

\subsection{Thomas-Fermi screening of the macroscopic electric fields}

\label{secmobile}

In the previous Sections we have formally justified our
prescription of suppressing the ${\bf G}=0$ component of the 
electrostatic potential when calculating the basic response
functions that enter the flexoelectric tensor.
Such a prescription is, however, well defined only in the 
context of a Taylor expansion in ${\bf q}$ around $\Gamma$;
it is, therefore, inappropriate to calculating the localized
representation of the response functions, introduced in 
Section~\ref{elec}.
In order to achieve a truly localized real-space representation 
of the charge density and polarization response to the displacement 
of an isolated atom, we shall follow the strategy of Martin~\cite{Martin},
and suppose that the problematic macroscopic fields are ``short-circuited''
by a very low density of mobile carriers superimposed to the insulating
crystal.
We shall demonstrate that, concerning the piezoelectric and 
flexoelectric properties of an arbitrary insulating crystal, the
two procedures lead to the same result.

The Poisson problem of Eq.~(\ref{Poisson}) in presence of mobile
charges can be rewritten as
\begin{equation}
\nabla^2 V({\bf r}) = - 4 \pi \left[ \rho^{\rm ext}({\bf r}) +
\rho^{\rm ind}({\bf r}) + \rho^{\rm free}({\bf r}) \right],
\end{equation}
where $V$ is the doubly-screened (i.e. both by $\rho^{\rm ind}$ 
and $\rho^{\rm free}$) potential, 
$\rho^{\rm ext}$ and $\rho^{\rm ind}$ have been defined in the 
previous Section, and $\rho^{\rm free}$ refers to the metallic 
carriers.
Within the Thomas-Fermi approximation, the carrier density is
related to the potential by
\begin{equation}
\rho^{\rm free}({\bf r}) = -\frac{k_0^2 V({\bf r})}{4 \pi},
\label{rhotf}
\end{equation}
where $k_0$ is the Thomas-Fermi screening wavevector.
In reciprocal space, the external perturbing charge and the
doubly-screened potential are then related by a linear problem in the
same form as Eq.~(\ref{Poisson}),
\begin{equation}
\bar{\Xi}_{\bf GG'}({\bf q}) V_{\bf G'}({\bf q}) = \rho^{\rm ext}_{\bf G}({\bf q}),
\end{equation}
but with a modified $\Xi$ matrix,
\begin{equation}
\bar{\Xi}_{\bf GG'}({\bf q}) = \frac{|{\bf q + G}|^2 + k_{\rm TF}^2}{4 \pi} \delta_{\rm GG'} 
- \Pi_{\bf GG'}({\bf q}).
%\label{vvv}
\end{equation}
We shall choose a value of $k_{\rm TF}$ that is much smaller than any reciprocal-space
vector ${\bf G}$ (except $\Gamma$), in order not to modify the electronic ground 
state of the unperturbed system. (This corresponds to choosing
a Thomas-Fermi screening length, $\lambda_{\rm TF}= 1/k_{\rm TF}$, much larger than 
any of the three primitive translation vectors in real space, i.e. a very
low-density gas of carriers.)
We shall now proceed to deriving the charge density and force response functions 
by following the same steps as in the previous Section. 
First, note that $\Xi$ and $\bar{\Xi}$ are essentially identical except for 
their head,
\begin{equation}
\bar{\Xi}_{00}({\bf q}) = \Xi_{00}({\bf q})  + \frac{k_0^2}{4 \pi},
\end{equation}
due to the assumption of small $k_{\rm TF}$. Therefore, if we suppress the 
${\bf G}=0$ term as we did earlier, we obtain the same response functions
at any ${\bf q}$. The full response functions in presence of the Thomas-Fermi
gas include a contribution from the macroscopic fields, which we shall 
evaluate in the following.
After a few steps of straightforward algebra, we obtain
\begin{equation}
\overline{\bm{\mathcal{E}}}^{\bf q}_{\kappa \alpha} = 
\frac{-i \, 4\pi {\bf q} \,  \overline{\rho}^{\bf q}_{\kappa \alpha} } {\bar{\xi}({\bf q})} \simeq
%\overline{\bm{\mathcal{E}}}^{\bf q}_{\kappa \alpha} = 
- \frac{4\pi}{\Omega} {\bf q} \, 
\frac{({\bf q} \cdot {\bf Z}^*_{\kappa})_\alpha} {k_{\rm TF}^2 + {\bf q} \cdot \bm{\epsilon} \cdot {\bf q}},
\end{equation}
where $\bar{\xi}$ relates to $\xi$ as 
$\bar{\Xi}_{00}$ relates to $\Xi_{00}$.
The macroscopic electric field is now manifestly analytic in ${\bf q}$, consistent
with the metallic screening mediated by the carrier gas.
%
%At zero-th and first order in ${\bf q}$, the field vanishes. Only at second order
%we have a finite contribution,
At the lowest orders in ${\bf q}$ we have
\begin{align}
\overline{\mathcal{E}}^{{\bf q}=0}_{\alpha, \kappa \beta} &= 0 \\
\frac{ \partial \overline{\mathcal{E}}^{\bf q}_{\alpha, \kappa \beta} }
     { \partial q_\gamma} \Big|_{{\bf q}=0} &= 0 \\
\frac{ \partial^2 \overline{\mathcal{E}}^{\bf q}_{\alpha, \kappa \beta} }
     { \partial q_\gamma \partial q_\lambda } \Big|_{{\bf q}=0} &= 
     -\frac{ 4 \pi}{ \Omega k_{\rm TF}^2 } 
      \left( \delta_{\alpha \gamma} Z_{\kappa, \lambda \beta} +
              \delta_{\alpha \lambda} Z_{\kappa, \gamma \beta} \right). \label{emac2}
\end{align}              
Note that the macroscopic electric field vanishes at $\mathcal{O}(q^0)$ 
and $\mathcal{O}(q^1)$. 
%the zero-th and first
%order in ${\bf q}$. 
This implies that, in presence of the screening carriers, inclusion
of the ${\bf G}=0$ component of the electrostatic potential has no influence on
the polarization and charge density response functions up to first order in ${\bf q}$.
In other words, $P^{(0,1)}_{\alpha, \kappa \beta}({\bf r})$ and 
$\rho^{(0,1)}_{\kappa \beta}({\bf r})$ are well defined. 
[In the context of the present discussion, we indicate a response 
function as \emph{well defined} if it enjoys the following property: 
By calculating it with the ``${\bf G} \neq 0$'' prescription (i.e. 
without metallic carriers, but by removing by hand the ${\bf G}=0$ 
component from the self-consistent electrostatic potential) one 
obtains the same result as in a ``TF'' calculation (i.e. with the 
${\bf G}=0$ electrostatic term included, but with the long-range 
fields suppressed by the metallic carriers in the ${\bf q}\rightarrow 0$ 
limit). 
Of course, a \emph{well defined} response function is also independent of the 
value of $k_{\rm TF}$.]
Furthermore, since an electric field induces a net polarization but not a net charge 
(at the same order in $q$), also the cell-average of the $\mathcal{O}(q^2)$ density,
$\overline{\rho}^{(2)}_{\kappa \beta}$ is well defined.
This is consistent with the claims of Ref.~\onlinecite{Martin}, that the dipolar 
and quadrupolar real-space moments of the induced charge density upon atomic
displacement are independent of $k_{\rm TF}$.
(Note that the aforementioned real-space moments coincide with 
$\Omega \overline{\rho}^{(1,\gamma)}_{\kappa \beta}$ and 
$\Omega \overline{\rho}^{(2,\gamma \lambda)}_{\kappa \beta}$, respectively.)
Therefore, the definition of the electronic response functions that are 
relevant for the piezoelectric case is unambiguous and poses no particular problem.

In the context of flexoelectricity, the fact that there is a nonzero $k_{\rm TF}$-dependent 
field at second order in $q$ might appear troublesome at first sight, as the flexoelectric 
polarization is precisely a $\mathcal{O}(q^2)$ effect.
%in the context of the flexoelectric response, as the latter ;
%however, we shall see in the following that it causes no ambiguities.
%
Indeed, the $\mathcal{O}(q^2)$ response functions $\rho^{(2)}_{\kappa \beta}({\bf r})$
and $P^{(2)}_{\alpha, \kappa \beta}({\bf r})$ are both affected by such
$k_{\rm TF}$-dependent field. (As we mentioned above, only the cell average of 
$\rho^{(2)}$ is well defined; note that the cell average of 
$P^{(2)}$ is not.) 
%
%
%
%(and, of course,
%$\overline{P}^{(2)}_{\alpha, \kappa \beta}$) are altered by the field.
%
Recall, however, that to calculate the flexoelectric tensor one never needs 
the \emph{individual} (i.e. $\kappa$-resolved) $P^{(2)}_{\alpha, \kappa \beta}$ 
functions -- only their \emph{sublattice sum} is relevant.
Since the macroscopic electric field, Eq.~(\ref{emac2}),
is proportional to the Born charge tensor, it is clear that its 
contribution vanishes (because of the acoustic sum rule) once the $\mathcal{O}(q^2)$ 
response functions are summed over $\kappa$.
Therefore, the functions 
\begin{align}
\rho^{(2,\gamma \delta)}_{\beta}({\bf r}) &= \sum_\kappa 
\rho^{(2,\gamma \delta)}_{\kappa \beta}({\bf r}), \\
P^{(2,\gamma \delta)}_{\alpha \beta}({\bf r}) &= \sum_\kappa 
P^{(2,\gamma \delta)}_{\alpha, \kappa \beta}({\bf r}),
\end{align}
are both well defined, and so is the total dynamical octupole tensor,
\begin{equation}
Q^{(3,\gamma \delta \lambda)}_{\beta} = 
\Omega \sum_\kappa \bar{\rho}^{(3,\gamma \delta \lambda)}_{\kappa \beta}.
\end{equation}
This conclusively proves that, when performing a long-wave expansion 
of the electronic response functions, one can work indifferently with
the ``${\bf G}\neq 0$'' and the ``TF'' prescription -- the calculated 
piezoelectric and flexoelectric tensors are identical.
%
%we obtain the same piezoelectric 
%and flexoelectric tensors lead to the same electronic response functions for both the  case. 
%In particular, these are relevant for the response of the crystal within 
%short-circuit electrical boundary conditions (as provided by the low-density 
%Thomas-Fermi gas).
As there is no ambiguity, we readily identify the latter as 
``fixed-$\mathcal{E}$''~\cite{Hong-13} electromechanical coefficients,
where the electric field $\mathcal{E}$ is assumed to be minus the gradient
of the macroscopic electrostatic potential.

\subsection{The reference potential issue}

\label{reference}

In the discussion of the Thomas-Fermi screening model we have
made an implicit assumption about the quantum-mechanical nature of 
the screening carriers, by writing the density of mobile charges,
Eq.~(\ref{rhotf}), as a function of the mean electrostatic potential.
This choice is not unique and needs to be properly justified, 
as the calculated values of the flexoelectric tensor components 
might depend on it. 
In this Section we shall briefly elaborate on this important point,
and show that there is indeed an ambiguity in the specification
of the electrical boundary conditions in the case of a strain-gradient
deformation.
Such ambiguity relates to a physical fact:
the breakdown of translational periodicity that is inherent to 
flexoelectric phenomena makes the notion of ``macroscopic electric 
field'' a bit more delicate than, e.g. in the piezoelectric case.
As we shall see in the following, in presence of a strain-gradient 
deformation the force acting on a charged particle \emph{depends}
on the nature of such particle, and not only on its charge. Hence,
the condition of ``zero macroscopic electric field'' depends on 
which type of test particle we choose as a probe to define the field.

Eq.~(\ref{rhotf}) refers to a free-electron parabolic band, whose
lower edge locally coincides with the (slowly varying) macroscopic 
electrostatic potential of the crystal. 
This does not appear very realistic in the general case of a 
lightly doped insulator or semiconductor. The carriers (e.g. 
electron or holes) typically occupy well-defined energy levels
in the band structure of the solid, rather than responding 
solely to electrostatic forces.
This implies that, in general, it would be more appropriate 
to replace the macroscopic electrostatic
potential $V({\bf r})$ in Eq.~(\ref{rhotf}) with the
energy level of the relevant band feature, e.g. the conduction
band minimum, $V_{\rm CBM}({\bf r})$ in the case of intrinsic
electron-like carriers.
%rather than 
%depending on the type of carrier that we
%are assuming to screen the macroscopic electric fields, 
%
%If we implement this modification in Eq.~(\ref{rhotf}), and
%we use it to recalculate the electronic (and ionic) response 
%functions, 
Under such a modified screening regime, the carriers will no longer 
enforce a flat electrostatic potential during a mechanical deformation, 
but rather a flat $V_{\rm CBM}({\bf r})$ (following up on the above 
example). 
Of course, this is not a concern in the piezoelectric case, 
where the bands always remain parallel since the periodicity
of the lattice is preserved in the deformed state.
This \emph{is}, however, an issue in the flexoelectric case, 
where a strain gradient inevitably produces a gradient
in the relative position of the band energies, via the 
so-called \emph{relative deformation potentials}.
Therefore, we inevitably obtain a different flexoelectric tensor, 
depending on what band feature we use as a reference for the
macroscopic field.

It is important to emphasize that the ambiguity described here is
not an artifact of the Thomas-Fermi screening model, but a physical
fact. To prove this, it is useful to translate the same arguments in 
the context of the ``${\bf G}\neq 0$'' prescription, which we have
established in this work as a simpler practical alternative to the 
``TF'' screening model. 
At the mathematical level, there is no fundamental reason to 
suppress the ${\bf G}= 0$ electrostatic term altogether. 
Strictly speaking, only the nonanaliticity associated with it needs to 
be removed.
We are therefore free to replace such ${\bf G}= 0$ component of
the electrostatic potential with an arbitrary \emph{analytic} 
function of ${\bf q}$ that respects the symmetry of the lattice.
Evidently, different choices of such a function will lead to 
different definitions of the flexoelectric tensor.
Since the relative deformation potentials are analytic~\cite{Resta-DP},
we can relate such a freedom to a \emph{band-structure term},
which describes the arbitrariness in the choice of the reference energy 
discussed above.

\section{Physical interpretation}

\label{discuss}

The goal of this section is to elaborate on the implications of the results
derived in this work, and to describe the microscopic mechanisms that 
contribute to the macroscopic flexoelectric response depending on crystal 
symmetry.
We shall exclusively focus on type-II tensors from now on, as they lend themselves
to a more intuitive physical interpretation in all cases.
The linear flexoelectric response to a type-II strain gradient is
\begin{equation}
\overline{P}_\alpha = \mu^{\rm II}_{\alpha \lambda,\beta \gamma } 
\varepsilon_{\beta \gamma, \lambda},
\end{equation}
where the total flexoelectric tensor, symmetric in $\beta \gamma$,  can be written, 
in full generality, as
\begin{equation}
\mu^{\rm II}_{\alpha \lambda,\beta \gamma} = 
\bar{\mu}^{\rm II}_{\alpha \lambda,\beta \gamma} + 
\mu^{\rm II,mix}_{\alpha \lambda,\beta \gamma} + 
\mu^{\rm II,latt}_{\alpha \lambda,\beta \gamma}.
\end{equation}
$\bar{\mu}^{\rm II}_{\alpha \lambda,\beta \gamma}$ is
the purely electronic (frozen-ion) response, which is active in all 
insulators~\cite{Resta-10,Hong-11}, regardless of symmetry or composition.
This term was discussed at length in Ref.~\onlinecite{Resta-10} and
Ref.~\onlinecite{Hong-11}, and we won't comment on it any further here.
$\mu^{\rm II,latt}_{\alpha \lambda,\beta \gamma}$ is the lattice-mediated 
contribution, analogous to the ``dynamical'' flexoelectric tensor discussed 
by Tagantsev~\cite{Tagantsev}, but expressed here in type-II form.
$\mu^{\rm II,mix}_{\alpha \lambda,\beta \gamma}$ is the remainder,
which is neither purely electronic in origin, neither lattice-mediated in 
the usual sense; we shall refer to it as ``mixed'' term henceforth.
In the following we shall discuss the explicit expressions of these latter two 
terms in the context of the theory developed so far, starting with
the more intuitive lattice-mediated part.

\subsection{Lattice-mediated contribution}

Based on the results of Section~\ref{longw}, the lattice-mediated flexoelectric
tensor is
\begin{equation}
\mu^{\rm II,latt}_{\alpha \lambda,\beta \gamma} = \frac{Z^*_{\kappa, \alpha \rho}}{\Omega} 
\widetilde{\Phi}^{(0)}_{\kappa \rho \kappa' \chi} \hat{C}^{\kappa'}_{\chi \lambda, \beta \gamma}.
\end{equation}
Recall that $Z^*_{\kappa, \alpha \rho}$ is the Born effective charge tensor of specie
$\kappa$, symmetric in the Cartesian indices $\alpha \rho$, and the 
mass-compensated force-response tensor $\hat{C}^{\kappa'}_{\chi \lambda, \beta \gamma}$
consists of three parts,
\begin{equation}
\hat{C}^{\kappa}_{\alpha \lambda, \beta \gamma}= \bar{C}^{\kappa}_{\alpha \lambda, \beta \gamma}
+ (\alpha \lambda, \beta \gamma)^{\kappa} - \frac{\Omega m_\kappa}{M} \mathcal{C}_{\alpha \lambda, \beta \gamma},
\label{mulatt}
\end{equation}

The mass-dependent contribution [third term on the right-hand side of 
Eq.~(\ref{mulatt})] is trivially
proportional to the elastic tensor, $\mathcal{C}_{\alpha \lambda, \beta \gamma}$,
and therefore uninteresting from the point of view of a microscopic analysis 
(see Section~\ref{longw} for details on the physical implications of this
term).

The contribution that depends on $\bar{C}^{\kappa}_{\alpha \lambda, \beta \gamma}$
is present on all compound crystals where the Born effective charges do not 
vanish, including simple rocksalt insulators such as MgO or NaCl.
The interpretation of this term is fairly simple. Consider, for example, a
rocksalt crystal with a longitudinal strain gradient along the (100) axis.
Each atomic plane will ``see'' a broken symmetry environment, with the two nearest 
neighboring planes located at slightly different distances. This, in turn, will 
produce (via the interatomic force constants) inequivalent longitudinal 
displacements of the two sublattices, and hence a macroscopic polarization 
oriented along (100).

The contribution depending on the round bracket can be readily understood
by recalling the explicit expression of the latter,
\begin{equation}
(\alpha \lambda, \beta \gamma)^{\kappa} = \Phi^{(1,\lambda)}_{\kappa \alpha, \kappa' \rho} 
\Gamma^{\kappa'}_{\rho \beta \gamma}.
\label{roundb}
\end{equation}
This contribution is nonzero only in compound crystals 
where $\Gamma^{\kappa'}_{\rho \beta \gamma} \neq 0$, i.e. 
(at least partially) ionic materials that undergo internal cell relaxations under a 
uniform applied strain.
(We shall exclude piezoelectric crystals for the time being, and restrict the
analysis to cases where such internal relaxations are nonpolar in character.)
Important members of this category are all centrosymmetric perovskite oxides
that are characterized by antiferrodistortive (AFD) tilts of the oxygen octahedral
network, e.g. SrTiO$_3$.
In these materials the local amplitude of the AFD order 
parameter linearly depends on strain, an effect that is known in
the literature as ``rotostriction''~\cite{morozovska-12a,morozovska-12b}.
A strain gradient can thus produce a gradient in the AFD order parameter, 
which in turn couples to the zone-center optical modes and generates a polarization.
Eq.~(\ref{roundb}) describes such a coupling in terms of the interatomic 
force constants of the crystal.
%[in a way that is described by
%This is In this context, simply states that 
%
Note that this mechanism was first proposed in the context of phenomenological 
theories~\cite{morozovska-12a,morozovska-12b} to explain the puzzling 
behavior~\cite{Pavlo} of the flexoelectric response of SrTiO$_3$ below its AFD 
transition temperature (105 K). 
%Recent experiments have revealed a 
% , and phenomenological studies have proposed an 
%explanation in terms of 
%

\subsection{``Mixed'' contribution}

The mixed response involves the first moment of the polarization response
to atomic displacements and the internal-strain response
tensor $\Gamma^\kappa_{\rho \beta \gamma}$,
\begin{equation}
\mu^{\rm II,mix}_{\alpha \lambda,\beta \gamma } =  
- P_{\alpha, \kappa \rho}^{(1,\lambda)} \Gamma^\kappa_{\rho \beta \gamma}.
\end{equation}
This term is nonzero only in crystals that display internal relaxations
under uniform strain, analogously to the lattice-mediated contribution depending
on the round brackets.
On the other hand, unlike the latter term, $\mu^{\rm II,mix}_{\alpha \lambda,\beta \gamma }$
does not involve the Born effective charges, and therefore can be (in principle)
present even in covalent crystals.
In fact, in diamond-structure semiconductors such as Si or Ge 
$\mu^{\rm II,mix}_{\alpha \lambda,\beta \gamma }$ was already
predicted and calculated from first-principles~\cite{Resta-DP}
in the framework of the theory of absolute deformation potentials
(which can be regarded as a precursor to the present theory of
flexoelectricity).
The effect, governed by the quadrupolar charge-density response to the
Raman-active optical mode of the diamond lattice (the latter responds
linearly to a shear strain), was found to be important in
both Si and Ge, giving a contribution that largely dominates the
(purely electronic) octupolar response~\cite{Resta-DP}.

%By symmetry, only the elements of the type 
%$\Gamma^\kappa_{x y z}$ (displacement of an atom along $x$ induced by
%a shear strain in the $yz$ plane) and  $P_{x, \kappa y}^{(1,z)}$ are 
%different from zero.
%
%As the crystal is not piezoelectric and there are two identical atoms
%related by symmetry, it must be that $P_{x, {\rm A} y}^{(1,z)} = -P_{x, {\rm B} y}^{(1,z)}$ 
%and $\Gamma^{\rm A}_{x y z} = -\Gamma^{\rm B}_{x y z}$.
%
%Finally, by using the relationship between charge and polarization moments,
%we have $P_{x, {\rm A} y}^{(1,z)} = Q_{{\rm A}y}^{(2,xz)}/2$. 
%
%Thus, this particular kind of flexoelectric response is governed by the 
%\emph{quadrupolar} moments of the electronic charge density that responds to 
%the displacement of an individual atomic sublattice, and by the \emph{piezoelectric}-like 
%internal-strain response to shear.
%
%This contribution is, strictly speaking, neither purely electronic in 
%origin, neither lattice-mediated in the usual sense, but an ``intermediate''
%type of coupling.
%
%Remarkably, that such an effect exists was discovered long before the
%recent surge of interest in flexoelectricity, in the context of the
%theory of deformation potentials~\cite{Resta-DP}.
%
%The quadrupolar-internal strain coupling 

\subsection{Piezoelectric materials}

Piezoelectrically active crystals deserve a separate discussion:
As we shall see in the following, they present an ambiguity in the 
definition of the flexoelectric tensor.
To see why this is the case, we rewrite $\mu^{\rm II}_{\alpha \lambda,\beta \gamma }$
by grouping the individual contributions in a slightly different way,
\begin{equation}
\mu^{\rm II}_{\alpha \lambda,\beta \gamma } = \tilde{\mu}^{\rm II}_{\alpha \lambda,\beta \gamma } +
e^\kappa_{\alpha \lambda \rho} \Gamma^\kappa_{\rho \beta \gamma},
\end{equation}
where in $\tilde{\mu}^{\rm II}_{\alpha \lambda,\beta \gamma }$ we have collected all 
contributions that do not depend on $\Gamma^\kappa_{\rho \beta \gamma}$, and
%that we have discussed in the context of the high-symmetry ionic insulators
%(electronic plus the lattice-mediated part due to the square brackets), and
\begin{equation}
e^\kappa_{\alpha \lambda \rho} = \frac{Z^*_{\kappa', \alpha \zeta}}{\Omega} \tilde{\Phi}^{(0)}_{\kappa' \zeta, \kappa'' \chi} 
\Phi^{(1,\lambda)}_{\kappa'' \chi, \kappa \rho} - P_{\alpha, \kappa \rho}^{(1,\lambda)}.
\end{equation}
The notation $e^\kappa_{\alpha \lambda \rho}$ is motivated by the
relationship
\begin{equation}
e_{\alpha \lambda \rho} = \sum_\kappa e^\kappa_{\alpha \lambda \rho},
\end{equation}
i.e. $e^\kappa_{\alpha \lambda \rho}$ can be thought as a ``sublattice-resolved 
piezoelectric coefficient''.
%
%(Note that, unlike its basis sum, the tensor $e^\kappa_{\alpha \lambda \rho}$
%is generally not invariant upon exchange of the last two Cartesian indices.
%For this same reason, $P_{\alpha, \kappa \rho}^{(1,\lambda)}$ cannot be expressed
%in terms of the quadrupolar moments of the induced charge, except in 
%high-symmetry crystals such as Si or Ge.)
%
%
%\subsubsection{Piezoelectric and ferroelectric materials}
%
%
%From the point of view of fundamental theory, the first point to be addressed 
%concerns the formal definition of the flexoelectric coefficient in a 
%piezoelectrically active material.
%
%A uniform strain gradient induces the following polarization,
%\begin{equation}
%\Delta \bar{P}_\alpha = \left( \tilde{\mu}^{\rm II}_{\alpha \lambda,\beta \gamma } +
%e_{\alpha \beta \gamma} r_\lambda +
%e^\kappa_{\alpha \lambda \rho} \Gamma^\kappa_{\rho \beta \gamma}  \right)
%\varepsilon_{\beta \gamma, \lambda},
%\end{equation}
%
%i.e. the nonvanishing piezoelectric tensor, $e_{\alpha \beta \gamma}$, introduces
%a polarization that increases linearly along the strain gradient axis, $\lambda$.
%
%At first sight this does not appear to be a major formal issue, as the two
%effects (piezoelectric and flexoelectric) may well coexist in the same phase
%and be individually well defined.
%
Now, recall the translational arbitrariness in the definition of the
internal-strain response tensor: $\Gamma^\kappa_{\rho \beta \gamma}$ is 
specified only modulo a $\kappa$-independent constant.
In nonpiezoelectric crystals such arbitrariness is harmless, as the 
basis sum of the $e^\kappa_{\alpha \lambda \rho}$ tensors vanishes.
Here, on the other hand, we have a clear ambiguity,
\begin{equation}
\Delta \mu^{\rm II}_{\alpha \lambda,\beta \gamma} = e_{\alpha \lambda \rho} \, \Delta \Gamma_{\rho \beta \gamma},
\end{equation}
where $\Delta \Gamma_{\rho \beta \gamma}$, of the dimension of a length,
reflects the aforementioned arbitrariness.

In a hand-waving way, one can relate this ambiguity to the difficulty of
calculating the dipole moment of a charged object -- the answer depends on
the choice of the origin.
Indeed, as we have argued in Section~\ref{secmartin}, strain gradients
may generate a net charge in a piezoelectric material, and the polarization 
is (loosely speaking) a dipole moment per unit volume.
It can be verified that, in presence of a noncentrosymmetric 
$\Gamma^\kappa_{\rho \beta \gamma}$
and a strain gradient, the precise point in the crystal where the local strain 
vanishes is not well defined, hence the origin dependence and the
ambiguity in the definition of $\mu^{\rm II}_{\alpha \lambda,\beta \gamma}$.

\subsection{Dependence on the static dielectric constant}

It has been pointed out, both in the context of experiments~\cite{Cross}
and phenomenological models~\cite{Tagantsev} that the flexoelectric
coefficients should be roughly proportional to the static dielectric 
constant of the material.
In the following we shall briefly comment on this statement in light of 
the formalism presented here.
To this end, it is useful to express the matrix $\widetilde{\Phi}^{(0)}$ as
\begin{equation}
\widetilde{\Phi}^{(0)}_{\kappa \alpha \kappa' \beta} = \frac{1}{\sqrt{m_\kappa}} 
\sum_{n}{}' \, \,
\frac{  \xi^n_{\kappa \alpha} \, \xi^n_{\kappa' \beta} }{\omega_n^2  }
\frac{1}{\sqrt{m_{\kappa'}}},
\label{phi0inv}
\end{equation}
where $\xi^n$ and $\omega^2_n$ are the eigenvectors and eigenvalues of the 
zone-center dynamical matrix,
$$%\begin{equation}
D^{(0)}_{\kappa \alpha \kappa' \beta} = \frac{1}{ \sqrt{m_\kappa m_{\kappa'} } } 
\Phi^{(0)}_{\kappa \alpha \kappa' \beta} = \sum_n \xi^n_{\kappa \alpha} \, \omega_n^2 \,  \xi^n_{\kappa' \beta} .
$$%\end{equation}
[Note the primed sum in Eq.~(\ref{phi0inv}), indicating that the zero-frequency
rigid translations are excluded.]
Using Eq.~(\ref{phi0inv}) we can rewrite the lattice-mediated flexoelectric 
tensor as
\begin{equation}
\mu^{\rm II,latt}_{\alpha \lambda,\beta \gamma} = \frac{1}{\Omega M_0} \sum_n{}' \, \, 
\frac{ Z^*_{\alpha n} \, \hat{C}_{n \lambda, \beta \gamma}}{\omega^2_n} ,
\end{equation}
where we have introduced the dynamical charge associated to the $n$-th mode~\cite{Antons:2005} ($M_0$
is an arbitrary mass constant),
$$%\begin{equation}
Z^*_{\alpha n} = \sum_{\kappa \rho}  Z^*_{\kappa,\alpha \rho} \sqrt{\frac{M_0 }{m_\kappa} } \xi^n_{\kappa \rho},
$$%\end{equation}
and the projection of the flexoelectric force-response tensor on the $n$-th mode eigenvector,
$$%\begin{equation}
\hat{C}_{n \lambda, \beta \gamma} = \sum_{\kappa \rho} \xi^n_{\kappa \rho} \sqrt{\frac{M_0 }{m_\kappa} }
\hat{C}^\kappa_{\rho \lambda, \beta \gamma}.
$$%\end{equation}
$\hat{C}_{n \lambda, \beta \gamma}$ describes the coupling between the strain gradient and
the individual zone-center optical modes; it can be thought, therefore, as a ``geometric field''
pushing the polar phonons out of their centrosymmetric equilibrium configuration in presence
of an inhomogeneous deformation.
The inverse frequency squared,  acts, as usual, as a restoring force, while $Z^*_{\alpha n}$
describe the polar activity of the phonon mode.
%As a closing remark, note that our theory is fully consistent with 
%the known relationship between the flexoelectric response and the 
%static dielectric constant of the material~\cite{Ma/Cross}.
%
%Indeed, our formula for the lattice-mediated response contains the inverse of the 
%zone-center dynamical matrix, $\tilde{\Phi}^{(0)}$, and the dynamical charges, $Z^*$, 
%as prefactors. 
%
%
%Hence, 
%if we assume~\cite{Tagantsev} that the lattice contribution dominates
%the effect, 
%Thus, the lattice-mediated flexoelectric response proceeds via the 
%transverse optical phonons at the $\Gamma$ point, whose individual 
%contribution is proportional to the inverse of the squared frequency 
%multiplied by the polarity, and further weighted by the force-response
%tensor.
%
This implies that materials with dielectrically ``soft'' optical 
modes are most likely to produce a large response, consistent with 
the experimental observations~\cite{Cross} and the conclusions of earlier 
phenomenological models.~\cite{Tagantsev}

%can be expressed in terms of the $\Gamma$-point transversal 

%$\bar{\mu}^{\rm I}_{12,12}$
%and $\bar{\mu}^{\rm I}_{12,12}$
% both $\bar{\mu}^{\rm I}$ and the square 
%brackets (and hence the overall flexoelectric tensor) have only 

\subsection{Relationship to the theory of deformation potentials}

In Section~\ref{reference} we have shown that there is
an ambiguity in the definition of the flexoelectric tensor, 
which can be traced back to the choice of an arbitrary reference
energy when imposing short-circuit electrical boundary conditions.
To make these considerations more quantitative, suppose that we 
choose a single-particle eigenvalue, $\epsilon_{n{\bf k}}$
($n$ is a band index and ${\bf k}$ is the crystal momentum; assume that the
eigenvalue is nondegenerate), as a reference for the flat-band condition. 
Then, the new flexoelectric 
tensor acquires an additional contribution (compared to the ``standard''
definition, based on the average electrostatic potential $\overline{V}$) 
that can be readily written in terms of bulk material properties,
\begin{equation}
\Delta \mu^{\rm II}_{\alpha \lambda, \beta \gamma} = % \frac{1}{4 \pi} 
\chi^{\rm st}_{\alpha \lambda} 
\frac{\partial V_{n{\bf k}} }{\partial \varepsilon_{\beta \gamma}},
\label{deltamu}
\end{equation} 
where $V_{n{\bf k}} = \epsilon_{n{\bf k}} / e + \overline{V}$
is the relative potential of $\epsilon_{n{\bf k}}$ with respect to
$\overline{V}$ (the quantity 
$\partial V_{n{\bf k}} / \partial \varepsilon_{\beta \gamma}$
is known as \emph{relative deformation potential}), and 
$$\chi^{\rm st}_{\alpha \lambda} = 
\frac{\epsilon^{\rm st}_{\alpha \lambda} - \delta_{\alpha \lambda}}{4 \pi}$$
is the static dielectric susceptibility of the material.
The physics behind Eq.~\eqref{deltamu} is easily understood: The relative
deformation potential induces an additional electric field, which is directed
along $\lambda$ (i.e. where the macroscopic strain undergoes a linear variation).
This electric field, in turn, induces a polarization via $\chi^{\rm st}_{\alpha \lambda}$.

Such an arbitrariness in $\bm{\mu}$ might seem disturbing at first sight, 
as it threatens the 
%as it threatens 
very interpretation of the flexoelectric tensor as a well-defined 
bulk property.
%
%Although in this paper we decided to focus mostly on bulk properties,
We shall see in the following that there is no such danger, and that
the arbitrariness always cancels out when $\bm{\mu}$ is 
applied to real physical problems.
Consider the case of a slab that is 
centrosymmetric at rest, and suppose to bend it; we have a macroscopic 
strain gradient that increases linearly along the direction perpendicular 
to the slab surfaces.
%(as in a bending experiment).
%
On general physical grounds, the total dipole moment induced by the
deformation must be a well-defined physical observable. How do we reconcile 
this fact with the aforementioned arbitrariness in the bulk $\bm{\mu}$-tensor?
Answers to this question must be looked for at the surface -- the only 
feature in the system which is neither bulk or vacuum. Hong and 
Vanderbilt~\cite{Hong-11} indeed observed that there are surface-specific 
contributions to the flexoelectric polarization of the slab, and that
these consist in the derivative of the surface work function with respect 
to a \emph{uniform} strain, $\partial \phi / \partial \varepsilon_{\beta \gamma}$.
The key point here is that the surface work function, $\phi$, \emph{suffers from the same
ambiguity as the flexoelectric tensor}: to define a surface potential offset one
needs to choose a reference energy inside the material. It is easy then to verify that the
arbitrariness exactly cancels out when calculating the total dipole moment of the
bent slab.
(Note that, in a uniform strain gradient, the contribution of the slab 
surface to the total dipole moment scales proportionally to the slab 
thickness.~\cite{Hong-11})
The message is that any attempt at quantitatively comparing the calculated 
bulk $\bm{\mu}$-tensor with experimental measurements is necessarily thwarted
by the inherent arbitrariness of the former; such ambiguity disappears
only when the surface-specific part is accounted for.

%hese do not vanish in 
%the thermodynamic limit, and that 
%
%In particular, these surface contributions 
%
%It is obvious that the work function is not unique, as it depends on
%the choice of a reference energy in the bulk (via a band-structure term).
%
%As the total dipole that is induced, e.g., by a bending deformation of the 
%slab must be a well-defined quantity, 
%independent of such an arbitrary 
%reference, 
%it is natural for the bulk flexoelectric tensor to have a similar
%ambiguity as the surface term, in such a way that the arbitrariness 
%cancels out when the two contributions are summed up.
%
%This does not mean that the bulk flexoelectric tensor is not well defined;
%it just means that the flexoelectric response of a finite object is
%ill-defined at the bulk level, as additional details on the surface
%band offset response are required in order to determine it.
%
%Throughout this work we have chosen the macroscopic electrostatic potential
%as a reference, both for simplicity and for continuity with earlier literature 
%works.
%
%It should be kept in mind, however, that such a choice is not unique,
%and its potential impact should be quantitatively assessed when 
%comparing the first-principles results to the experimental measurements.

As the present work is exclusively concerned with macroscopic bulk effects,
we will not pursue the discussion of surface-related issues any further here.
Instead, in the remainder of this Section we shall discuss a genuine
bulk phenomenon where the above arbitrariness is potentially worrisome,
i.e. the flexoelectric response induced by a sound wave.
Obviously, in an acoustic phonon there are no surfaces: the bulk flexoelectric 
tensor (possibly combined with other bulk properties) must give a complete
description of the electrical response at order $\mathcal{O}(q^2)$.
Furthermore, such description must be unambiguous: the charge density and 
lattice response to a phonon is uniquely determined by a set of well-defined 
quantum numbers. It is, therefore, necessary to prove that the arbitrariness
%;
%therefore, it is important to verify that the arbitrariness in $\mu$ does 
%not affect the 
$\Delta \bm{\mu}$ causes no harm in this respect. 

Consider a long-wave acoustic
phonon of wavevector ${\bf q}$. The strain-gradient tensor is given by
$\eta^{\bf q}_{\beta, \gamma \delta} = -U_\beta q_\gamma q_\delta$,
where $U_\beta$ is an eigenvector of the sound-wave equation, Eq.~\eqref{acoustic}.
The macroscopic polarization associated to such a strain gradient would be, in 
short-circuit boundary conditions,
\begin{equation}
P^{\rm SC}_\alpha = \mu^{\rm II}_{\alpha \lambda,\beta \gamma} 
\varepsilon^{\bf q}_{\beta \gamma,\lambda},
\label{psc}
\end{equation}
where $\varepsilon^{\bf q}_{\beta \gamma,\lambda}$ is related to 
$\eta^{\bf q}_{\beta, \gamma \delta}$ via Eq.~\eqref{eqeta2}.
In an acoustic phonon, however, the electrical boundary conditions (EBC)
differ from the short-circuit (SC) fixed-$\mathcal{E}$ EBC that are implicit 
in Eq.~\eqref{psc}, in that a zero electric displacement field, $D$,
is enforced along the propagation direction, $\hat{\bf q}={\bf q}/q$. 
Then, the total polarization reads
\begin{equation}
P_\alpha = P^{\rm SC}_\alpha + \chi^{\rm st}_{\alpha \lambda} \mathcal{E}_\lambda,
\label{eoc1}
\end{equation}
where $\mathcal{E}_\lambda$ is the longitudinal electric field 
that arises from the open-circuit, or fixed-$D$, EBC,
\begin{equation}
\bm{\mathcal{E}} = -4 \pi \,\hat{\bf q} 
\frac{ \hat{\bf q} \cdot {\bf P}^{\rm SC}}{ \hat{\bf q}\cdot \bm{\epsilon}^{\rm st} \cdot \hat{\bf q}}.
\label{eoc2}
\end{equation}
It follows by combining Eq.~\eqref{eoc1} and Eq.~\eqref{eoc2} that the longitudinal 
component of the total polarization reads
\begin{equation}
\hat{\bf q} \cdot {\bf P} = \frac{ \hat{\bf q} \cdot {\bf P}^{\rm SC}}
{\hat{\bf q}\cdot \bm{\epsilon}^{\rm st} \cdot \hat{\bf q}},
\end{equation}
confirming that the longitudinal component of the displacement field 
${\bf D} = \bm{\mathcal{E}} + 4 \pi {\bf P}$ indeed vanishes.
[Note that in general, for an anisotropic dielectric, the open-circuit electric
field in Eq.~\eqref{eoc1} induces components of ${\bf P}$ that are also directed 
\emph{perpendicular} to the propagation direction; this observation is important
for what follows.]

What happens now if we choose a different reference for the definition of the
flexoelectric tensor? The induced short-circuit 
polarization acquires a term due to 
$\Delta \mu^{\rm II}_{\alpha \lambda, \beta \gamma}$, as defined in
Eq.~\eqref{deltamu}.
Along the longitudinal direction, $\hat{\bf q}$, it is easy to verify that 
the only consequence of $\Delta \bm{\mu}^{\rm II}$ is a \emph{redefinition} of 
the macroscopic electric field,
\begin{equation}
\mathcal{E}^{n{\bf k}}_\lambda = 
\mathcal{E}_\lambda - 
\frac{\partial V_{n{\bf k}} }{\partial \varepsilon_{\beta \gamma}}
\varepsilon^{\bf q}_{\beta \gamma,\lambda}.
\label{adp}
\end{equation}
This result can be interpreted as follows. The acoustic phonon perturbation
breaks the periodicity of the lattice along the propagation direction. Therefore,
along $\hat{\bf q}$ one no longer expects the band energies (or other
references such as the average electrostatic potential) to be parallel to each 
other. Instead, their relative position varies along $\hat{\bf q}$, proportionally
to the local strain, via the band-structure term 
$\partial V_{n{\bf k}} / \partial \varepsilon_{\beta \gamma}$.
Nevertheless, the \emph{absolute} variation of each individual band, 
$\mathcal{E}^{n{\bf k}}_\lambda$, is a well-defined
bulk property, and is exactly given by the electrostatic contribution $\mathcal{E}_\lambda$ 
plus the corresponding band-structure part, according to Eq.~\eqref{adp}.
Thus, we formally identify the quantity $\mathcal{E}^{n{\bf k}}_\lambda$ 
as an \emph{absolute deformation potential}~\cite{Resta-DP}
(ADP), hereby extending the scopes of the ADP theory to a general 
nonpiezoelectric crystal lattice, including
ionic solids (the latter were excluded from the analysis of Ref.~\onlinecite{Resta-DP}).

The only remaining task in this Section is to prove that the polarization 
along the direction normal to ${\bf q}$ is unaffected by the arbitrariness 
$\Delta \bm{\mu}$.
Along the direction normal to the propagation direction (which we shall indicate 
as a vector $\hat{\bf w}$ such that $\hat{\bf w} \cdot \hat{\bf w}=1$ and
$\hat{\bf w} \cdot \hat{\bf q}=0$), the short-circuit 
polarization acquires a term due to $\Delta \bm{\mu}^{\rm II}$,
\begin{equation}
\hat{\bf w} \cdot \Delta {\bf P}^{\rm SC} = \hat{w}_\alpha \chi^{\rm st}_{\alpha \lambda} 
\frac{\partial V_{n{\bf k}} }{\partial \varepsilon_{\beta \gamma}} 
\varepsilon^{\bf q}_{\beta \gamma,\lambda}.
\end{equation}
Upon inspection of Eq.~\eqref{adp} it is straightforward to verify that the
additional contribution to the open-circuit field, $\mathcal{E}^{n{\bf k}}_\lambda-\mathcal{E}_\lambda$,
produces a polarization [via Eq.~\eqref{eoc1}] that exactly cancels
$\hat{\bf w} \cdot \Delta {\bf P}^{\rm SC}$. 
This demonstrates that $\hat{\bf w} \cdot {\bf P}$, unlike $\hat{\bf w} \cdot {\bf P}^{\rm SC}$, 
is well defined and independent of the arbitrary reference used to define the flexoelectric tensor,
concluding our proof.

\section{Conclusions}

\label{concl}

We have performed a rigorous derivation of the full flexoelectric tensor
in an arbitrary crystalline insulator. 
Based on this result, we have discussed a number of topics relevant for
the physics of the flexoelectric effect, in particular concerning the 
electrical boundary conditions, the relationship between the static
and dynamic response, and the microscopic mechanisms that may be at
play in a variety of materials classes. 

We expect this work to open several exciting avenues for future research.
From the materials design point of view, the first priority is to apply 
the present method to perform first-principles calculations of real materials, 
and understand what mechanisms are most promising for delivering a large 
response.
(At the time of writing, we are aware of an
independent work~\cite{Hong-13} where the full flexoelectric properties of
several cubic materials were calculated from first principles,
by means of methodologies that are similar to those developed here.)
On the methodological front, it will be interesting to work out the 
analytic derivation of the Sternheimer equation in the long-wave
limit, and hence avoid the finite-difference derivation of the
response quantities.
Also, achieving a first-principles implementation of the quantum-mechanical 
current density operator would be desirable, in order to access the full 
microscopic polarization response functions.
Preliminary work along these directions is under way.

\section*{Acknowledgments}

%I gratefully acknowledge Raffaele Resta and David Vanderbilt 

This work was supported 
by DGI-Spain through Grants No. MAT2010-18113 and No. CSD2007-00041.
% (MS); 
%by the European Commission through the project EC-FP7,
%Grant No. NMP3-SL-2009-228989 ``OxIDes'' (PhG and MS); and by
%the Francqui Foundation through a Research Professorship (PhG).
%
%We thankfully acknowledge the computer resources, 
%technical expertise and assistance provided by the 
%Red Espa\~nola de Supercomputaci\'on (RES) and by the 
%Supercomputing Center of Galicia (CESGA).

\appendix*

\bibliography{flexo,max-feb28}

\end{document}